\documentclass[useAMS,usenatbib]{mn2e}

\usepackage{epsfig}
\usepackage{amsmath}
\usepackage{color}
\usepackage{ulem}
\usepackage{amssymb}    

\newcommand{\Msol}{\rm\,M_{\odot}}

\newcommand{\Mhalo}{\rm\,M_{halo}}
\newcommand{\rh}{\rm\,r_{h}}
\newcommand{\Gyr}{\rm\,Gyr}

\newcommand{\Mpc}{\rm\,Mpc}

\newcommand{\kms}{\rm\,km\,s^{-1}}
\newcommand{\Mr}{\rm\,M_{r}}
\newcommand{\fr}{\rm\,f_{red}}
\newcommand{\dfr}{\rm\,\Delta(f_{red})}
\newcommand{\mur}{\rm\,\mu_{red}}
\newcommand{\dmur}{\rm\,\Delta(\mu_{red})}
\newcommand{\mub}{\rm\,\mu_{blue}}
\newcommand{\dmub}{\rm\,\Delta(\mu_{blue})}
\newcommand{\sigmar}{\rm\,\sigma_{red}}
\newcommand{\dsigmar}{\rm\,\Delta(\sigma_{red})}
\newcommand{\sigmab}{\rm\,\sigma_{blue}}
\newcommand{\dsigmab}{\rm\,\Delta(\sigma_{blue})}
\newcommand{\dmu}{\rm\,\Delta(\mu)}
\newcommand{\dsigma}{\rm\,\Delta(\sigma)}
\newcommand{\driro}{\rm\,\Sigma_{r_i,r_o}}
\newcommand{\dsca}{\rm\,\Sigma_{0,0.5}}
\newcommand{\dscb}{\rm\,\Sigma_{0,1}}
\newcommand{\dscc}{\rm\,\Sigma_{0.5,1}}
\newcommand{\dscd}{\rm\,\Sigma_{1,2}}
\newcommand{\dsce}{\rm\,\Sigma_{2,3}}

\newcommand{\dscj}{\rm\,\Sigma_{0,2}}
\newcommand{\Nnriro}{\rm\,N^{n}_{r_i,r_o}}
\newcommand{\Csriro}{\rm\,C^{s}_{r_i,r_o}}
\newcommand{\Nunriro}{\rm\,N^{un}_{r_i,r_o}}
\newcommand{\Nuni}{\rm\,N^{un}_{0,3}}
\newcommand{\Npriro}{\rm\,N^{p}_{r_i,r_o}}
\newcommand{\Nsriro}{\rm\,N^{s}_{r_i,r_o}}
\newcommand{\Csca}{\rm\,C^{s}_{0,0.5}}
\newcommand{\Cscb}{\rm\,C^{s}_{0,1}}
\newcommand{\Csci}{\rm\,C^{s}_{0,3}}

\newcommand{\Pl}{\rm\,P_{L}}
\newcommand{\rdiv}{\rm\,r_{div}}

\title[Multiscale environment and galaxy colours]
{A multiscale approach to environment and its influence on the colour distribution of galaxies}

\author[D.~J.~Wilman, S.~Zibetti et al.]
{D.~J.~Wilman$^{1,4}$,S.~Zibetti$^{2}$,T.~Budav\'ari$^{3}$\\
$^1$Max-Planck-Institut f\"ur Extraterrestrische Physik, Giessenbachstra\ss e, D-85748 Garching, Germany.\\
$^2$Max-Planck-Institut f\"ur Astronomie, K\"onigstuhl 17, D-69117 Heidelberg, Germany.\\
$^3$Department of Physics and Astronomy, The Johns Hopkins University, 3701 San Martin Drive, Baltimore, MD 21218, USA\\
$^4$email: dwilman@mpe.mpg.de}

\begin{document}

\maketitle

\begin{abstract}
  We present a multiscale approach to measurements of galaxy density,
  applied to a volume-limited sample constructed from the Sloan
  Digital Sky Survey Data Release 5 (SDSS DR5). We populate a rich
  parameter space by obtaining independent measurements of density on
  different scales for each galaxy. This parameterization purely
  relies on observations and avoids the implicit assumptions involved,
  e.g., in the construction of group catalogues. As the first
  application of this method, we study how the bimodality
  in galaxy colour distribution (specifically $u-r$) depends on
  multiscale density.  The $u-r$ galaxy colour distribution is
  described as the sum of two gaussians (a red and a blue one) with
  five parameters: the fraction of red galaxies ($\fr$) and the
  position and width of the red and blue peaks ($\mur$, $\mub$,
  $\sigmar$ and $\sigmab$). Galaxies mostly react to their smallest
  scale ($<0.5\Mpc$) environments: in denser environments red galaxies
  are more common (larger $\fr$), redder (larger $\mur$) and with a
  narrower distribution (smaller $\sigmar$), while blue galaxies are
  redder (larger $\mub$) but with a broader distribution (larger
  $\sigmab$).  There are residual correlations of $\fr$ and $\mub$
  with $0.5 - 1\Mpc$ scale density, which imply that total or partial
  truncation of star formation can relate to a galaxy's environment on
  these scales.  Beyond $1\Mpc$ ($0.5\Mpc$ for $\mur$) there are no
  positive correlations with density.  However $\fr$ {\it
    anti-correlates} with density on $>2\Mpc$ scales at fixed density
  on smaller scales, and $\mur$ anti-correlates with density on
  $>1\Mpc$ scales.  We examine these trends qualitatively in the
  context of the halo model, utilizing the properties of haloes within
  which the galaxies are embedded, derived by \citet{Yang07} and
  applied to a group catalogue.  This yields an excellent description
  of the trends with multiscale density, including the
  anti-correlations on large scales, which map the region of accretion
  onto massive haloes.  Thus we conclude that galaxies become red only
  once they have been accreted onto haloes of a certain mass.  The
  mean colour of red galaxies $\mur$ depends positively only on
  $<0.5\Mpc$ scale density, which can most easily be explained if
  correlations of $\mur$ with environment are driven by metallicity
  via the enrichment history of a galaxy within its subhalo, during
  its epoch of star formation.

\end{abstract}

\begin{keywords}
methods: statistical - galaxies: evolution - galaxies: haloes - galaxies: statistics - galaxies: fundamental parameters - galaxies - stellar content
\end{keywords}

\section{Introduction}\label{sec:intro}

\subsection{Motivation}

The evolution of galaxies is heavily intertwined with the growth of
the dark matter dominated structure in which they live, as baryons
react to their local gravitational potential.  It is therefore one of
the most challenging goals of observational astronomy to explain the
continuously evolving galaxy population in the context of a
hierarchical universe.  Most fundamentally perhaps, galaxies exhibit a
distinct bimodality of properties which correlates strongly with the
properties of the embedding potential.  This is true whether the
potential is measured on galaxy scales via properties such as
luminosity, stellar or dynamical mass, or on larger scales via
measurement of the galaxies' local environment.  In particular,
galaxies living in deeper potential wells are more likely to have
formed stars at earlier times, both in terms of the average stellar
age \citep[e.g.][]{Thomas05,Smith06} and the last episode of star
formation \citep[and thus a lower fraction are continuing to form
stars, e.g.][]{Lewis02,Baldry04,Balogh04Ha,Balogh04,Kauffmann04}.
Such galaxies are apparently unable to restart star formation, since
they have no stable access to a suitable reservoir of gas which can
cool to form a star-forming disk \citep[e.g.][]{Gallagher75,Caon2000}.
Simultaneously the morphology of these galaxies must evolve from the
typical rotation-dominated spiral or irregular morphologies of
star-forming galaxies to form a pressure dominated elliptical, or a
bulge plus smooth disk lenticular
\citep[e.g.][]{Dressler80,Dressler97,Wilman09}.  The precise nature of
these transformations remains controversial, despite evidence from
simulations and observations for various mechanisms which can explain
some or all of these observations.  Galaxy mergers are especially
attractive in the context of a hierarchical Universe, and can explain
both morphological changes and the rapid exhaustion of cold gas,
although additional physics may be required to keep a galaxy from
reforming a gas disk \citep[e.g.][]{Springel05,Hopkins09,Johansson09}.
Stripping or evaporation of cold and/or hot gas associated to an
infalling galaxy will take place within a dense hot intra-cluster
medium \citep[e.g.][]{GunnGott72,Chung09,LTC}, although it is unclear
whether these processes can be important within lower density
environments \citep{Kawata08,McCarthy08,Jeltema08}. Tidal interactions
between galaxies can drive gravitational instabilities within
galaxies, with gas loss or triggering of star formation as likely
consequences. Numerous fast interactions (harrassment) can also drive
morphological evolution \citep[e.g.][]{Moore99}.

To statistically describe the evolutionary path of galaxies, it is
necessary not only to take a firm grasp of the relevant physical
processes but also to place robust observational constraints on
precisely how the galaxy population reacts to its environment.  It has
been customary in this field to use neighbouring galaxies as test
particles in order to `measure' the local environment of each
galaxy. One computes a `local density' by simply counting neighbours
within some fixed radius and velocity range or, almost equivalently,
by computing the distance to the N$^{th}$ nearest neighbour, where N
is typically chosen to be 5 or 10 (\citet{Dressler80,Balogh04Ha}, see
\citet{Cooper05,Kovac09} for a discussion of these and more complex
methods).  As with so many traditional measures, this measurement is
fairly arbitrary in nature \footnote{N is typically selected to obtain
  sufficient S/N whilst retaining a ``local'' measurement.}, and it is
not easy to see precisely how it relates to the underlying density
field of galaxies and dark matter.

The two point correlation function or power spectrum of galaxies on
the other hand, measure the excess probability over random of finding
two galaxies with a given separation. The comparison of galaxy 
subsets selected by e.g. colour, provides a measure of the relative 
bias of these galaxy types as a function of the scale of pair 
separation, diagnosing the scale at which one class dominates 
relative to another.
\citep[e.g.][]{Norberg01,Budavari03,Zehavi05}.  Marked correlation
functions weight the galaxies in each pair by a given property, 
normalizing by the unweighted correlation function 
\citep[e.g.][]{Skibba06}. 
This examines how galaxies with large or small values of the chosen
property are clustered. 
The bias or marked correlation as a function of scale
are intimitely related to the details of how galaxy properties depend
on their environments.

These statistics simply provide the mean behaviour of galaxies by
tracing their overdensity on different scales which, however, are
themselves closely correlated with one another. A better approach
would be to examine how galaxy properties correlate with overdensity
on one scale for a fixed overdensity on another, independently
computed scale.  Now that such a large, homogeneous sample of galaxies
is available with the Sloan Digital Sky Survey (SDSS), there have been
a few attempts at isolating such behaviour on different scales.
\citet{Kauffmann04}, \citet{Blanton06} and \citet{Blanton07} conclude
that galaxies only react to their environments on $<1\Mpc$ scales,
with no significant residual dependence on larger scale overdensity.
This contradicts the apparent dependence on large scales found by
\citet{Balogh04Ha}, which may have resulted from low signal to noise
measurement of density on small scales \citep{Blanton06}. These works,
however, contained correlated measurements of density on both scales.

\subsection{A New Method}

In this paper we approach this problem in a different way.  Firstly,
we parameterize the environment using annular, non-overlapping
measurements of density on different scales to ensure that they are
(formally) independent. Secondly, we consider the $u-r$ colour of
galaxies and utilize the remarkable fact that its distribution is
bimodal and can be modelled as the sum of two gaussians.  This relates
to the bimodality of galaxy properties, since galaxies dominated by
old stellar populations have red $u-r$ colours and actively
star-forming galaxies are blue. \citet{Baldry04} initially
demonstrated the power of this technique, illustrating the low
chi-squared values produced with such a fit to $u-r$ SDSS colours, and
demonstrating that the luminosity dependence of galaxy colours was
strong and easily parameterized using this model.  \citet{Balogh04}
and \citet{Baldry06} have since extended this method, with a powerful
demonstration of the strong and fundamental dependence of the fraction
of red galaxies on environment (measured using the distance to the
$4^{th}$ and $5^{th}$ nearest neighbours) and stellar mass.  The
parameterization of the double gaussian fit (the fraction of red
galaxies $\fr$, and the position and width of each peak) provides five
quantities which can be interpreted in terms of changes in the
spectral energy distribution and the potential physical trigger, and
which can be correlated with environment measured on different scales.

\subsection{Expectations from the Halo Model}

Finally, to ease the physical interpretation of our results, we would
like to know how measurements of density map onto the dark matter
dominated `cosmic web'.  To this goal we rely on the models that
describe the clumpy distribution of dark matter in the Universe in
terms of haloes, i.e. the potential wells in which galaxies form,
infall, virialize and interact with one another and with the
intra-cluster/group medium (ICM/IGM) \citep[e.g.][]{White78, Cooray02,
  Diaferio01}.  The halo model (which applies on both galaxy and
group/cluster scales) is a simplification of reality, but has been
hugely successful in describing the dynamics and large scale structure
statistics of galaxies \citep[e.g.][]{Rubin83,Berlind03}.  In
particular, the two point correlation function appears to demonstrate
an inflection at $\sim2\Mpc$ which is attributed to the switch from
sub-halo scales (the one halo term) to halo-halo scales (the two halo
term) \citep[e.g.][]{Zehavi04,Zehavi05,Cooray06}.  Models of galaxy
formation and evolution pay homage to the halo model by assuming
galaxies to care about their environment only in terms of their
embedding dark matter halo \citep[e.g.][]{Kauffmann93,Cole00}.  This
puts a natural scale on the expected dependence of galaxy properties
on their environments: they should not care about scales beyond their
halo.  However simulations also show that the formation (collapse)
time of a halo depends upon the overdensity on larger scales
\citep[e.g.][]{Sheth04,Gao05}. In this case, one might expect galaxies
to have evolved further in a region of the Universe which is more
overdense on large scales.

\subsection{Structure of the paper}

Section~\ref{sec:sample} introduces the sample used in this paper,
selected from the SDSS Data release 5 (DR5).  In
Section~\ref{sec:density} we describe in detail the method used to
compute the density of galaxies on different scales, paying close
attention to the completeness corrections required to ensure a robust
measurement in the presence of incompleteness in SDSS DR5.

Multiscale density information for the 73662 galaxies of our
photometrically complete sample are available for public use on 
request to the authors.

The method of fitting a double gaussian model is described in
Section~\ref{sec:fitting}.  Our results are presented in
Section~\ref{sec:results}.  Firstly we examine the simple dependence
of the galaxy colour distribution on luminosity and density on a
single scale in Section~\ref{sec:deplumdens}.  Then, in
Section~\ref{resultsscale}, we present our main results in which
the importance of different scales are tested independently.  This
forms the purely observational part of the paper.

In Section~\ref{sec:interpretation} we utilize a catalogue of
embedding haloes (groups) which has been assigned to SDSS galaxies
within the context of a halo model and with a few implicit assumptions
\citep{Yang07}.  We examine the dependence of halo-based properties
such as mass and halo-centric radius on the density computed on
different scales and compare this to the way in which the galaxy
colour distribution depends upon the same measurements of density.
What emerges is a remarkably consistent picture, in which the halo
model is extremely successful at explaining the colours of galaxies:
smaller scale density can drive some aspects of colour evolution, but
star forming galaxies do not seem to feel the halo environment prior
to infall.  This exercise should provide the first, qualitative step
towards a fully quantative, model-independent description of the way
in which galaxies trace their environment.  Models can then be
constrained independently of the assumptions that are required in the
construction of group catalogues or modeling of the correlation
function. The rich parameter space of multiscale densities will help
overcoming degeneracies and difficulties in interpreting single
aperture measurements of local density.

In this paper we assume the cosmological parameters:
$(\Omega_M,\Lambda,h) = (0.3,0.7,0.75)$.

\section{Sample}\label{sec:sample}

We base our analysis on the Fifth Data Release \citep[][DR5]{SDSS_DR5}
of the Sloan Digital Sky Survey \citep[][SDSS]{SDSS}. We define an
initial sample of galaxies with a valid spectroscopic redshift
determination, which are part of the so-called `main galaxy sample'
(i.e. all extended sources with petrosian dereddened magnitude
brighter than 17.77 in $r$-band and mean surface brightness within
half-light radius $\mu_r\leq 23.0$~mag~arcsec$^{-2}$, see
\citealt{strauss_etal02}).  This sample covers 5713 square degrees.
We further restrict the selection to the redshift range $0.015\leq z$:
the lower limit is chosen such that local deviations from the Hubble
flow negligibly affect distance estimates\footnote{In doing so we also
  avoid photometric problems arising with galaxies of large apparent
  size.}.

For this sample we compute the two photometric quantities relevant to
the present work, namely the rest-frame de-reddened Petrosian $r$-band
(absolute) magnitude and the rest-frame model colour $u-r$.
De-reddening is based on the estimate of foreground Galactic
extinction derived by \citet{schlegel_dust}, while the transformation
to rest-frame quantities is performed using the software {\sc
  k-correct} \citep[v4.1.3][]{Blanton07kcor}, based on the five-band
$u,g,r,i,z$ spectral energy distribution \citep[see][for details on
the SDSS photometric system]{Fukugita96}. By choosing the petrosian
magnitude as a base and proxy for the total galaxy luminosity we
provide that the SDSS selection criteria simply propagate in our
sample selection and binning, and at the same time we avoid redshift
dependent biases in the estimate of the total luminosity (this derives
from the definition of petrosian magnitude, \citealt{petrosian76}). On
the other hand, as noted in \citet{EDR}, the so-called SDSS model
magnitudes are best suited to derive colours, as they are computed
with a weighted aperture photometry scheme common to all five bands,
which ensures results that are independent on the wildly different
signal-to-noise ratio in each band.  The choice of the $u-r$ colour
among all possible combinations of the five SDSS pass-bands appears
optimal to study galaxy bimodality \citep[as shown, e.g.,
in][]{Baldry04}: the physical reason is that $u-r$ provides a long
wavelength leverage (therefore it is very sensitive to the shape of
the optical SED, depending on star formation history, metallicity and
dust extinction) across the 4000\AA-break (which is the strongest
spectral feature at optical wavelengths and responds mainly to changes
in stellar population age).

\begin{figure}
   \centerline{\includegraphics[width=0.49\textwidth]{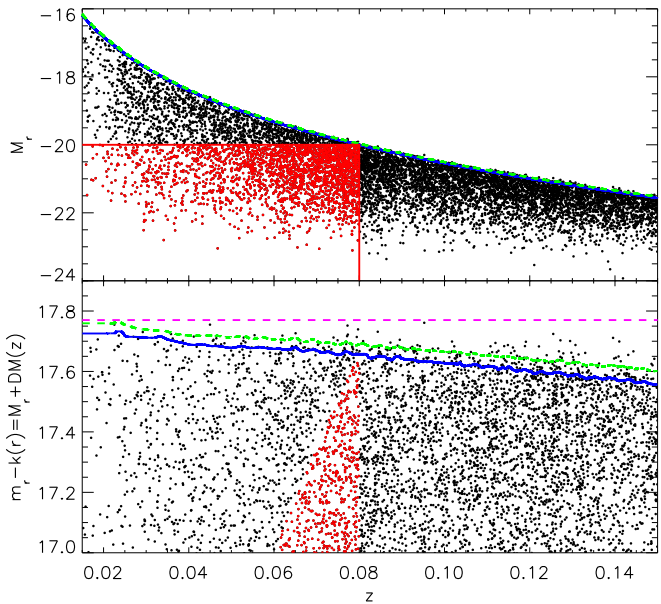}}
   \caption{Sample selection in {\bf \it top:} dereddened Petrosian
     luminosity $M_r$ vs redshift and {\bf \it bottom:} k-corrected
     dereddened Petrosian magnitude $m_r$ vs redshift for a random
     selection of galaxies in the sample.  Galaxies in the
     volume-limited sample ($z\leq0.08$ and $M_r \leq -20.0$) are
     highlighted in red.  The blue solid line corresponds to the
     95$^{th}\%$ile of the k-correction distribution at the magnitude
     limit of $m_r=17.77$. The dashed green line is equivalent, only
     for the 50$^{th}\%$ile of the k-correction distribution, whilst
     the dashed magenta line illustrates the magnitude limit (no
     k-correction).}
   \label{figure:selection}
 \end{figure}

The upper panel of Figure~\ref{figure:selection} illustrates the
Petrosian r-band luminosity $M_r$ as a function of redshift $z$ for a
random subsample of our galaxies. The spectroscopic selection
threshold at a luminosity corresponding to apparent magnitude
$m_r=17.77$ is a strong function of redshift, and a weak function of
SED type at given redshift (k-correction).  This is better illustrated
in the lower panel where we show the k-corrected magnitude not
corrected for the distance modulus, as a function of redshift.  To
assess the effect of the k-correction on the luminosity limit, the
percentiles of the k-correction distribution are computed as a
function of redshift, and are applied at the magnitude limit. The
solid blue and dashed green lines show the 95$^{th}$ and 50$^{th}$
percentiles respectively, which clearly become larger with redshift as
the k-corrected magnitude limit diverges from the non-corrected limit
of $m_r=17.77$ (magenta dashed line).

 To study environment, it is important that the luminosity limit for
 each galaxy and its neighbours (which will be used to estimate the
 local environmental density) is the same, regardless of distance.
 Therefore, a volume-limited sample with a constant luminosity limit
 as a function of redshift is required.  Similar to \citet{Balogh04},
 we find that a suitable compromise between volume and depth can be
 achieved by limiting to $z\leq0.08$ and $M_r \leq -20.0$.  Galaxies
 meeting these criteria are plotted red in
 Figure~\ref{figure:selection}, and within the region delineated by
 the solid red lines in the luminosity-redshift plane (upper panel).
 All galaxies with k-corrections less than the 95$^{th}$ percentile of
 the distribution, and Petrosian r-band luminosities $M_r \leq -20.0$
 will be potential spectroscopic targets out to $z=0.08$, providing a
 measure of environment.  The most extreme galaxies with the upper 5\%
 of k-corrections may still be missed at redshifts very close to
 $z=0.08$. However these will by definition be extremely few and with
 perhaps the least reliable k-corrections.

The volume-limited sample contains 92082 galaxies, with a median
redshift of 0.066.

\section{Measurement of density on different scales}\label{sec:density}

The goal of this paper is to test the correlation between the galaxy
colour bimodality and the environmental density on different scales,
which we measure as follows. A cylindrical annulus, centred on each
galaxy in the volume-limited sample, is used to measure the local
projected density of neighbouring galaxies above the limiting
luminosity between an inner and outer radius and within a fixed
rest-frame velocity range.  By selecting non-overlapping annuli we can
extract complimentary information about the overdensity of galaxies on
different scales, which will only correlate with each other because
the underlying density field correlates on different scales, coupled
with the effects of projection and redshift distortions.  For an
annulus with inner radius $r_i$ and outer radius $r_o$, the surface
density $\driro$ is given by:

\begin{equation} 
\driro = \frac{\Nnriro}{\pi(r_o^2-r_i^2)}
\label{equation:density}
\end{equation}

where $\Nnriro$ is the completeness-corrected number of neighbours
brighter than our limit, within the rest-frame velocity range $\pm
dv$, and with a projected distance from the primary galaxy $r$ where
$r_i\leq r \leq r_o$.  For the purposes of this paper we choose
neighbours brighter than the volume-limited sample depth of
$M_r=-20.0$ and within $dv=1000kms^{-1}$ of the primary galaxy.  This
is equivalent to that used by \citet{Balogh04}, sampling
$\pm\sim1\sigma$ for massive clusters in velocity space, or a redshift
space cylinder of length $\pm\sim14\Mpc$.  On large scales this traces
the non-linear regime for high densities, whilst voids remain
underdense ($\sim1\%$ of galaxies have zero neighbours within $r_o =
3\Mpc$).

\subsection{Selection and Correction for Completeness}
 
In order to estimate local galaxy densities properly, selection
effects must be taken into account and completeness corrections
applied.  In the first place, we request the photometric coverage of
the region used to compute local density to be highly complete.  We
compute the photometric completeness within a $3\Mpc$ circle centred
on each galaxy in the volume-limited sample, at the redshift of the
galaxy.  Using the same tools that also power the spatial computations
within the National Virtual Observatory's Footprint Service%
\footnote{{\it http://voservices.net/footprint}} \citep{budavari07},
we perform the formal algebra of the relevant spherical shapes on the
sky. The 3Mpc circle is first intersected with the sky coverage of the
DR5 survey that yields the observed area. Furthermore we have to
censor problematic regions due to nearby bright stars, bleeding,
satellite trails and missing fields. The Sloan collaboration provides
spherical polygons of these {\it masks} that one has to exclude by
subtracting their shapes from the observed area. In addition, we also
censor out bad seeing patches that are worse than 1.7$\arcsec$ in the
r band.  Using the Spherical Library \citep{budavari09}, we compute
the exact shape of every cell, and analitically calculate its
area. The completeness is then just the ratio of this area to the full
circle.

The resultant cumulative distribution of photometric completeness
shows two regimes with 80\% of the galaxies having completeness above
98.645\%. This is an acceptable loss of area which will only have
minimal effects on our density estimates.  The remaining 20\% of
galaxies have much less complete photometry within 3$\Mpc$.  This is
where the edge effects of the SDSS coverage become important.  We keep
the 80\% of galaxies with better than 98.645\% photometric
completeness, for which the residual incompleteness can be attributed
to regions removed due to bright stars, satellite / asteroid trails
and imaging defects.  This leaves us with a sample of 73662 galaxies.

Local density estimates are obviously affected by the spectroscopic
incompleteness of the survey, which is likely to depend upon
environment.  In fact, due to fiber collision effects (i.e. fibers
could not be assigned to targets within 55\arcsec\ of each other)
spectroscopic observations are biased against galaxies in denser
regions of the sky.  For example, $\sim10\%$ of galaxies with
$\dsca\geq23\Mpc^{-2}$ are missing spectra for half their potential
neighbours.  It is possible to correct statistically for the
spectroscopic incompleteness effects by computing a correction factor
for the number of neighbours, based upon the fraction of
targets\footnote{In this paper ``Target'' refers to a galaxy which
  meets the SDSS magnitude limit of $m_r=17.77$} with a valid redshift
measurement.

The spectroscopic completeness correction factor is computed on a
range of different scales:

\begin{equation}
\Csriro = \frac{\Npriro}{\Nsriro}
\end{equation}

where $\Npriro$ is the total number of photometrically identified
spectroscopic targets within the annulus and $\Nsriro$ is the number
with redshift. This factor corrects for the incomplete sampling of
galaxies within dense regions, up-weighting the number of neighbours
in equation~\ref{equation:density} such that:

\begin{equation}
\Nnriro = \Csriro \times \Nunriro
\end{equation}

where $\Nunriro$ is the uncorrected number of redshift space
neighbours within our radial, velocity and luminosity constraints.

\begin{figure*}[p]
  \centerline{\includegraphics[width=0.45\textwidth]{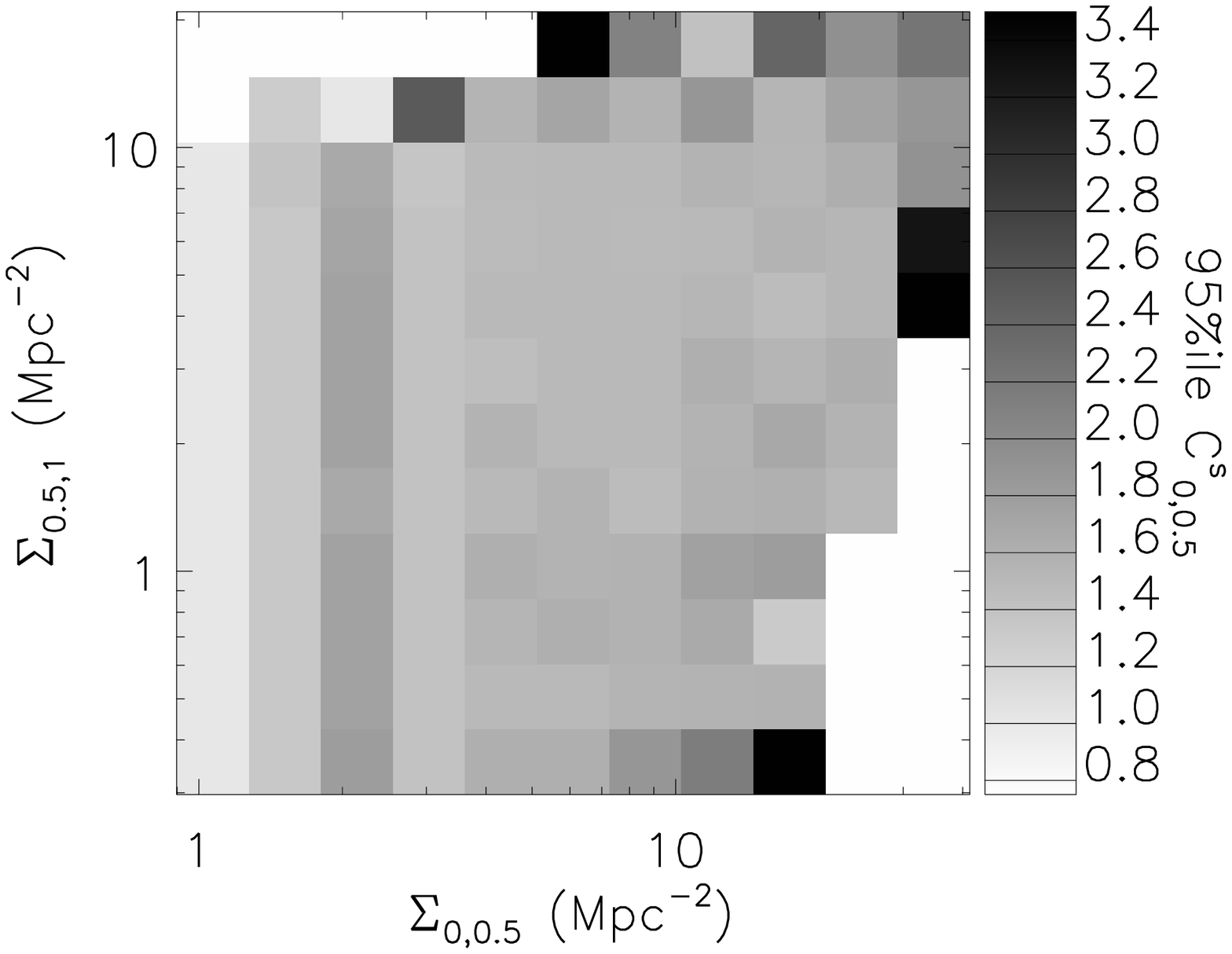}
               \includegraphics[width=0.45\textwidth]{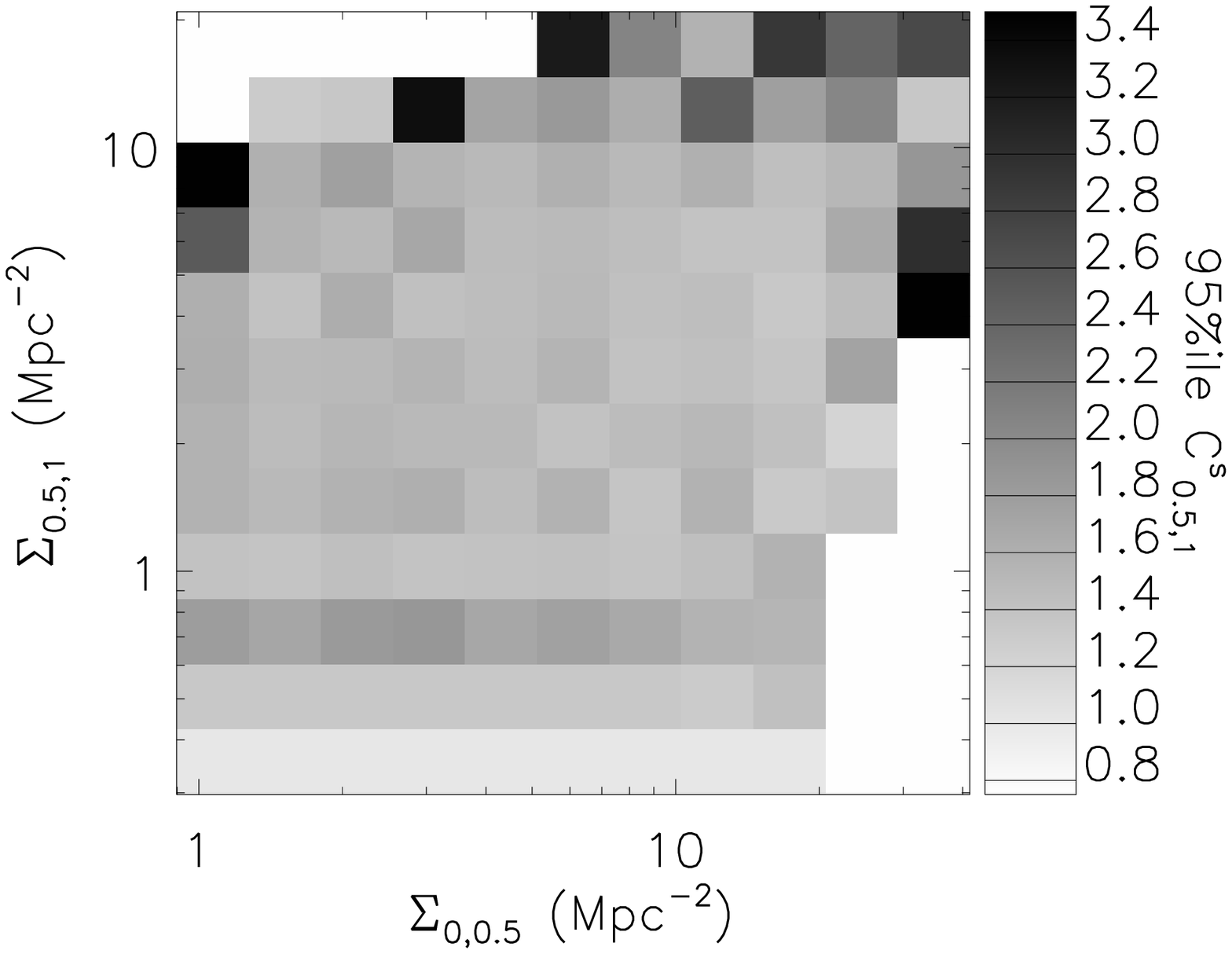}}
             \caption{95\%ile of the completeness corrections $\Csca$
               and $\Cscb$ as a function of $\dsca$ and $\dscc$.
               Larger completeness corrections are required at high
               density, where fiber allocation is difficult.
               Quantized neighbours lead to ``stripes'' of high $C^s$
               in particular lower density bins.}
  \label{figure:completeness}
\end{figure*}

Figure~\ref{figure:completeness} illustrates how $\Csca$ and $\Cscb$
depend upon $\dsca$ and $\dscb$.  We take the 95\%ile of the
distribution to examine the largest corrections in each bin of
density.  As expected, the largest corrections exist at high density
due to fiber collision effects, although at low densities the
quantized number of neighbours combined with finite binning leads to
artificial selection of high completeness galaxies in particular bins
(striping).

Galaxies with no known neighbours ($\Nunriro=0$) will automatically
also have no neighbours in the completeness-corrected density
($\driro=0$), regardless of the completeness correction
($\Csriro$). In theory this means a galaxy with many neighbours can
still have $\driro=0$ if all true neighbours have no redshift
information.  However, only $\sim4\%$ of galaxies with one known
neighbour within $3\Mpc$ ($\Nuni=1$) have completeness correction
larger than 50\% ($\Csci>1.5$). Therefore completeness corrections are
typically not large for galaxies in low density regions, and it is
unlikely that the signal from galaxies with $\driro=0$ is dominated by
galaxies which truly have significantly larger densities.

\section{Fitting Method}\label{sec:fitting}

We parameterize the galaxy colour distribution as a double gaussian,
as illustrated by \citet{Baldry04}, \citet{Balogh04} and
\citet{Baldry06}, to examine the dependence of the galaxy colour
distribution on galaxy luminosity and environment on different scales.
The colour distribution $\Phi(C)$ for a given sample of $N_T$ galaxies
 is thus
fit by the following function:

\begin{equation}
  \Phi(C) = \frac{\fr N_T}{\sqrt{2\pi}\sigmar}e^{\frac{-(C-\mur)^2}{2\sigmar^2}} + \frac{(1-\fr)N_T}{\sqrt{2\pi}\sigmab}e^{\frac{-(C-\mub)^2}{2\sigmab^2}}
\end{equation}
where the five relevant parameters are: the fraction of red galaxies
$\fr$ and the mean colours and widths of the red and blue peaks
$\mur$, $\mub$, $\sigmar$ and $\sigmab$.

Our fitting procedure for any given sample is based upon the one
outlined in Section~4 of \citet{Baldry04}. Here we summarize it and
describe the details specific to our analysis.

The galaxies are binned in colour, with a bin width of 0.1 magnitudes.
The fit is limited to the range $0.75\leq u-r \leq3.0$.

The double gaussian model is then fit to the binned colour
distribution using a Levenberg-Marquardt least-squares
fit\footnote{This was implemented using the mpcurvefit procedure
  within idl, written by Craig B. Markwardt.}. In doing this, each
colour bin is weighted by the inverse variance, computed as the square
of the Poisson error plus a softening factor of two
($W=\frac{1}{N+2}$). Bins containing no galaxies are weighted
equivalently to bins with one galaxy ($W=\frac{1}{3}$).

Low number statistics can produce noisy colour distributions, and so
additional constraints need to be placed upon the fits, and initial
estimates of the parameters given, in order to prevent the fitting
algorithm from converging to false minima.  Most importantly, we only
fit samples of at least 25 galaxies, which extensive testing shows
recovers sensible results.  Initial estimates of $\mu$ and $\sigma$
are taken from the fit parameterized in Table~1 of \citet{Baldry04}.
An estimate of $\fr=0.5$ is applied to the brightest bin with
estimates for each successive bin of luminosity taking the value
computed for the previous bin.  Once each luminosity bin is split into
bins of density, the initial parameter estimates are set to the values
computed for the full bin.  The results are insensitive to the precise
values selected.  The blue peak is constrained such that $\mub$ cannot
be less than 0.2$mag$ bluer than the initial estimate of $\mur$. This
ensures that the two peaks remain separate entities, and do not switch
places.  Finally, we also constrain the widths ($\sigmar$, $\sigmab$)
to be within the range $0.05\leq\sigma\leq0.8$.  This avoids
pathological solutions in cases where the colour distribution exhibits
large noise spikes or unusually large number of galaxies in the tails
of the distribution.

In all cases good fits are achieved, yielding reduced chi-squared
values near or below unity.  As discussed by \citet{Baldry04}, the
softening factor of two privides an approximation to the uncertainties
involved for low number counts with a Poisson distribution.
Inevitably this approximation, combined with the finite weighting of
bins containing no counts, leads to frequent cases of reduced
chi-squared significantly less than one.  In the following analysis
errors on the $\fr$ are computed purely statistically, and are
insensitive to (but consistent with) the fitting errors.  Errors on
the other parameters are estimated from fitting.  In these cases it
was necessary to multiply the error on each parameter by
$\sqrt(\chi_{red}^2)$ since the errors on these parameters are
overestimated for $\chi_{red}^2<1$ (propogated from overestimated
errors on the datapoints).  Whilst this is clearly an approximate
correction, we note that the fit errors on these parameters are only
applied in Figures~\ref{figure:redlum} and \ref{figure:bluelum}, and
that our main conclusions are based instead on the correlation with
density of properties computed in independent bins.

\section{Results}\label{sec:results}

\subsection{Dependence on Luminosity and Density}\label{sec:deplumdens}

A large number of galaxies per luminosity bin are required to examine
the multiscale density dependence of galaxy properties within that
bin.  Therefore we limit our analysis to the luminosity range
$-21.5\leq\Mr\leq-20.0$.

\begin{figure*}
  \centerline{\includegraphics[width=0.95\textwidth]{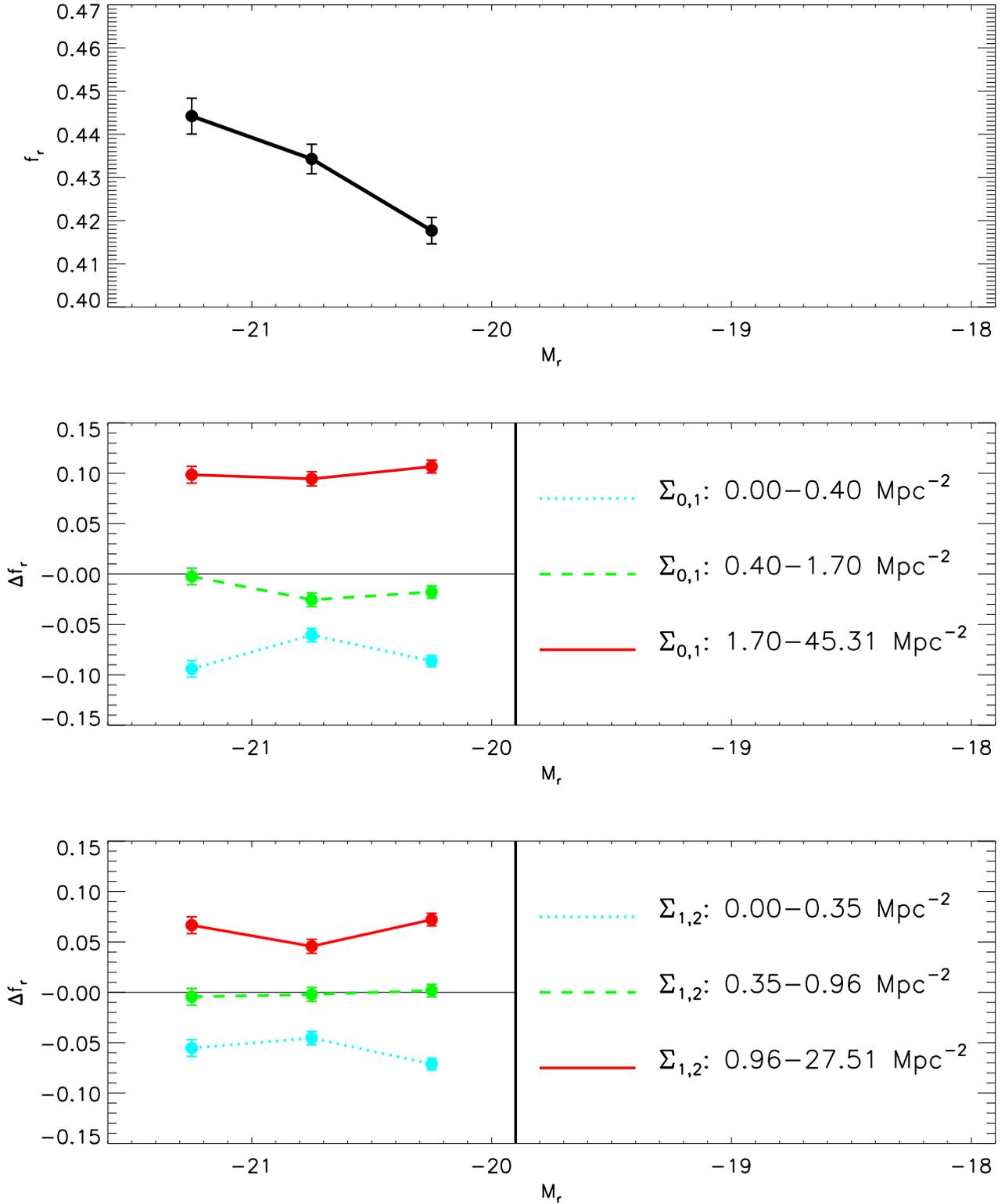}}
  \caption{{\it Top:} $\fr$ as a function of luminosity, $\Mr$, for the full
    volume-limited sample in the luminosity range of interest.
    {\it Middle:} Residuals for different bins of $\dscb$. 
    {\it Bottom:} and in
    different bins of larger scale density, $\dscd$.  Errors are
    computed using the \citet{Gehrels86} approximation, combining
    Poissonian and Binomial statistical errors.  For double gaussian
    fits to a large number of galaxies this approximates the fitting
    error, whilst for smaller numbers the statistical error is
    larger.}
  \label{figure:fredlum}
\end{figure*}

\begin{figure*}
  \centerline{\includegraphics[width=0.95\textwidth]{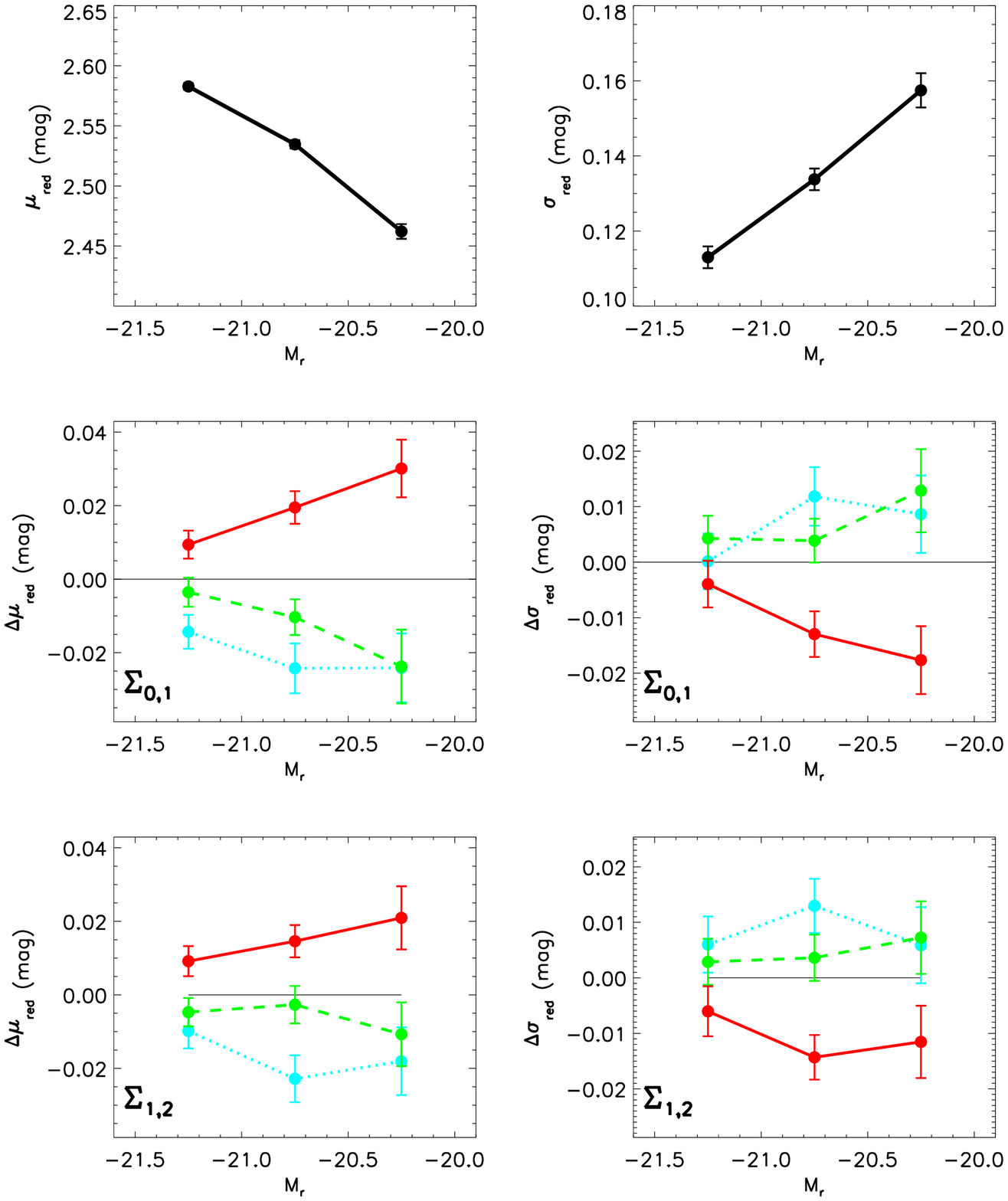}}
  \caption{{\it Top:} $\mur$ and $\sigmar$ as a function of luminosity,
    $\Mr$, for the full volume-limited sample in the luminosity range
    of interest.  {\it Middle:} Residuals for different bins of $\dscb$.
    {\it Bottom:} and in different bins of larger scale density,
    $\dscd$. Errors are pure fitting errors. Key as in
    Figure~\ref{figure:fredlum}.}
  \label{figure:redlum}
\end{figure*}

\begin{figure*}[p]
  \centerline{\includegraphics[width=0.95\textwidth]{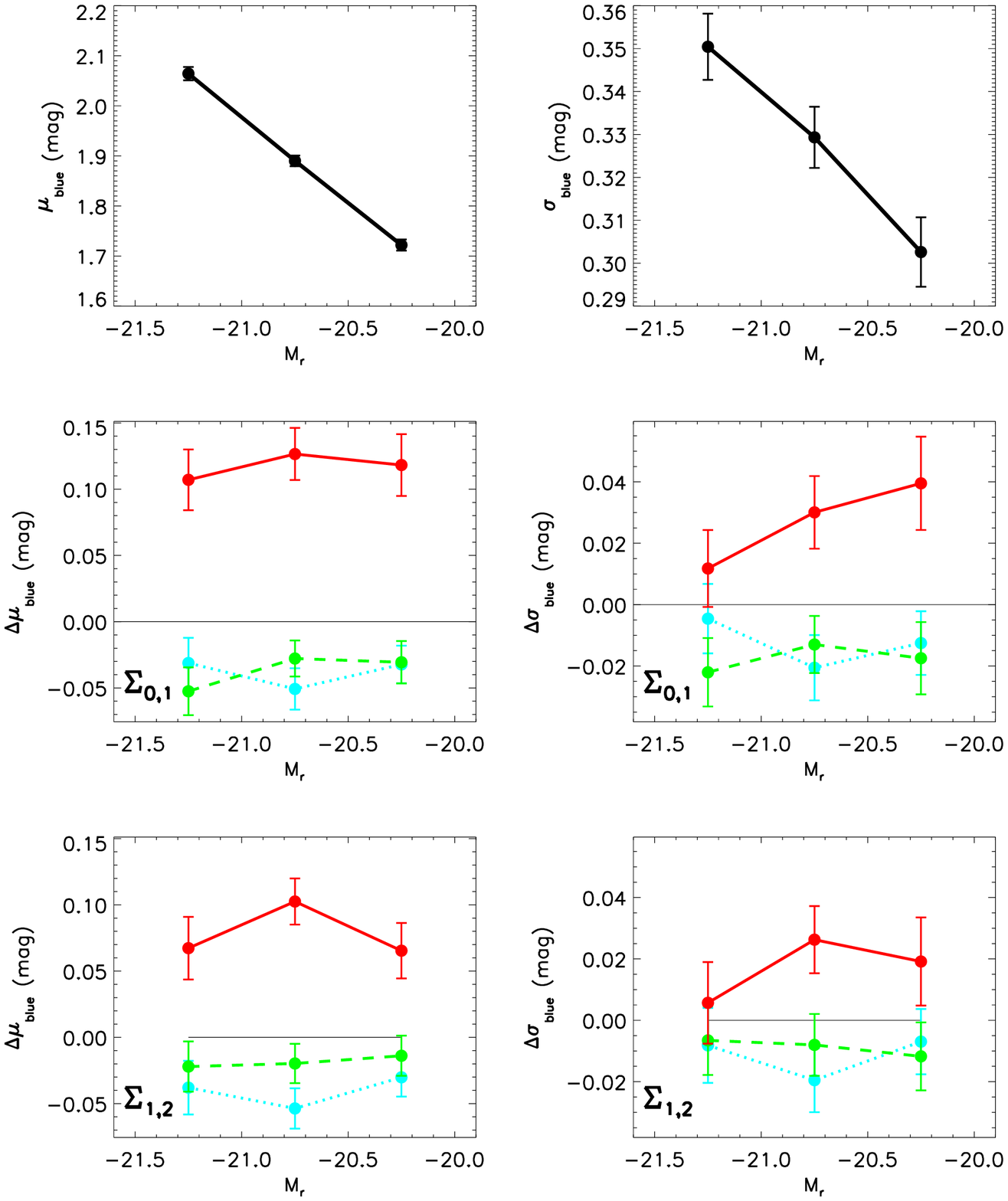}}
  \caption{{\it Top:} $\mub$ and $\sigmab$ as a function of luminosity,
    $\Mr$, for the full volume-limited sample in the luminosity range
    of interest.  {\it Middle:} Residuals for different bins of $\dscb$.
    {\it Bottom:} and in different bins of larger scale density,
    $\dscd$. Errors are pure fitting errors. Key as in
    Figure~\ref{figure:fredlum}.}
  \label{figure:bluelum}
\end{figure*}

The top panels in Figures~\ref{figure:fredlum}, \ref{figure:redlum}
and \ref{figure:bluelum} illustrate the dependence of the five
parameter family ($\fr$, $\mur$, $\mub$, $\sigmar$, $\sigmab$) on
luminosity, $\Mr$, as resulting from the fitting procedure applied to
the volume-limited sample within this luminosity range.  We confirm the
results of \citet{Baldry04} indicating that all five parameters are
strong functions of luminosity: lower luminosity galaxies display a
lower fraction of red galaxies, bluer mean colours for both peaks and
a broader red peak but narrower blue peak than do more luminous
galaxies within the range considered. Small differences from
\citet{Baldry04} result from improved k-corrections (a later version
of the {\it k-correct} code was used). There is no measurable
dependence of colour ($\mur$ or $\mub$) for constant luminosity
galaxies on redshift, indicating that the k-corrections are consistent
across our small redshift range.  We note that, for the luminosity
range sampled by the volume-limited sample, the fraction of red
galaxies $\fr$ is only weakly dependent on luminosity, while both
$\mu$ and $\sigma$ show strong variation for both peaks.  This
luminosity dependence of these properties within the volume-limited
sample is consistent with that obtained for a larger sample with
variable redshift limit as a function of galaxy luminosity.

Residual quantities ($\dfr$, $\dmur$, $\dmub$,
$\dsigmar$, $\dsigmab$) are defined for each subsample (sub) as
follows:

\begin{equation}
\dfr = \fr(\Mr,sub) - \fr(\Mr)
\end{equation}
\begin{equation}
\dmu = \mu(\Mr,sub) - \mu(\Mr)
\end{equation}
\begin{equation}
\dsigma = \sigma(\Mr,sub) - \sigma(\Mr)
\end{equation}

The middle and lower panels of Figures~\ref{figure:fredlum},
\ref{figure:redlum} and \ref{figure:bluelum} show how these quantities
depend upon luminosity for three bins of density on $<1\Mpc$ scales
($\dscb$, middle panels) and $1-2\Mpc$ scales ($\dscd$, lower panels).
These density bins contain the upper, middle and lower thirds of the
density distribution on these scales, and the whole volume-limited
sample is represented in black.

The density dependence of the colour distribution on both small and
large scales is extremely clear in these figures:

\begin{itemize} 
\item{The fraction of red galaxies is a strong function of
    environment, with a difference in $\fr$ between the most and
    least dense thirds of the distribution ranging from 0.15 to
    0.19 and from 0.09 to 0.15 for densities computed on $<1\Mpc$
    and $1-2\Mpc$ scales, respectively. This is almost independent
    of luminosity in the range $-21.5\leq\Mr\leq-20.0$.}
\item{The characteristics of the red peak are also highly significant
    functions of environment, although the dynamic range is smaller
    than for the luminosity dependence. The difference in the mean
    colour $\mur$ for the most and least dense thirds of the
    distribution on $<1\Mpc$ scales is in the range 0.02 to 0.06
    mag. In the most overdense regions, in going to fainter luminosity
    $\dmur$ increases (becomes redder), whilst the relative width of
    the peak $\dsigmar$ simultaneously decreases. A similar range but
    with less obvious luminosity dependence is observed using
    densities on $1-2\Mpc$ scales.}
\item{The blue peak properties $\mub$ and $\sigmab$ are more strongly
    correlated with environment than the equivalent red peak
    properties. A difference in $\mub$ of 0.15 to 0.18 mag between the
    most and least dense bins on $<1\Mpc$ scales is present at all
    luminosities. The difference is still significant but slightly
    smaller using density on $1-2\Mpc$ scales. $\sigmab$ is also
    dependent on density in the sense that the blue peak appears
    broader for galaxies in denser environments, possibly more so at
    fainter luminosities.}
\end{itemize}

\begin{figure*}
      \centerline{\includegraphics[width=0.3\textwidth]{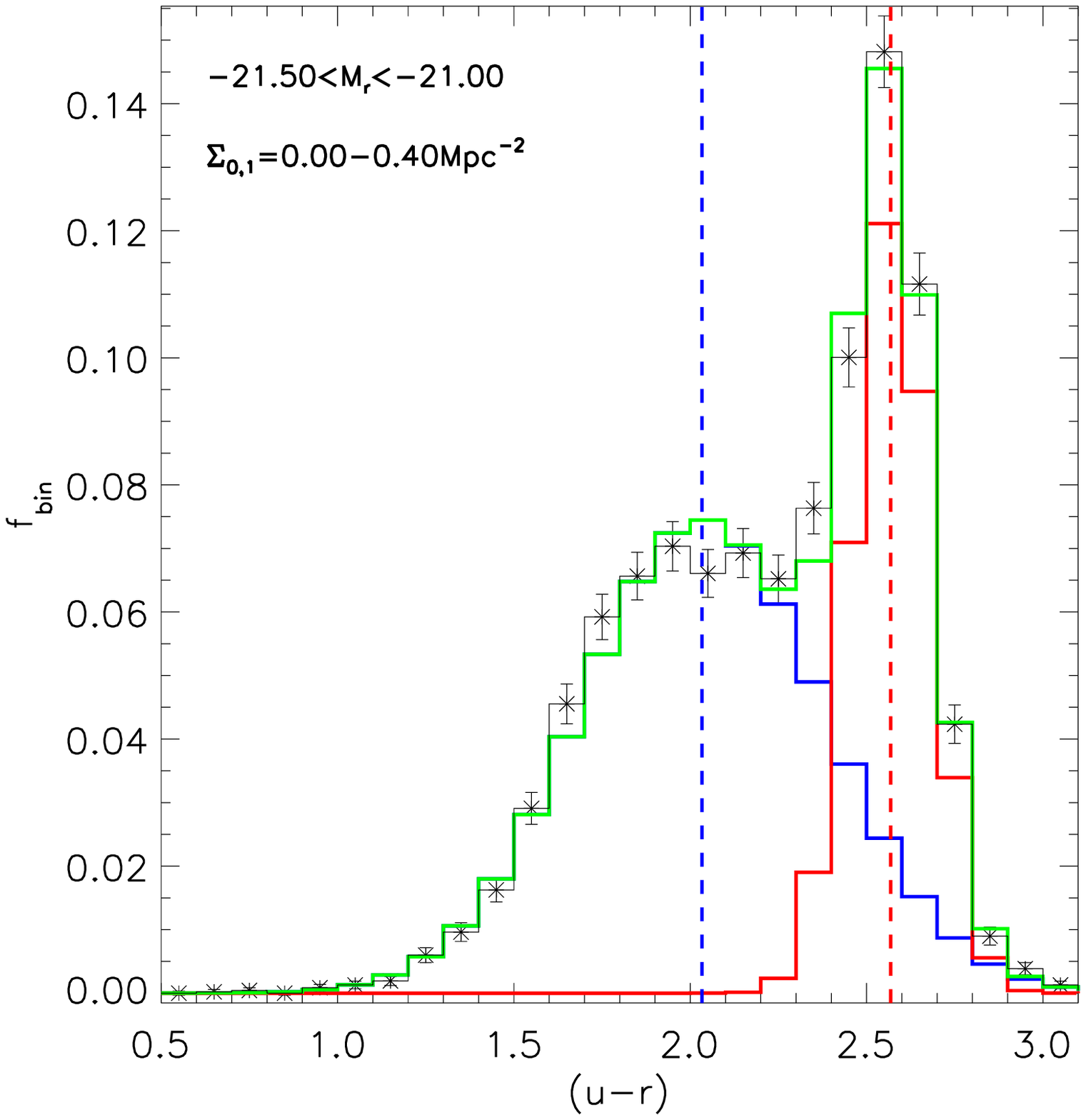}
                  \hspace{0.02\textwidth}
                          \includegraphics[width=0.3\textwidth]{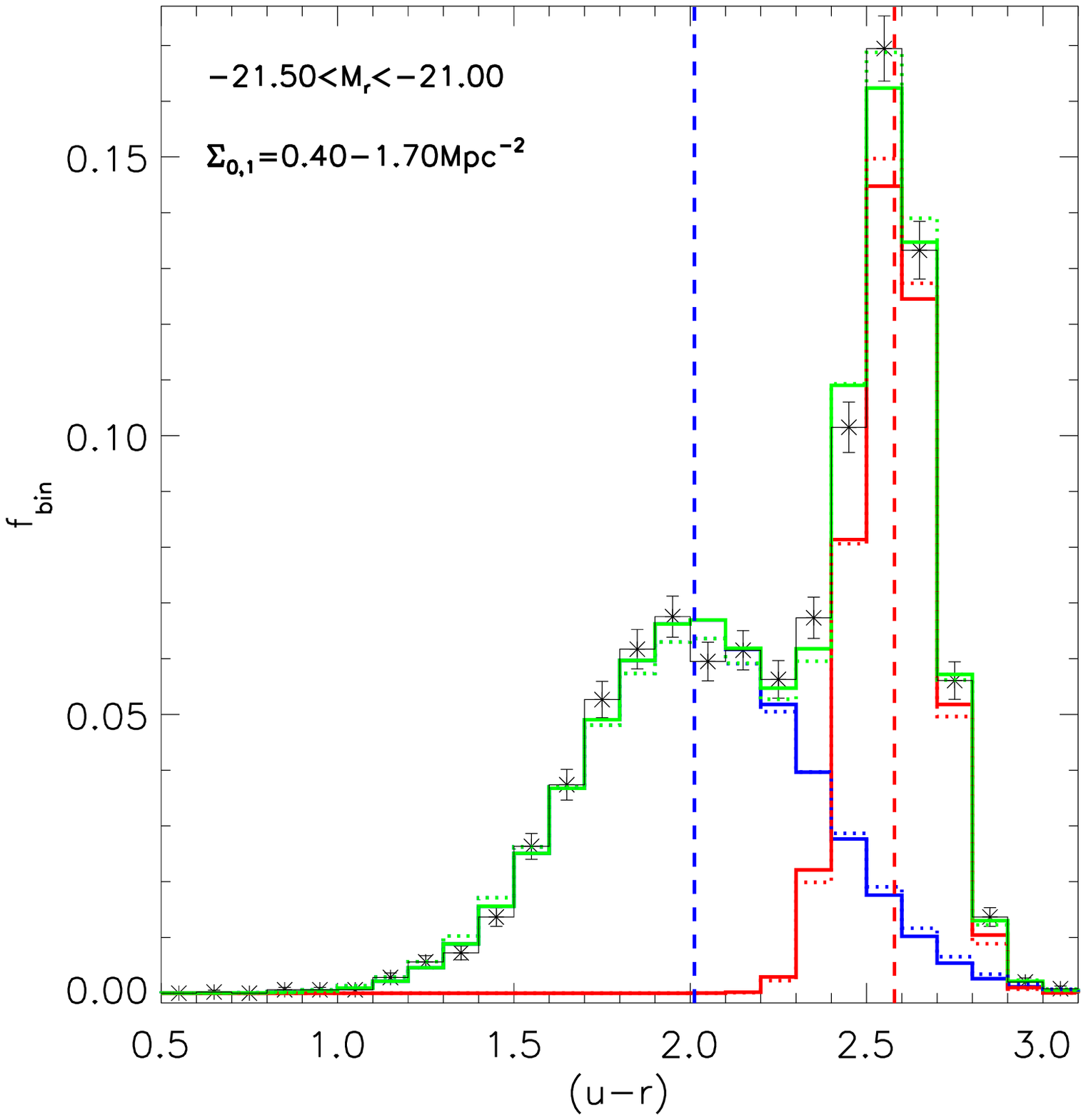}
                  \hspace{0.02\textwidth}
                          \includegraphics[width=0.3\textwidth]{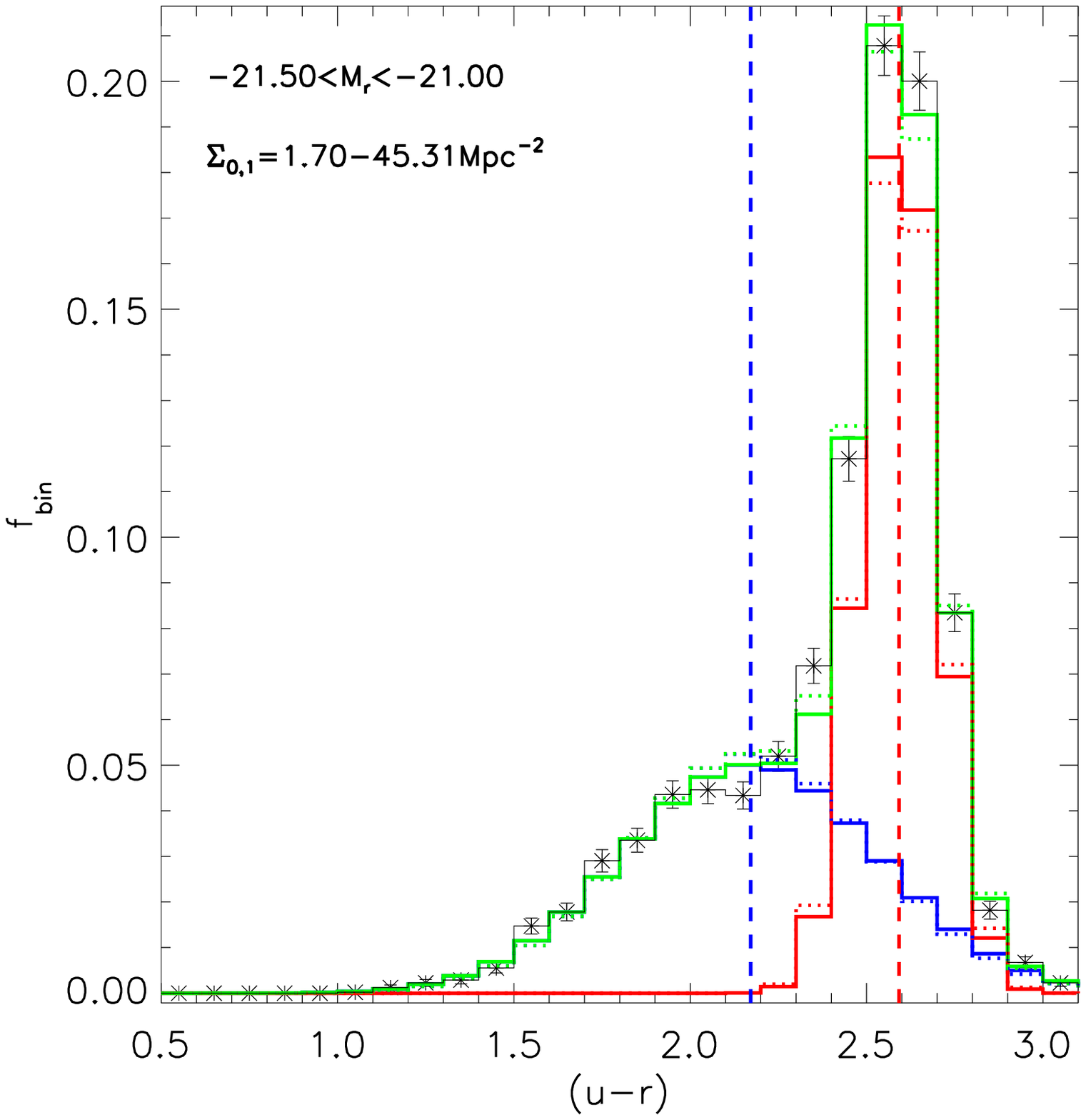}}
      \centerline{\includegraphics[width=0.3\textwidth]{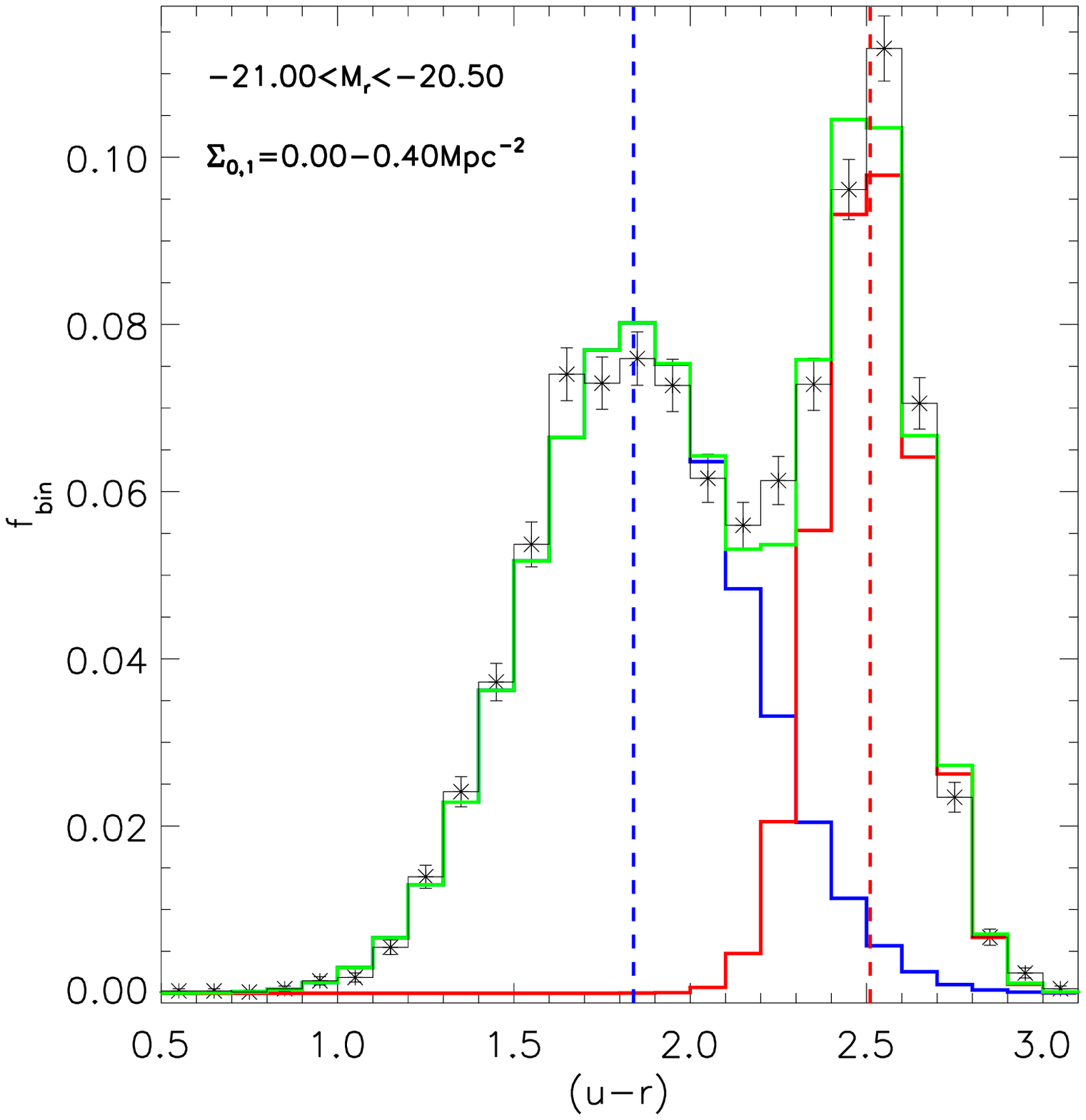}
                  \hspace{0.02\textwidth}
                          \includegraphics[width=0.3\textwidth]{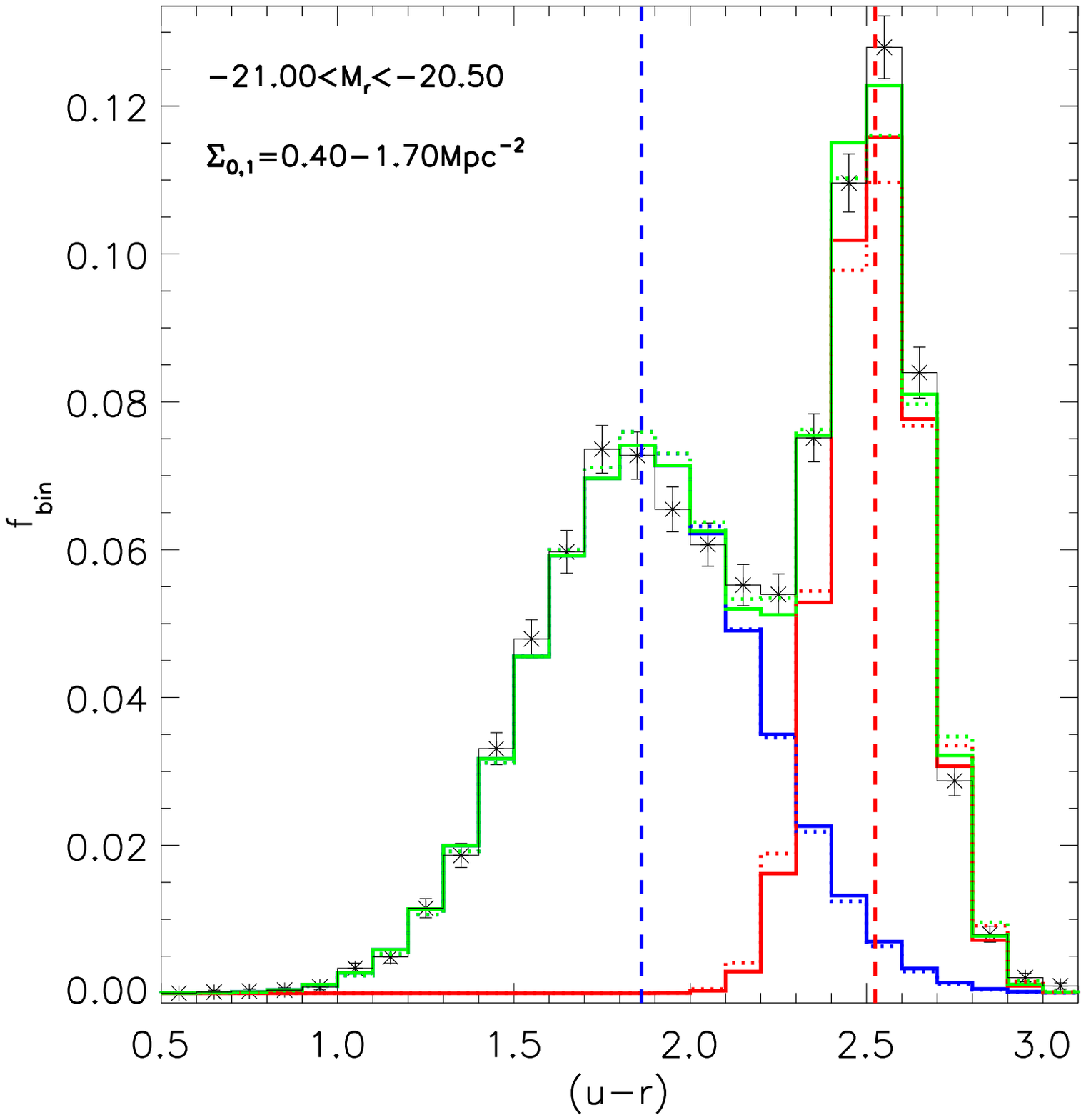}
                  \hspace{0.02\textwidth}
                          \includegraphics[width=0.3\textwidth]{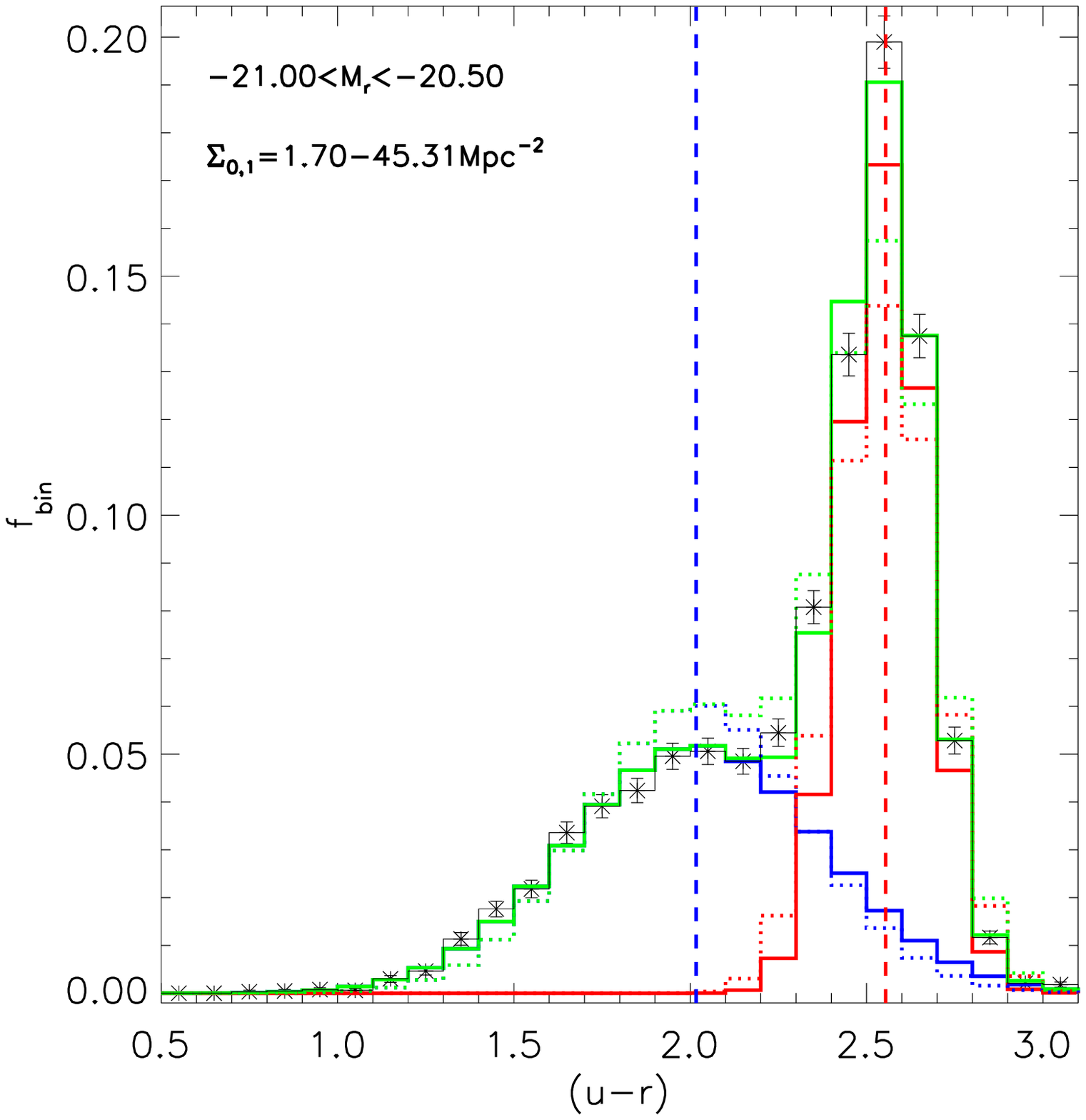}}
      \centerline{\includegraphics[width=0.3\textwidth]{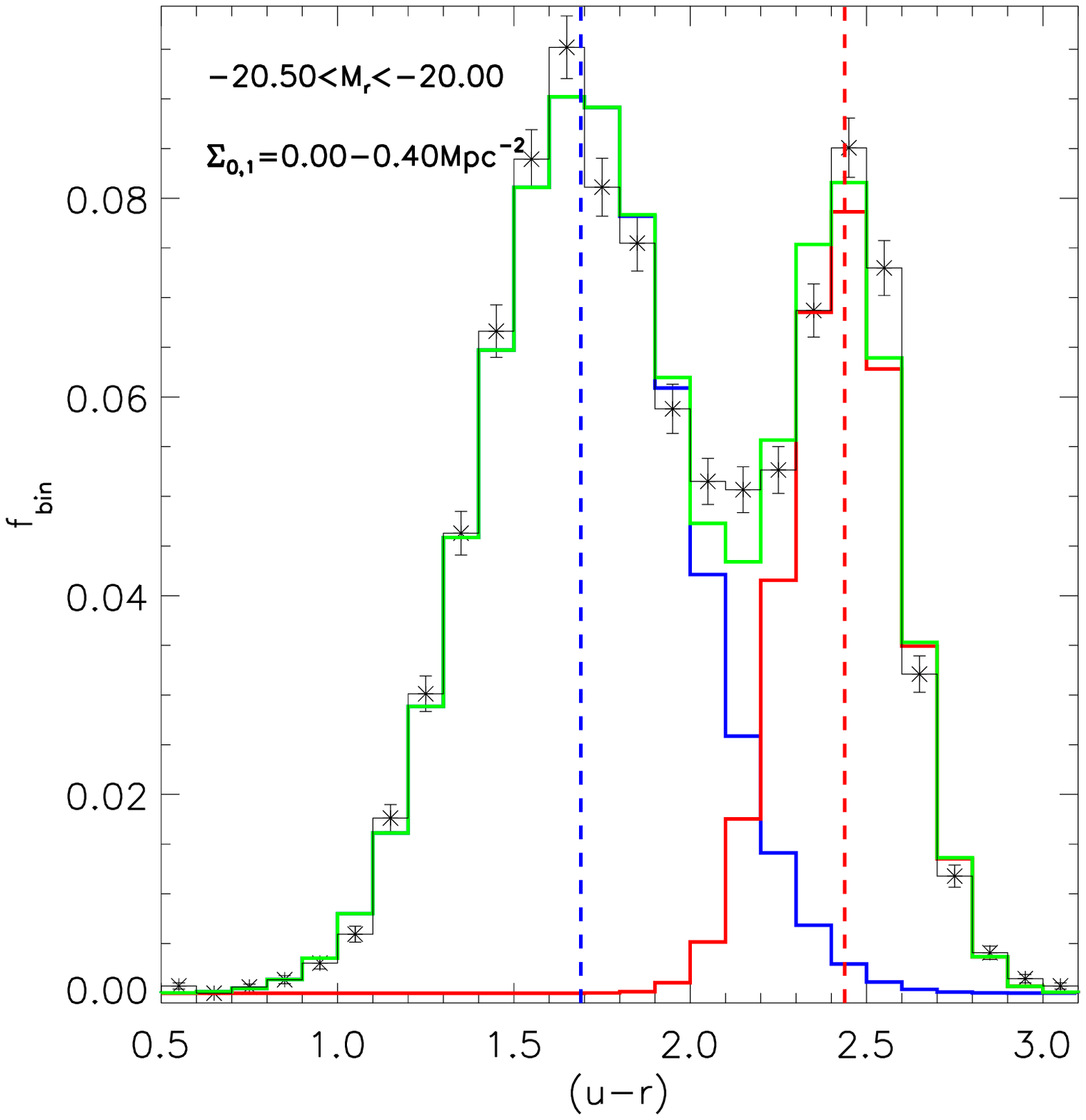}
                  \hspace{0.02\textwidth}
                          \includegraphics[width=0.3\textwidth]{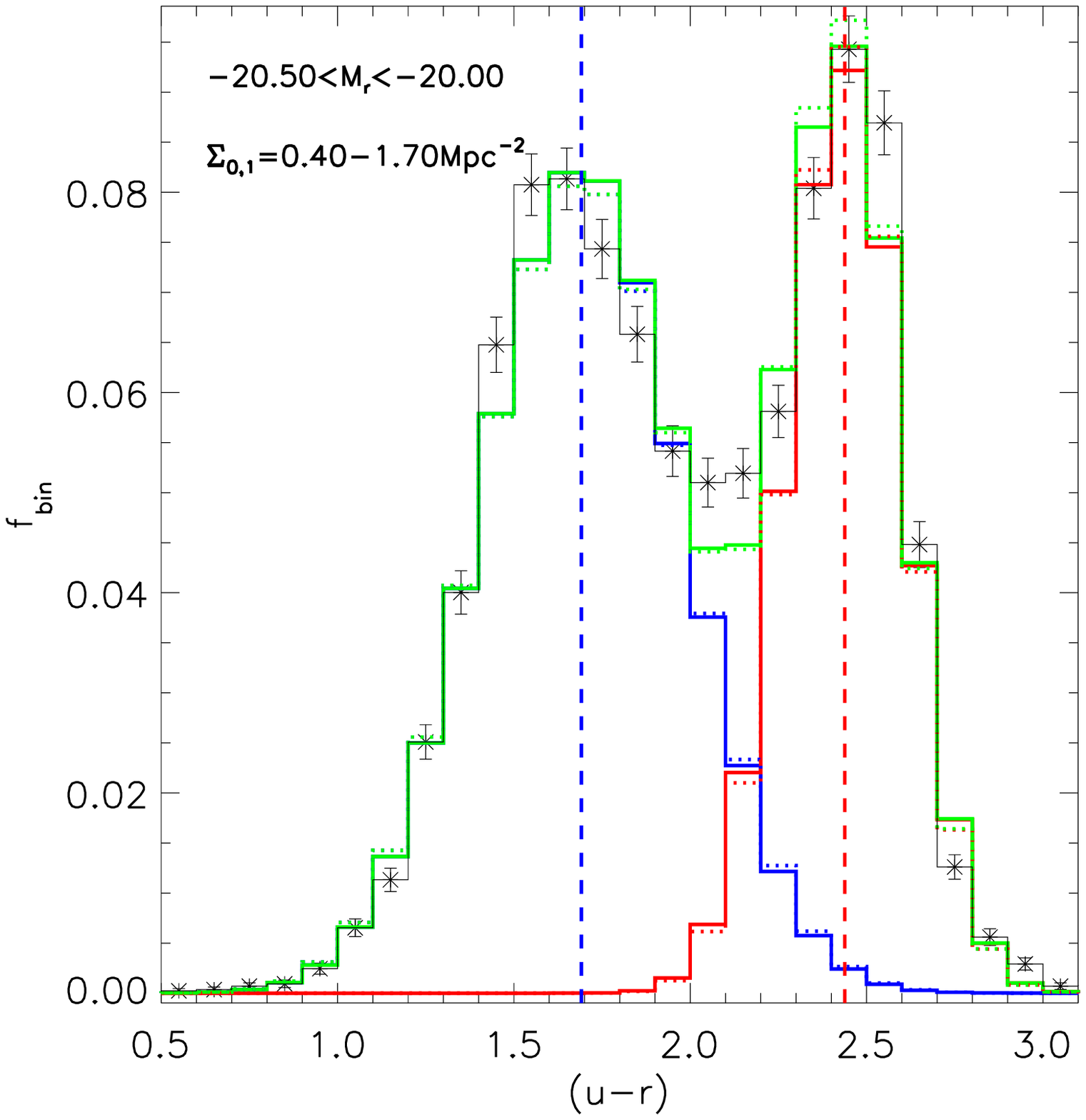}
                  \hspace{0.02\textwidth}
                          \includegraphics[width=0.3\textwidth]{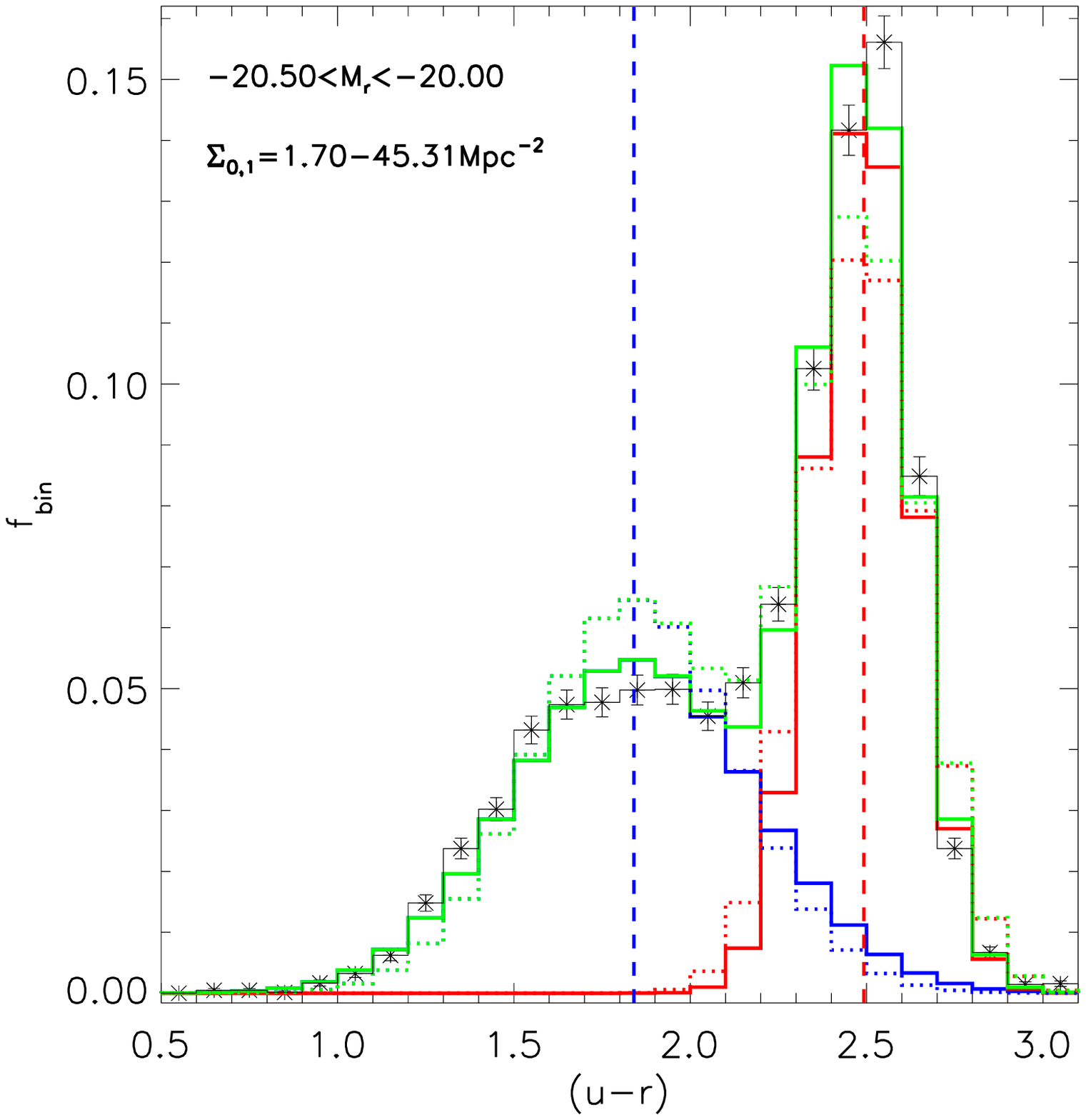}}
                        \caption{$u-r$ colour distribution of galaxies
                          (points and black solid line) as a function
                          of luminosity and density, with double
                          gaussian fits (red$+$blue peak$=$green
                          line).  Top to bottom: Decreasing luminosity
                          $\Mr=-21.5$ to $-20.0$ in bins of $0.5$mag
                          width.  Left to right: Underdense to
                          overdense, split evenly into bins by
                          $\dscb$.  Overplotted is the mean colour for
                          each peak (vertical lines), indicating that
                          both peaks move redwards going to higher
                          density.  Also overplotted (dotted lines) is
                          the sum of two gaussians with equivalent
                          parameters, except the widths
                          ($\sigmab$,$\sigmar$) are fixed to the value
                          computed for the lowest density bins (left
                          panels). The deviation in the highest
                          density bins illustrates the dependence of
                          $\sigma$ on density (strongest in the
                          faintest bin, lower right).}
      \label{figure:egfitslumdens}
\end{figure*}

To demonstrate the reality of these trends, we show the actual colour
distributions with overplotted fits in
Figure~\ref{figure:egfitslumdens}.  The strong luminosity dependence
of $\mub$, $\sigmab$, $\mur$ and $\sigmar$ and the density dependence
of $\fr$ are immediately obvious.  To better illustrate the
differences, the values of $\mub$ and $\mur$ are marked as
dashed vertical lines. Similarly the density dependence of $\sigmab$
and $\sigmar$ can be seen by comparing the actual fit with the
fit as it would be if $\sigma$ was fixed to the values computed for
the lowest density bin (overplotted dotted lines).  It is apparent,
especially in the faintest, highest density bin (lower right) and for
$u-r$ near $\mub$ and $\mur$ that a broader blue peak and narrower red
peak are required.  
\citet{Masjedi06} find that very nearby neighbours (within
$\sim 20\arcsec$) significantly influence the measured luminosity of
galaxies: however model colours are likely to be much more
robust. In any case, whilst $\sigmab$ increases at high density,
this is unlikely to be related to increased photometric errors:
$\sigmar$ should be as influenced but decreases, whilst the
$\sim 20\%$ drop in the amplitude of the blue distribution (bottom
right panel of Figure~\ref{figure:egfitslumdens}, compare solid and dashed lines)
is too strong to be caused by such a rare population of very
close neighbours. 
It must be noted that for such well populated
samples there are small but mildly significant deviations from a
perfect double gaussian distribution, visible as assymetries in the
distribution or an excess of galaxies at intermediate colours. Whilst
these deviations are potentially interesting, they are so small that
they can and should be ignored in the context of this paper.

The similar dependence of $\fr$ on density across this luminosity
range (i.e., no luminosity dependence for $\dfr$) is in qualitative
agreement with \citet{Balogh04}, although they appear to measure a
larger dynamic range for $\fr$ as a function of the surface density
within the $5^{th}$ nearest neighbour, $\Sigma_5$, at fixed
luminosity. Binning more finely in density, we find that the increase
in $\dfr$ for galaxies in the highest density third is mainly
contributed by a small fraction of galaxies ($\sim15\%$) with $\dscb
\gtrsim 4.0\Mpc^{-2}$. It is also noticeable that the trends of
$\dmur$ and $\dmub$ with density are quantitatively comparable to the
\citet{Balogh04} results, with more extreme values when more extreme
(rare) environments are isolated.

A detailed interpretation of our results will be left for
Section~\ref{sec:interpretation}.  For now we note that
Figure~\ref{figure:fredlum} suggests that suppression of star
formation leads to the formation of red galaxies preferentially in
environments characterized by a high local ($<1\Mpc$ scale) density.
The importance of different scales will be addressed in
Section~\ref{resultsscale}.  The dependence of $\mur$ and $\sigmar$ on
environment could relate to differences in either/both age or/and
metallicity of the stellar population, although intragalactic dust in
galactic and group scale haloes can account for part of the effect
\citep[$\lesssim 0.01$~mag,][]{Menard09,McGee09b}.  A plausible
scenario is that less luminous galaxies in dense environments will
typically be satellites, which may have experienced early suppression
of their star formation.  Such galaxies will have an older and purer 
underlying population than similar galaxies in lower density
environments, leading to redder colours and smaller scatter
(i.e. smaller $\sigmar$).  Meanwhile the redder, broader blue peak in
dense environments suggests a more disturbed and (on average) less
active recent star formation history compared to those in less dense
environments, although internal dust extinction may also play a role.

\subsection{Scale Dependence}\label{resultsscale}

\begin{figure*}
              \centerline{\includegraphics[width=0.3\textwidth]{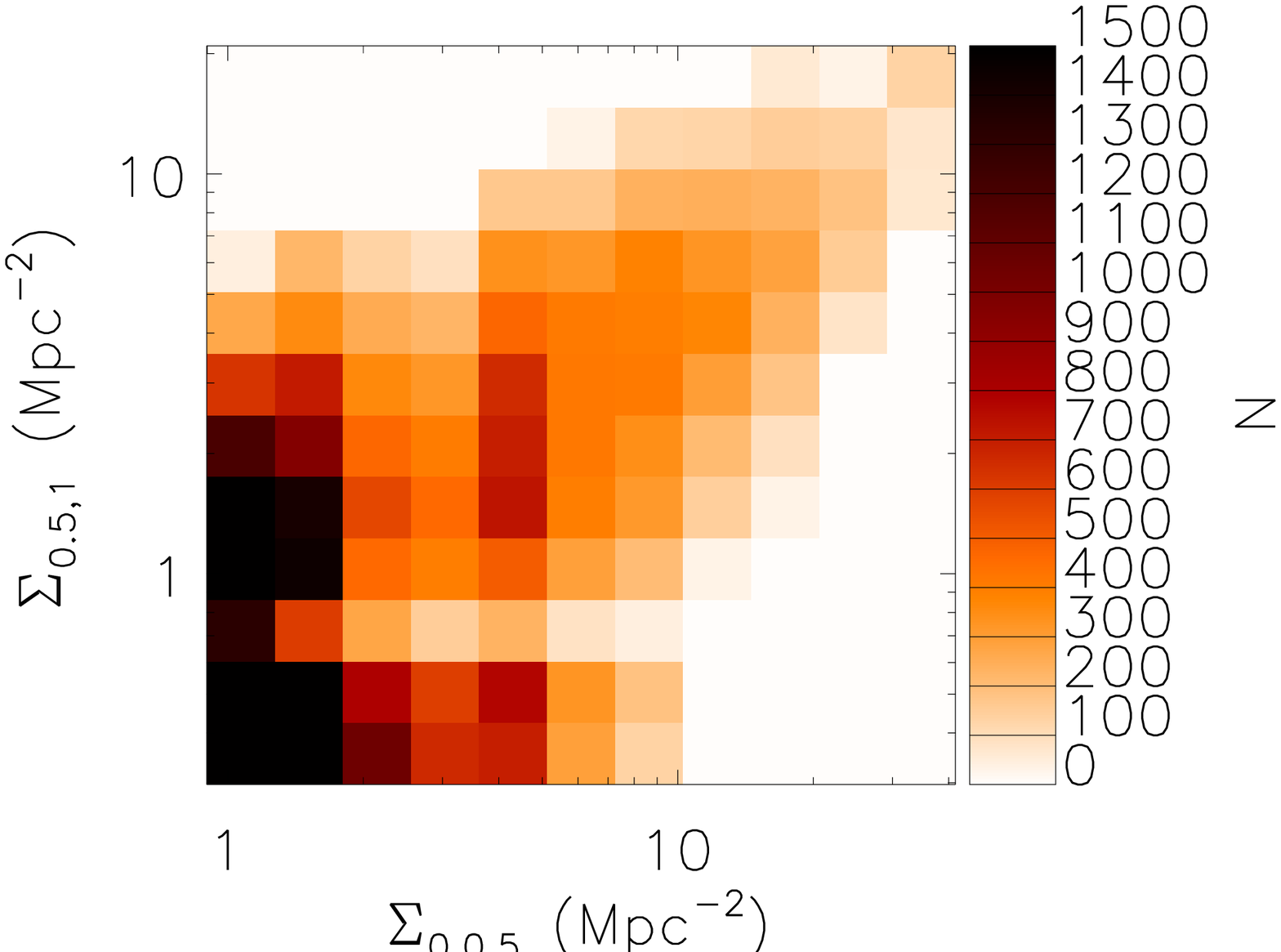}
                          \hspace{0.02\textwidth}
                          \includegraphics[width=0.3\textwidth]{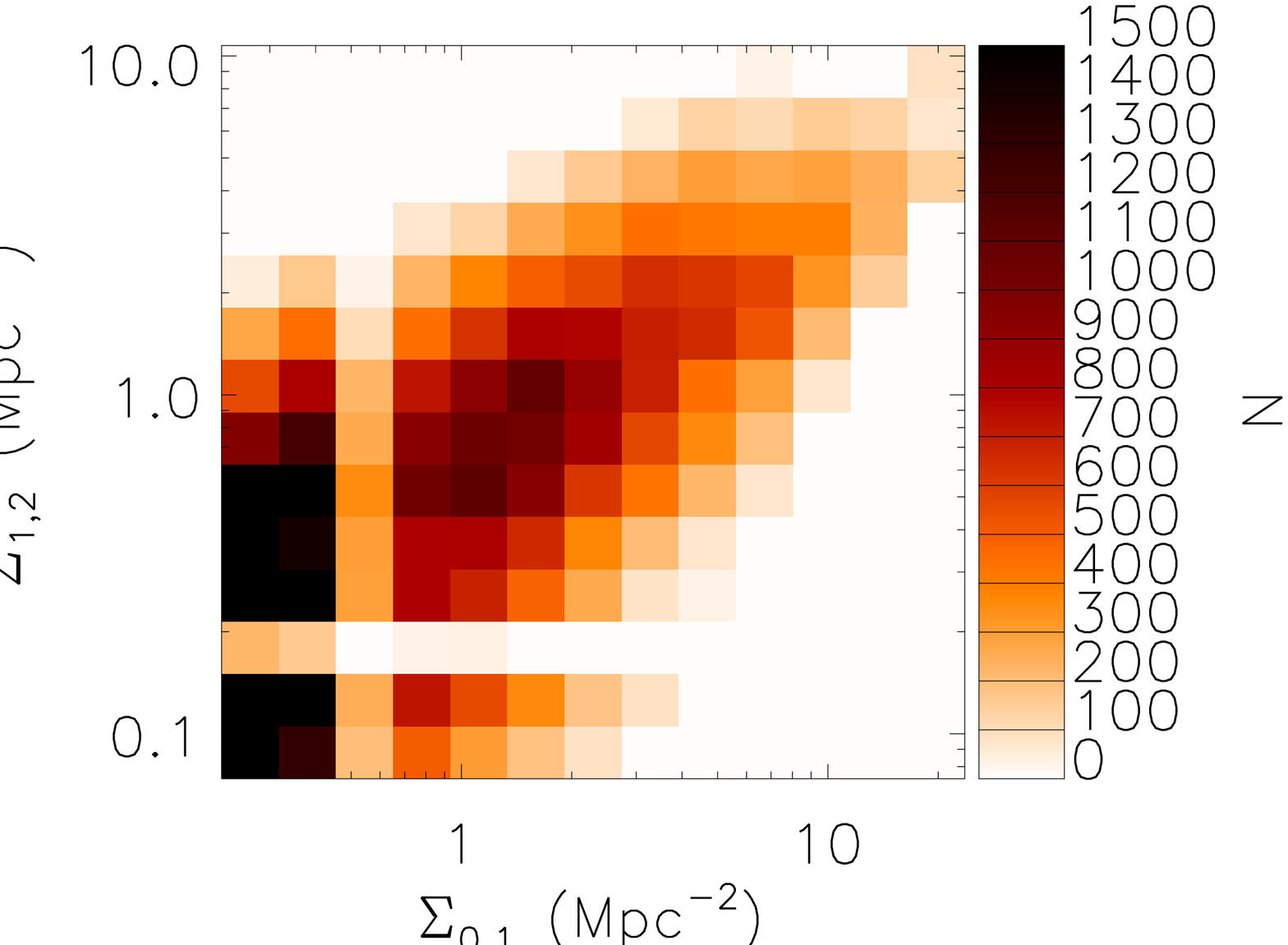}
                          \hspace{0.02\textwidth}
                          \includegraphics[width=0.3\textwidth]{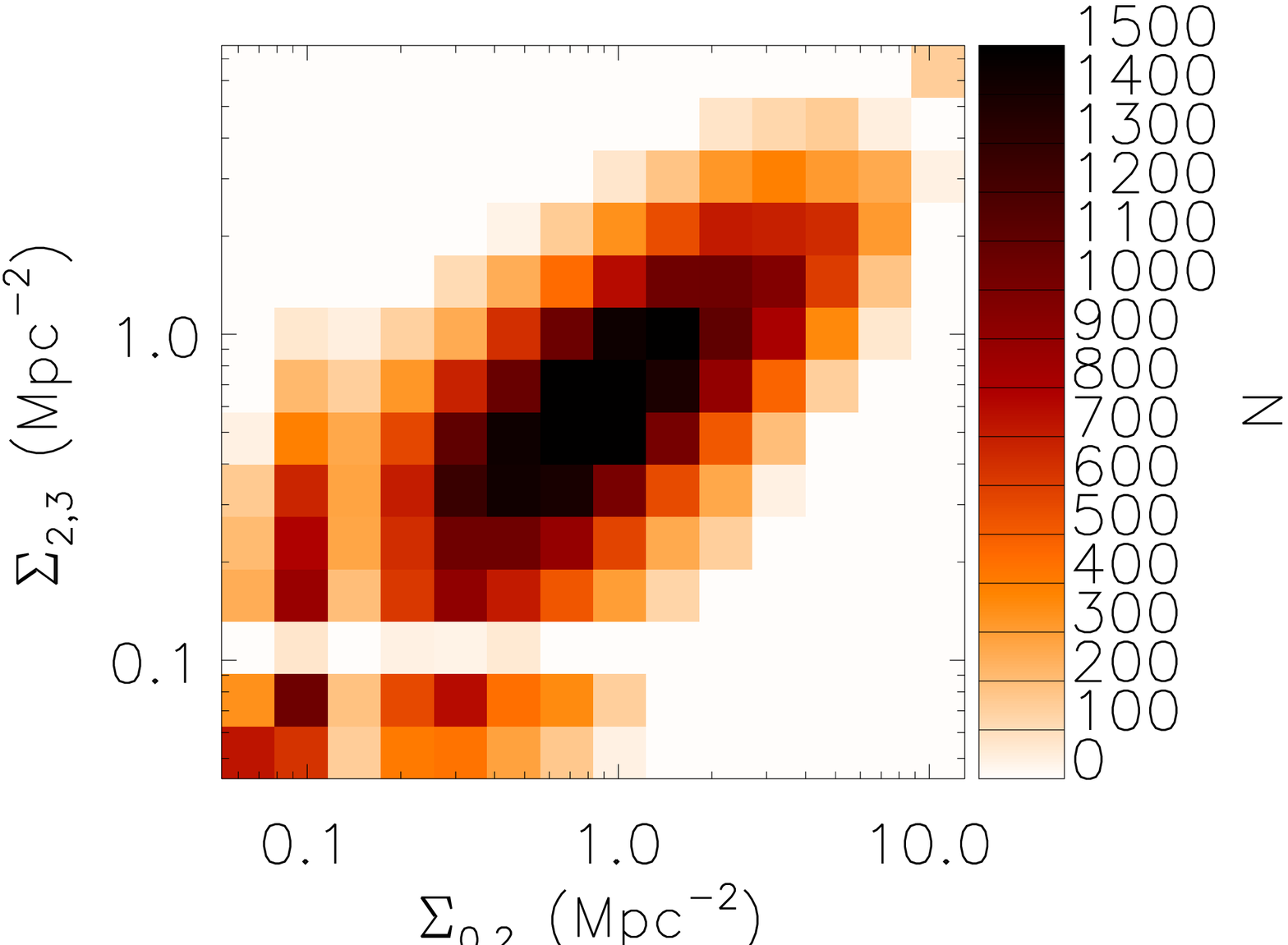}}
                        \caption{Number of $\-21.5<\Mr<-20.0$ galaxies
                          binned in two independently measured scales
                          of density.  {\it Left:} on scales $\dsca$ and
                          $\dscc$; {\it Centre:} on scales $\dscb$ and
                          $\dscd$; {\it Right:} on scales $\dscj$ and
                          $\dsce$. Small and large scale densities are
                          strongly correlated with each other.}
  \label{figure:Nscale}
\end{figure*}

Figures~\ref{figure:fredlum}, \ref{figure:redlum} and
\ref{figure:bluelum} illustrate that whilst the colour distribution of
galaxies depends upon density on $<1\Mpc$ scales, $\dscb$, it also
depends (to a smaller extent) on a totally independent measurement of
density on larger scales, $\dscd$.  However, although $\dscd$ is {\it
  measured} independently of $\dscb$, the two quantities are
nonetheless strongly correlated.  This is illustrated in
Figure~\ref{figure:Nscale}, in which we show the distribution of
$-21.5<\Mr<-20.0$ galaxies on density computed on the different scales
that we will consider in our analysis. Binning is logarithmic in
density, with less than 50\% density increase within each bin on
either scale except the highest and lowest density bins, which have
infinity and 0 as upper and lower boundaries, respectively.  The left
panel examines the distribution in the $\dsca$ versus $\dscc$ plane,
whilst successively larger scales are examined in the $\dscb$ versus
$\dscd$ and $\dscj$ versus $\dsce$ planes in the central and right
panels respectively.

The strong correlation between small and large scale densities at
$<3\Mpc$ probes the non-linear regime of collapse which amplifies
overdensities via infall of galaxies onto structures.  For massive
systems this correlation is further enhanced, particularly on small 
scales, by:

\begin{itemize} 
\item{The presence of galaxies within the same virialized structure;}
\item{Projection effects: physical separation $\geq$ projected separation;} 
\item{Redshift distortions (the ``finger of God effect''): the true
    physical separation from neighbours within $\pm1000\kms$ will be a
    strong function of environment, with different contributions of
    galaxies inside and outside the local halo (in group catalogue
    construction, galaxies outside the local group halo are known as
    "interlopers").  However, a cylindrical depth of $\pm1000\kms$
    optimally reduces these effects with a fixed aperture
    \citep{Cooper05}.}
\end{itemize}


To study how the colour distribution of galaxies is affected by
density on different scales we perform the following analysis.  First
we bin the galaxies in the volume limited sample according to their
luminosity, in bins of 0.5 mag over the range
$\-21.5<\Mr<-20.0$. Within each luminosity bin, galaxies are binned
further according to the density of their environment computed in two
spatial scales. Finally, the colour distribution of galaxies within
each bin of luminosity and density on two scales is parameterized
using the double gaussian model. Fitting of the characteristic
parameters is performed independently for each bin (provided that at
least 25 galaxies are included) and residuals are computed to remove
the luminosity dependence ($\dfr$, $\dmur$, $\dsigmar$, $\dmub$,
$\dsigmab$). In order to improve the signal-to-noise of the parameter
distributions as a function of densities, we average the values of
$\Delta$ in the three luminosity bins, weighted by the number of
contributing galaxies.



\begin{figure*}
              \centerline{
                \includegraphics[width=0.333\textwidth]{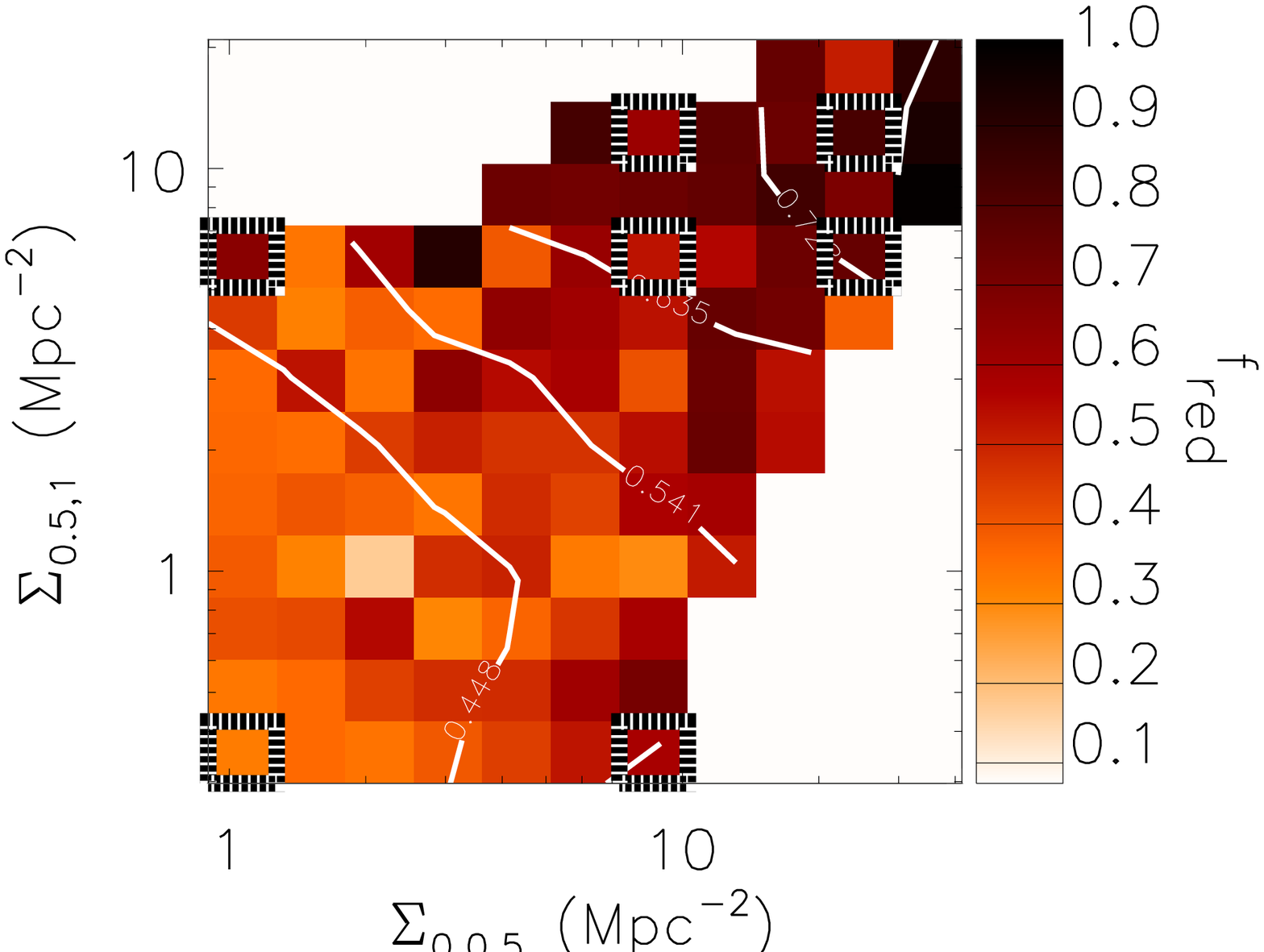}
                \includegraphics[width=0.333\textwidth]{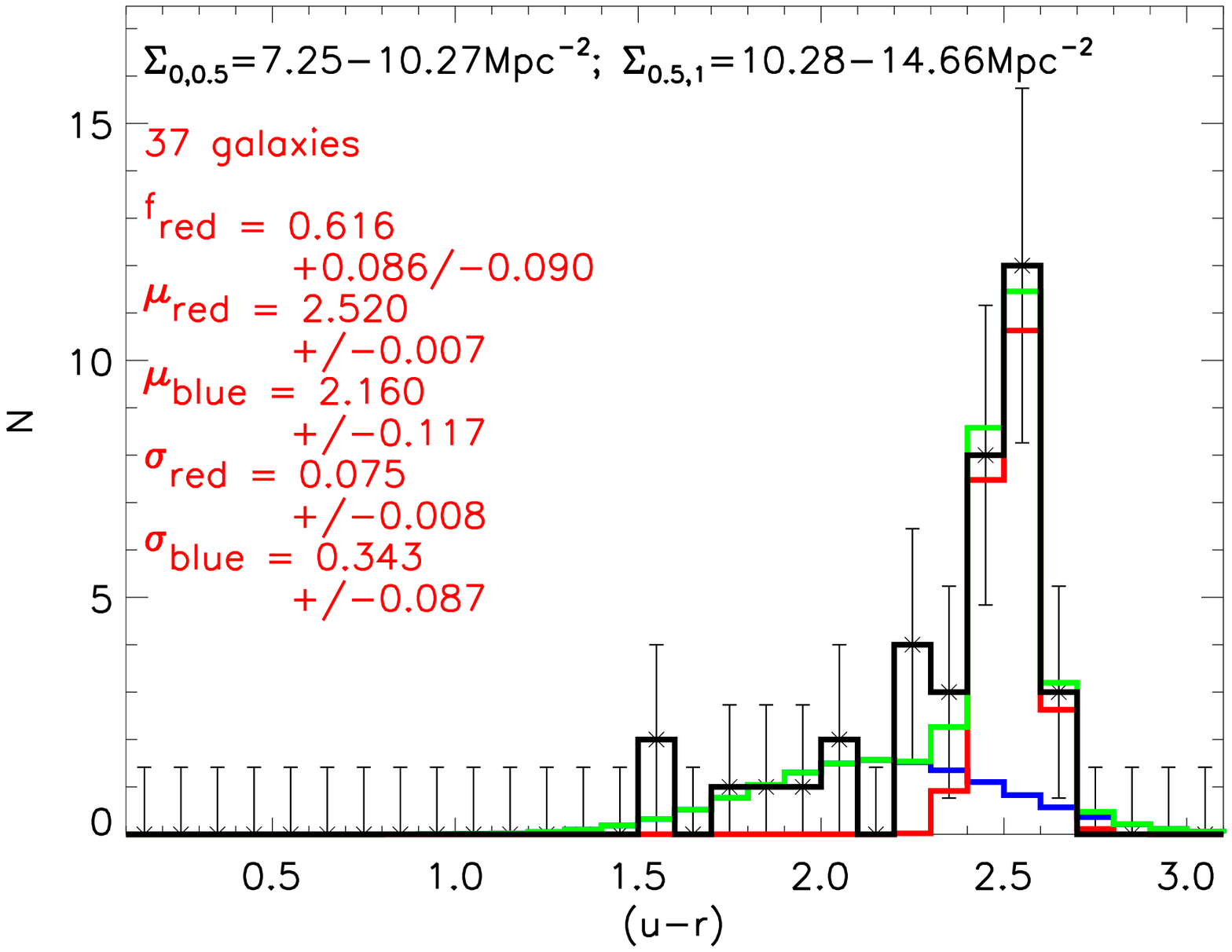}
                \includegraphics[width=0.333\textwidth]{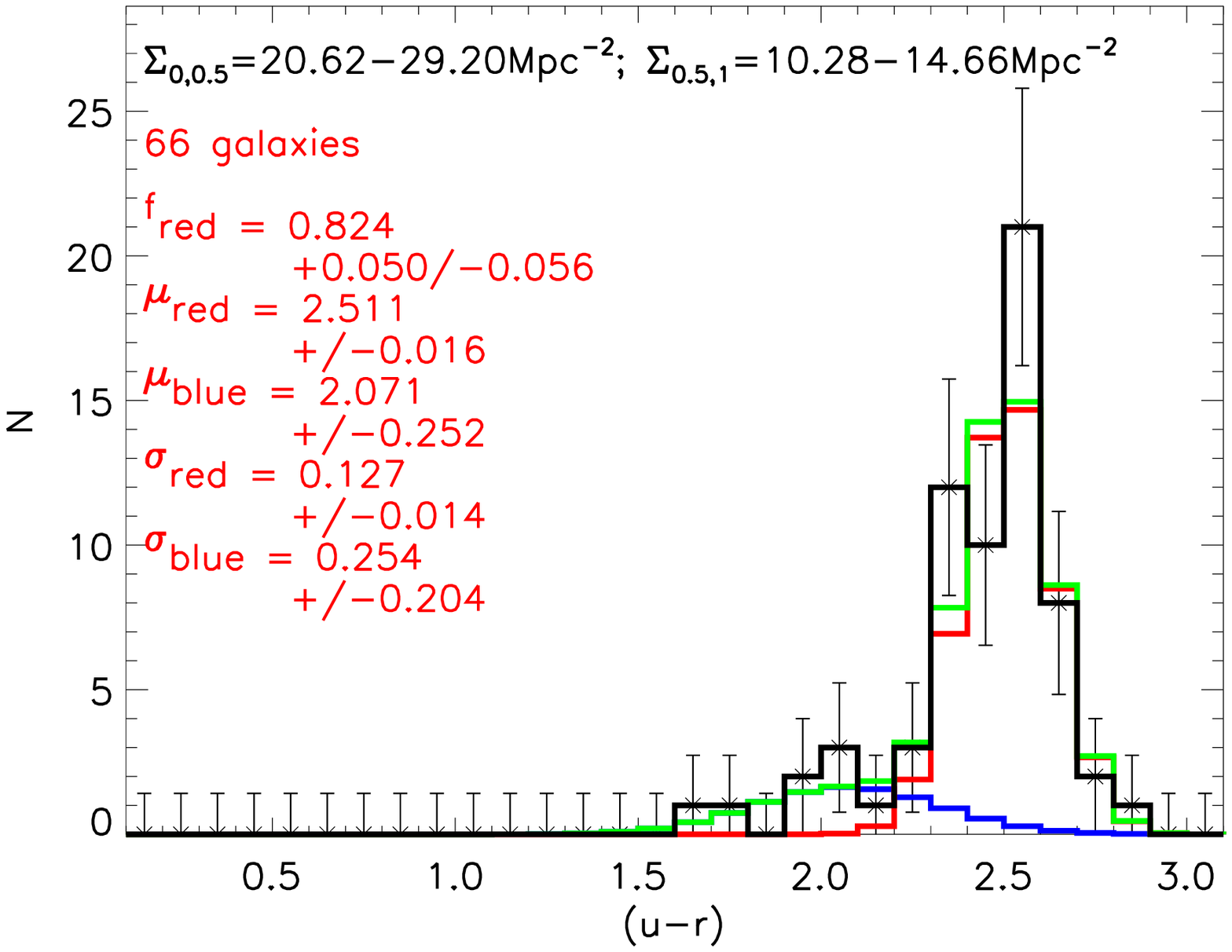}
}
\vspace{0.02\textheight}
              \centerline{
                \includegraphics[width=0.333\textwidth]{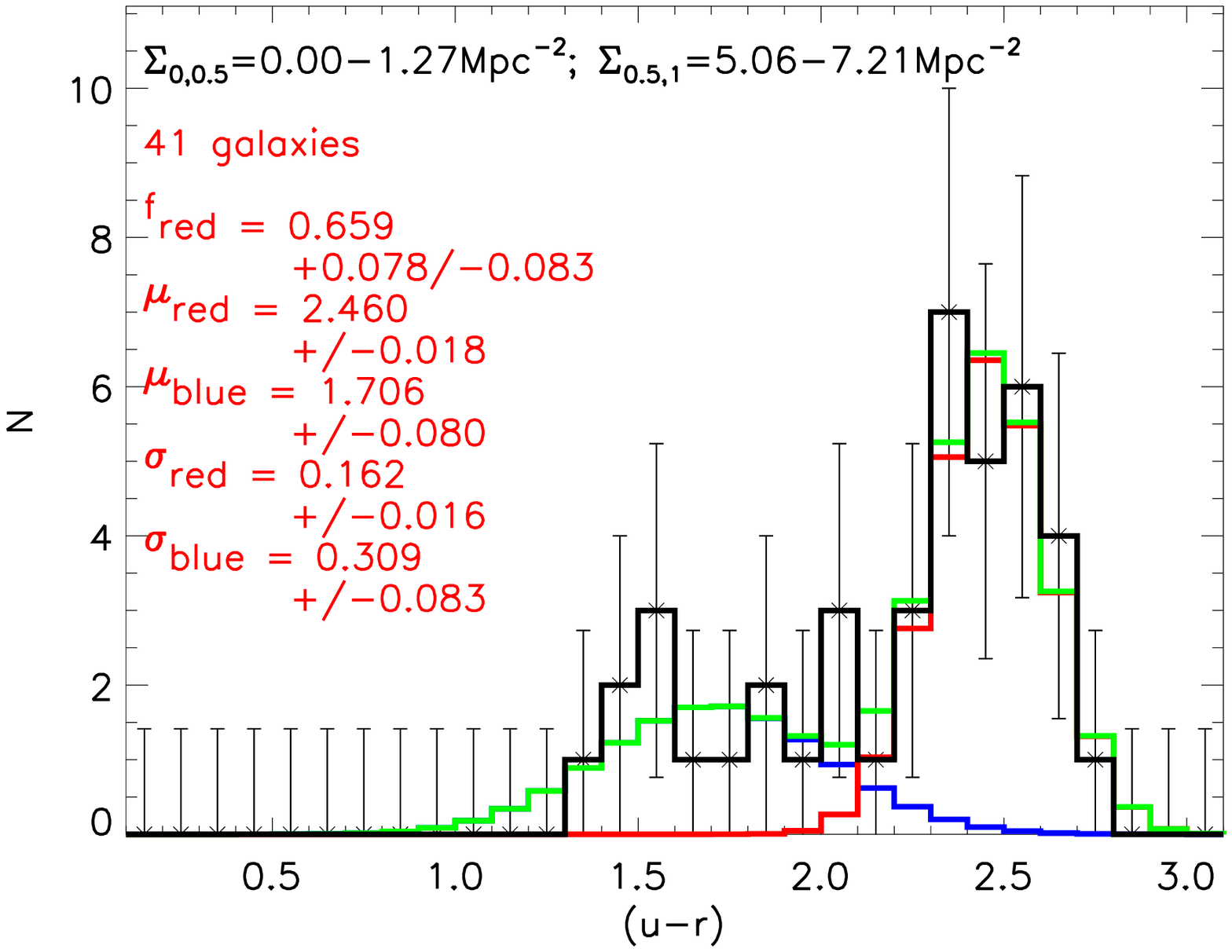}
                \includegraphics[width=0.333\textwidth]{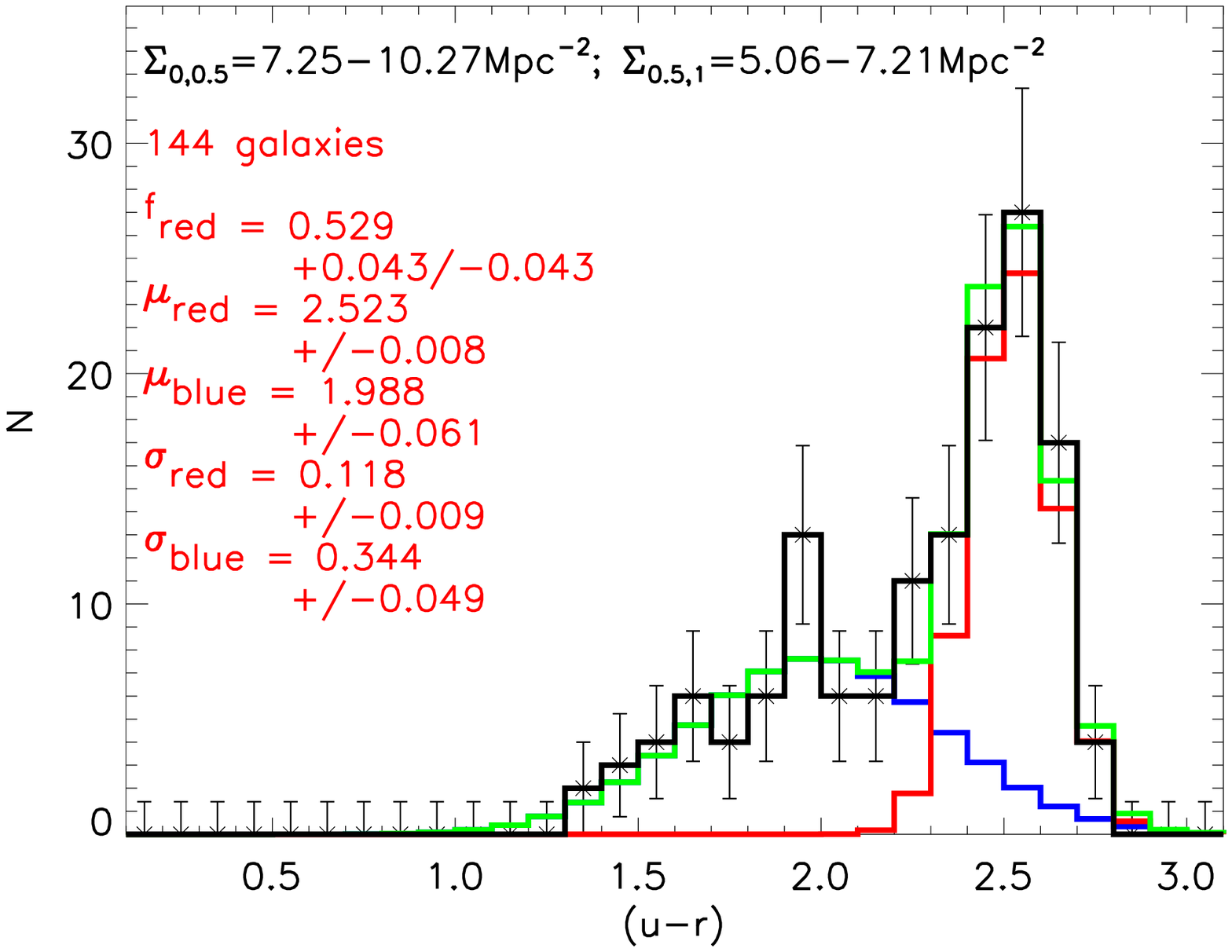}
                \includegraphics[width=0.333\textwidth]{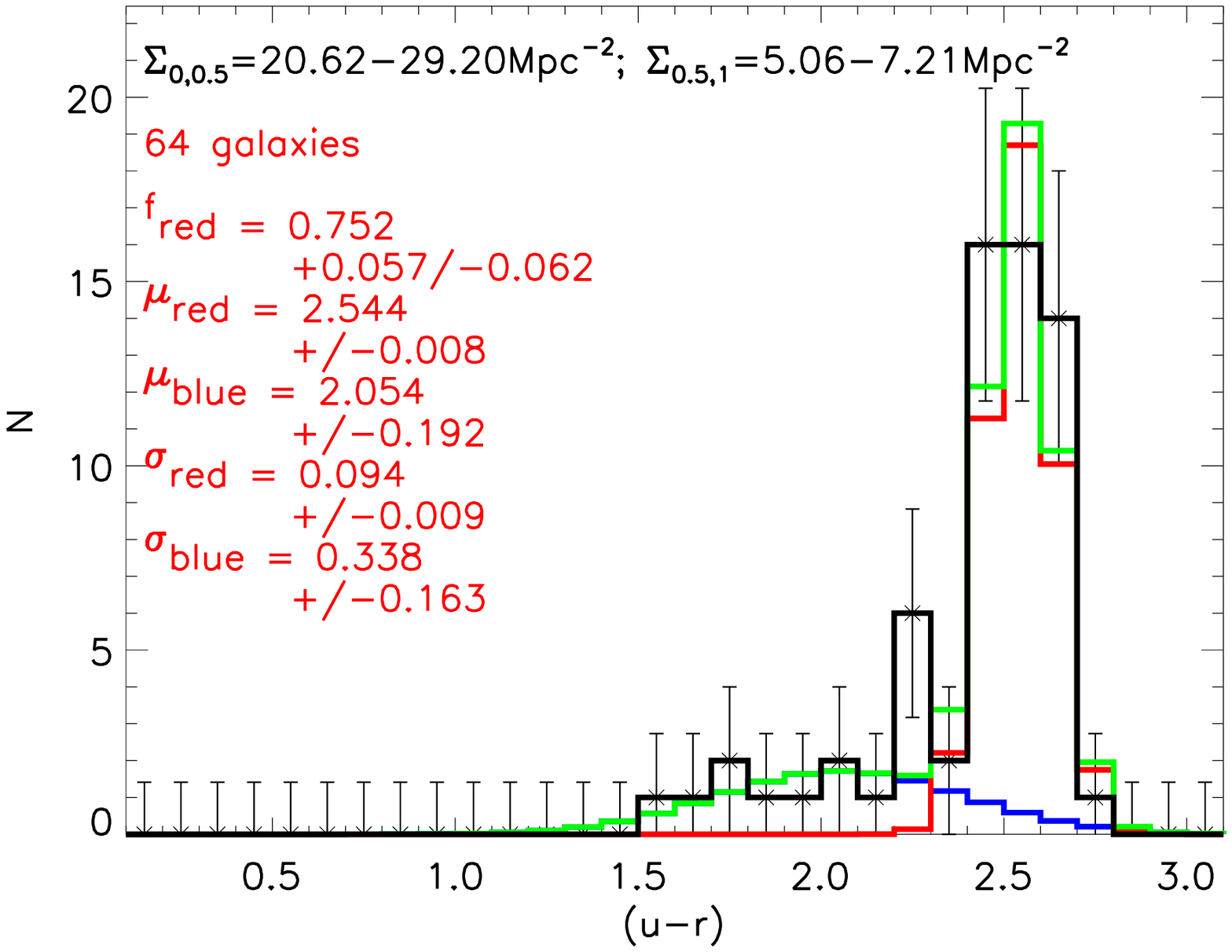}
}
\vspace{0.02\textheight}
              \centerline{
                \includegraphics[width=0.333\textwidth]{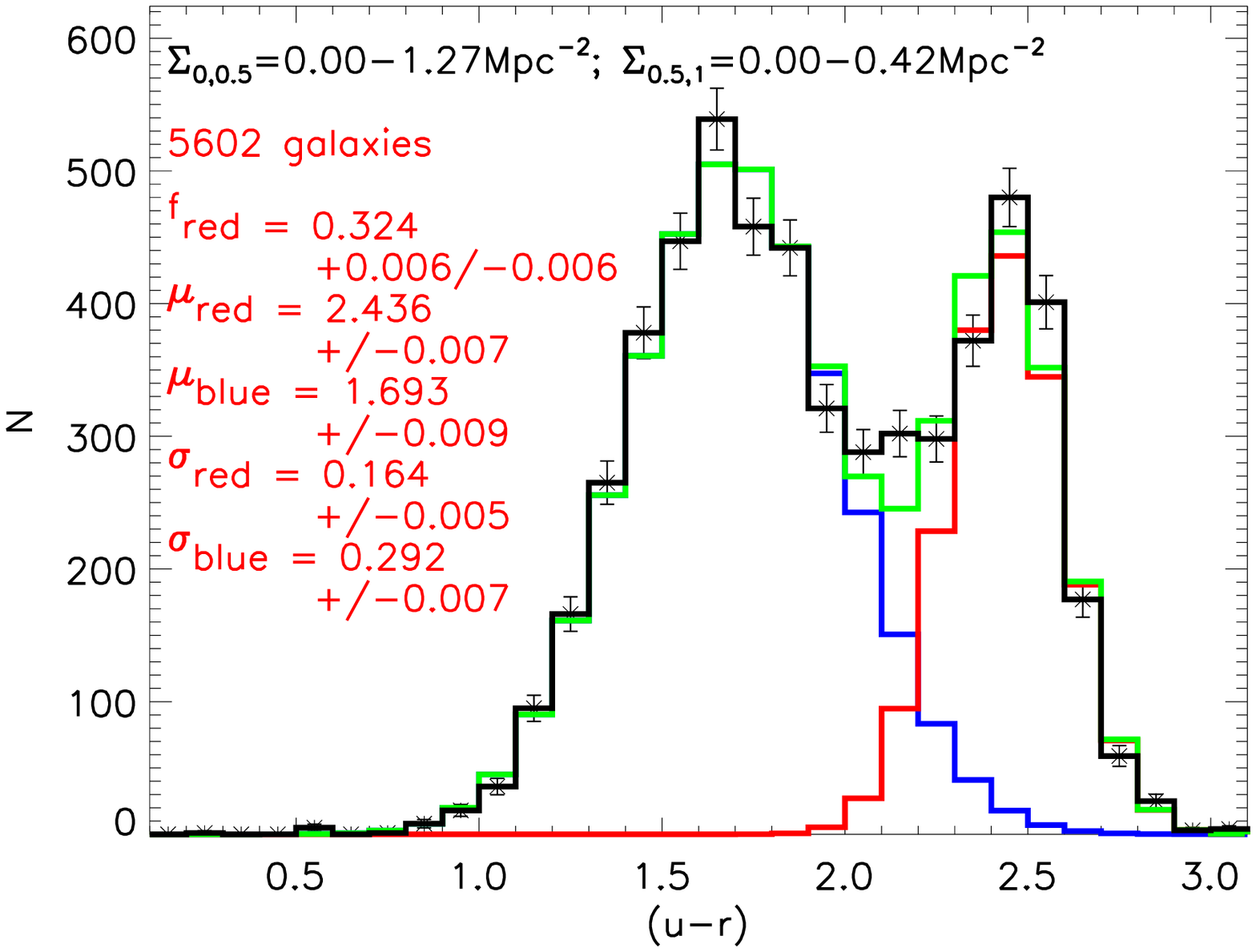}
                \includegraphics[width=0.333\textwidth]{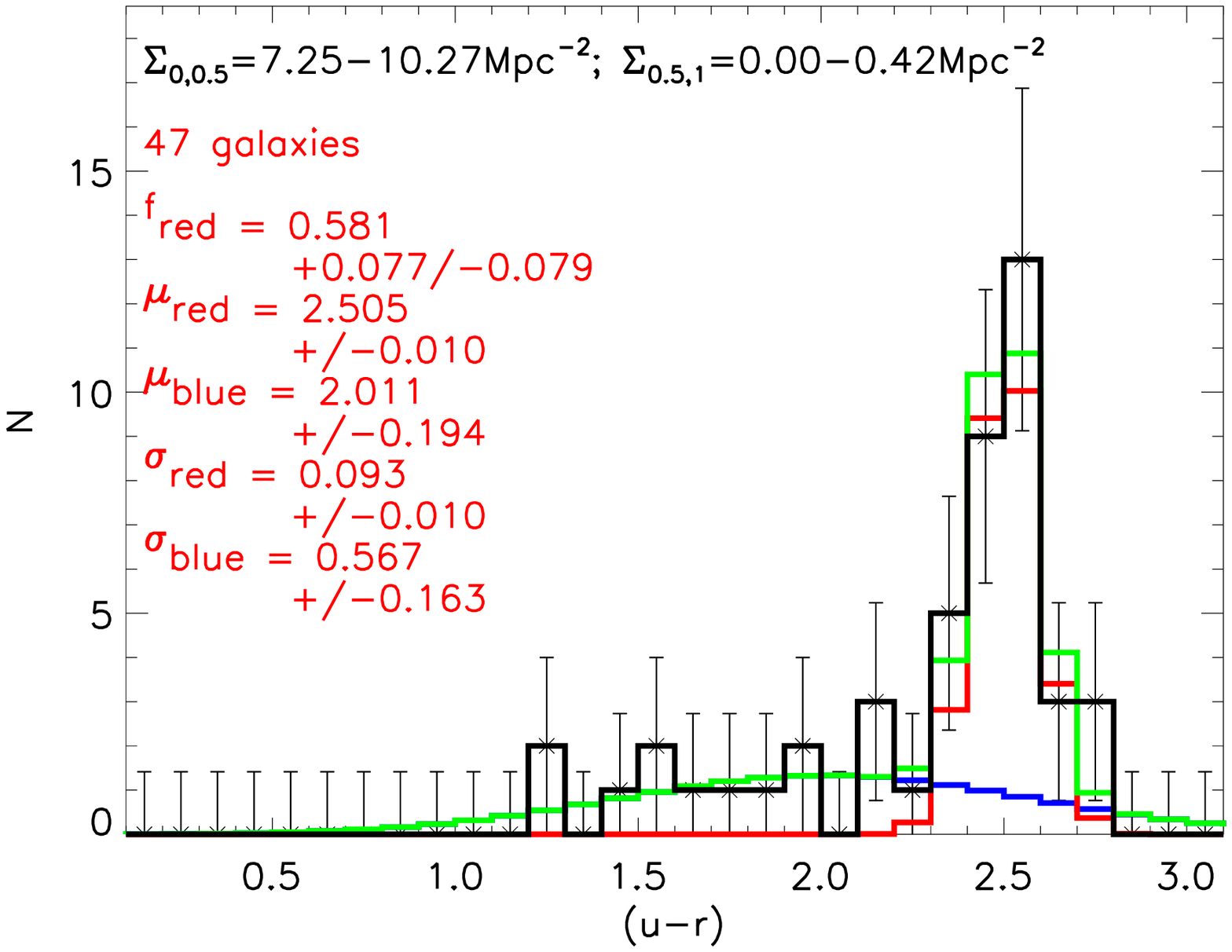}
                \hspace{0.333\textwidth}
}
\caption{Example double gaussian fits to the colour distribution of
  $-20.5\leq\Mr\leq-20.0$ galaxies binned by density on scales $\dsca$
  and $\dscc$.  {\it Top left:} The density dependence of the red galaxy
  fraction, $\fr$ for galaxies on scales $\dsca$ and $\dscc$, as the
  left-hand panel of Figure~\ref{figure:fredscale}, except limited to
  galaxies in the $-20.5\leq\Mr\leq-20.0$ luminosity bin.  Other
  panels: Fits to the $u-r$ colour distribution for galaxies in the
  bins indicated using black and white squares in the top-left panel.
  Panels refer to the bin in the equivalent position in the
  density-density plane.  The colour distribution (points and black
  line) is in each case fit by a double-gaussian (green line) which is
  the combination of blue and red peaks (represented by blue and red
  lines).  $\sqrt(N+2)$ errors are assigned to the counts in each bin
  of colour.  The number of contributing galaxies and the derived
  parameters are also indicated in each panel.  }
  \label{figure:egfit2d}
\end{figure*}

We illustrate that the fits thus produced are sufficiently good in
Figure~\ref{figure:egfit2d}.  To do so we must select a single
luminosity bin, $-20.5\leq\Mr\leq-20.0$. The top-left panel shows that
the fraction of red galaxies, $\fr$ can clearly be traced as a
function of density computed on 0--0.5 $\Mpc$ vs 0.5--1 $\Mpc$ scales.
To illustrate fitted populations at a variety of multiscale densities,
and in particular including bins containing close to our minimum
number of galaxies (25), we select the seven bins from this
density-density plane as indicated by black and white squares.  For
each of these bins the full double gaussian fit to the $u-r$ colour
distribution is shown in the other panels.  Most importantly, both
peaks can be seen in each panel, and the fits are as good as the data
allows. For bins containing few galaxies, especially those
with few blue (or red) galaxies, the fit parameters are uncertain.
Whilst this is formally accounted for in fitting errors (see 
Section~\ref{sec:fitting}), errors on $\fr$ are computed statistically 
sing the \citet{Gehrels86} approximation, combining
Poissonian and Binomial statistical errors. For double gaussian
fits to a large number of galaxies this approximates the fitting
error, whilst for smaller numbers the statistical error is
larger. 
However, the scatter between different, independent bins in the 
density-density plane provides the best estimate of the "true" error: 
this is consistent with the statistical error, and is reduced once 
the three luminosity bins are coadded.



\citet{Kauffmann04},\citet{Blanton06} and \citet{Blanton07} have all claimed that galaxy
properties exhibit little dependence on density beyond $1\Mpc$.
Beyond this separation, the correlation function of blue or red
galaxies is consistent within $5\%$ of that expected if the type of
galaxy only responds to its halo mass and halocentric position
\citep{Blanton07}. Our choice of pairs of small and large scales,
  whose effects are to be compared (namely 0--0.5 $\Mpc$ vs 0.5--1, 0--1
  vs 1--2, 0--2 vs 2--3), is designed to provide complimentary
information. In each case we have selected an ``interior'' (circular)  
annulus and a larger scale annulus which meet but do not overlap. 
This provides independently
measured quantities, but means that any signal on scales just beyond
that measured by the small scale should be picked up on the large
scale. To facilitate discussion of these scales, we introduce the 
division radius, $\rdiv$ which is equal to 0.5, 1 and 2$\Mpc$ respectively 
for our three pairs of small and large scales. 
The $\dscb$ versus $\dscd$ plane examines the influence of
$>1\Mpc$ scales.  Smaller scales ($<0.5\Mpc$) trace local
substructure. This is examined in the $\dsca$ versus $\dscc$ plane.
Conversely, the $\dscj$ and $\dsce$ plane probes the largest scales
measured.  If galaxies only correlate with the properties of their
embedding halo, then one might expect large scale density only to
matter where those scales trace galaxies within the same halo (division 
radius, $\rdiv \lesssim 2r_{200}$, a typical halo
diameter). This occurs at $\Mhalo\sim2\times10^{13}\Msol$
($5\times10^{12}\Msol$,$8\times10^{13}\Msol$) for $\rdiv = 1\Mpc$
($0.5\Mpc$,$2\Mpc$).



\begin{figure*}
              \centerline{\includegraphics[width=0.3\textwidth]{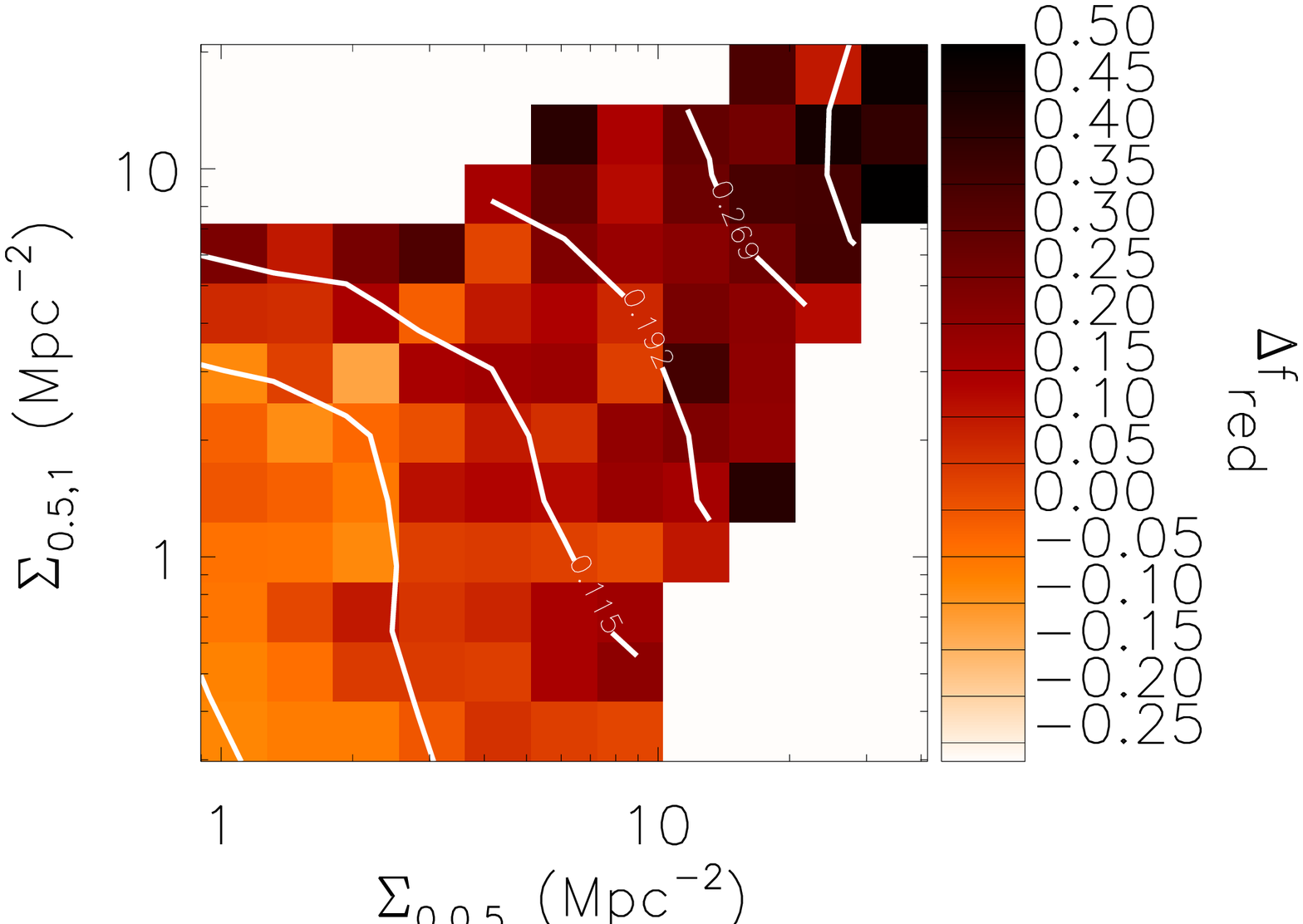}
                          \hspace{0.02\textwidth}
                          \includegraphics[width=0.3\textwidth]{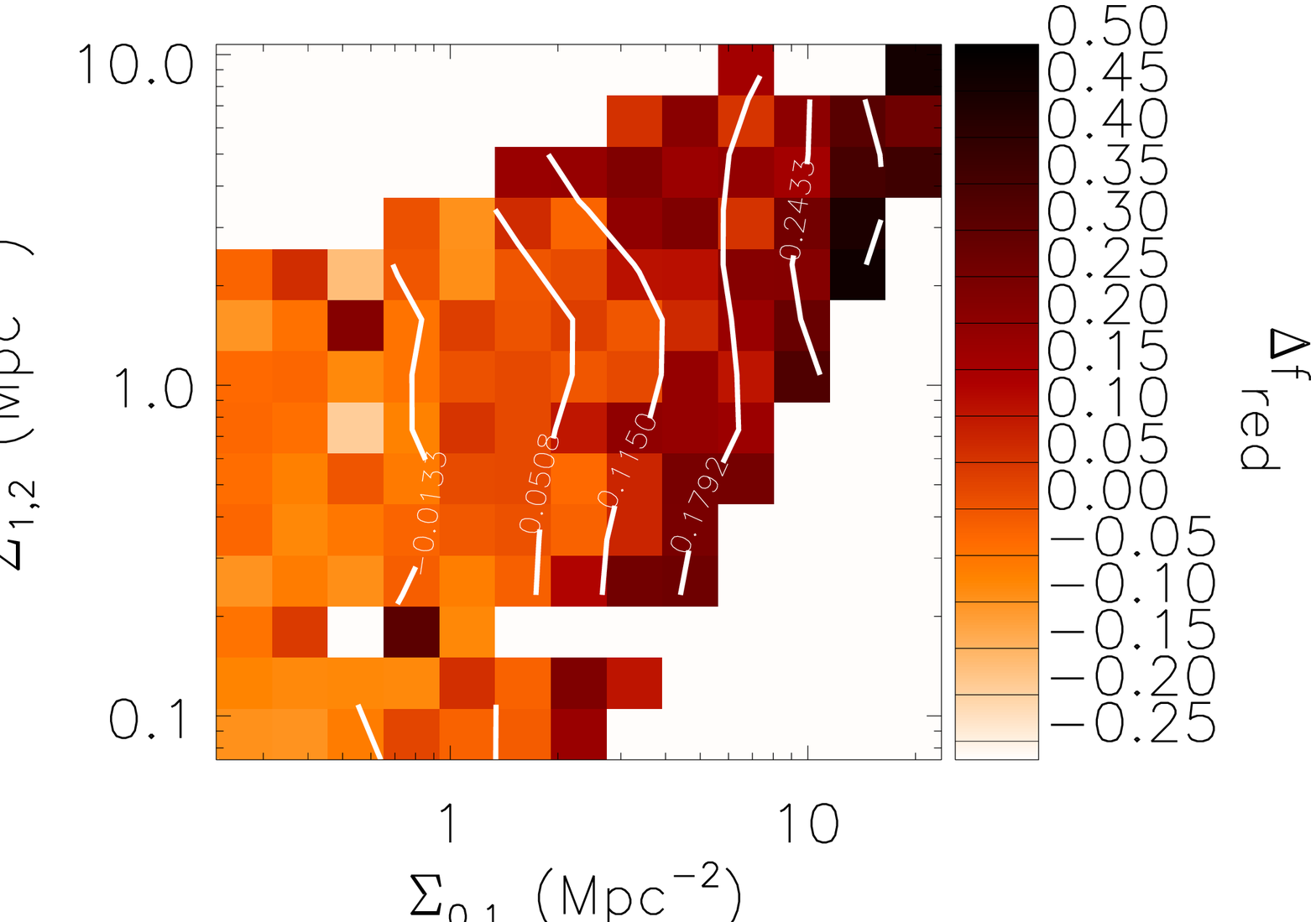}
                          \hspace{0.02\textwidth}
                          \includegraphics[width=0.3\textwidth]{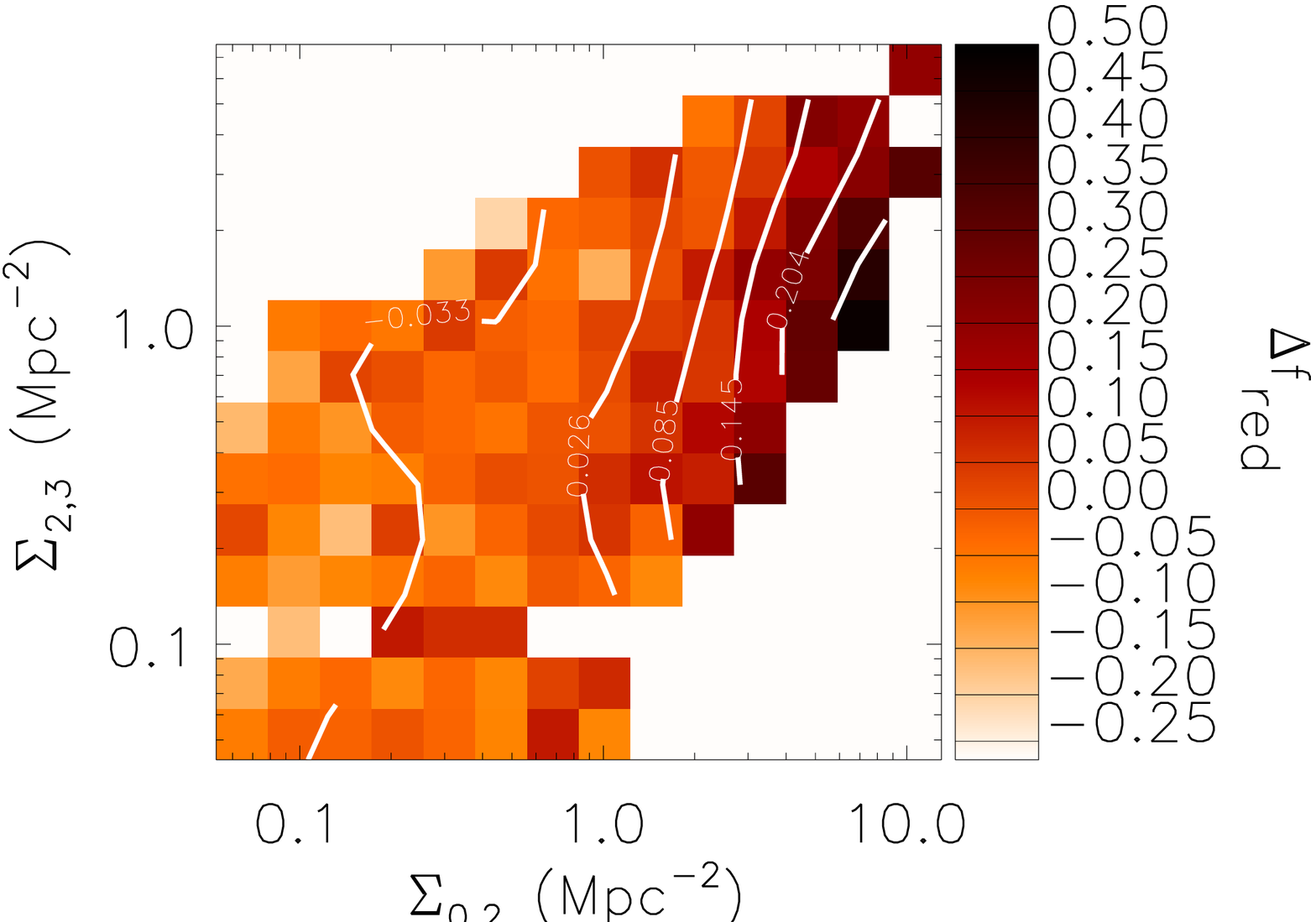}}
  \caption{The density dependence of the red galaxy fraction, $\dfr$, for $\-21.5<\Mr<-20.0$ galaxies binned in two scales of density. 
{\it Left:} on scales $\dsca$ and $\dscc$; {\it Centre:} on scales $\dscb$ and $\dscd$; {\it Right:} on scales $\dscj$ and $\dsce$.
Almost vertical contours illustrate that $\dfr$ depends mostly on small scale density.}
  \label{figure:fredscale}
\end{figure*}

\begin{figure*}
              \centerline{\includegraphics[width=0.3\textwidth]{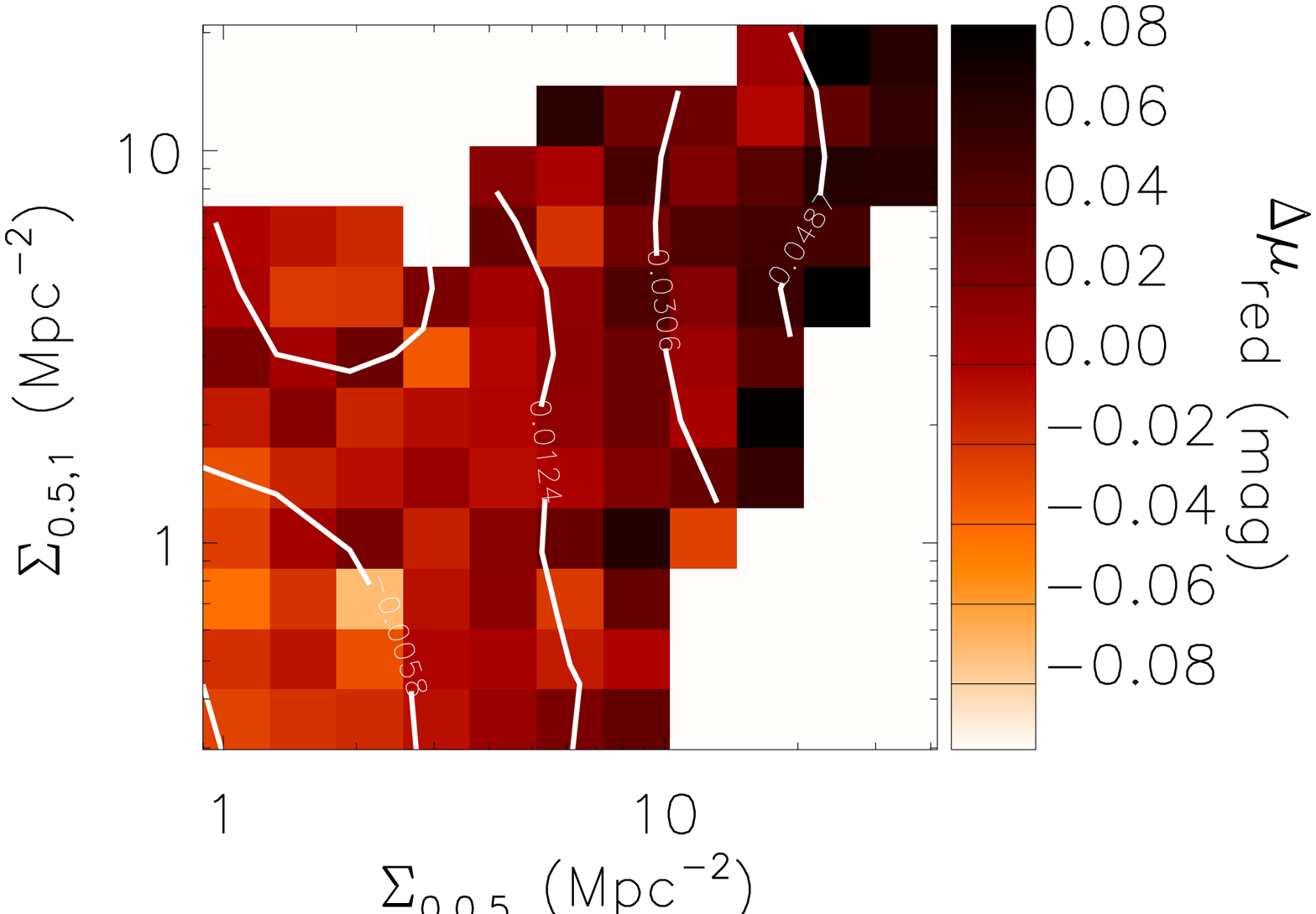}
                          \hspace{0.02\textwidth}
                          \includegraphics[width=0.3\textwidth]{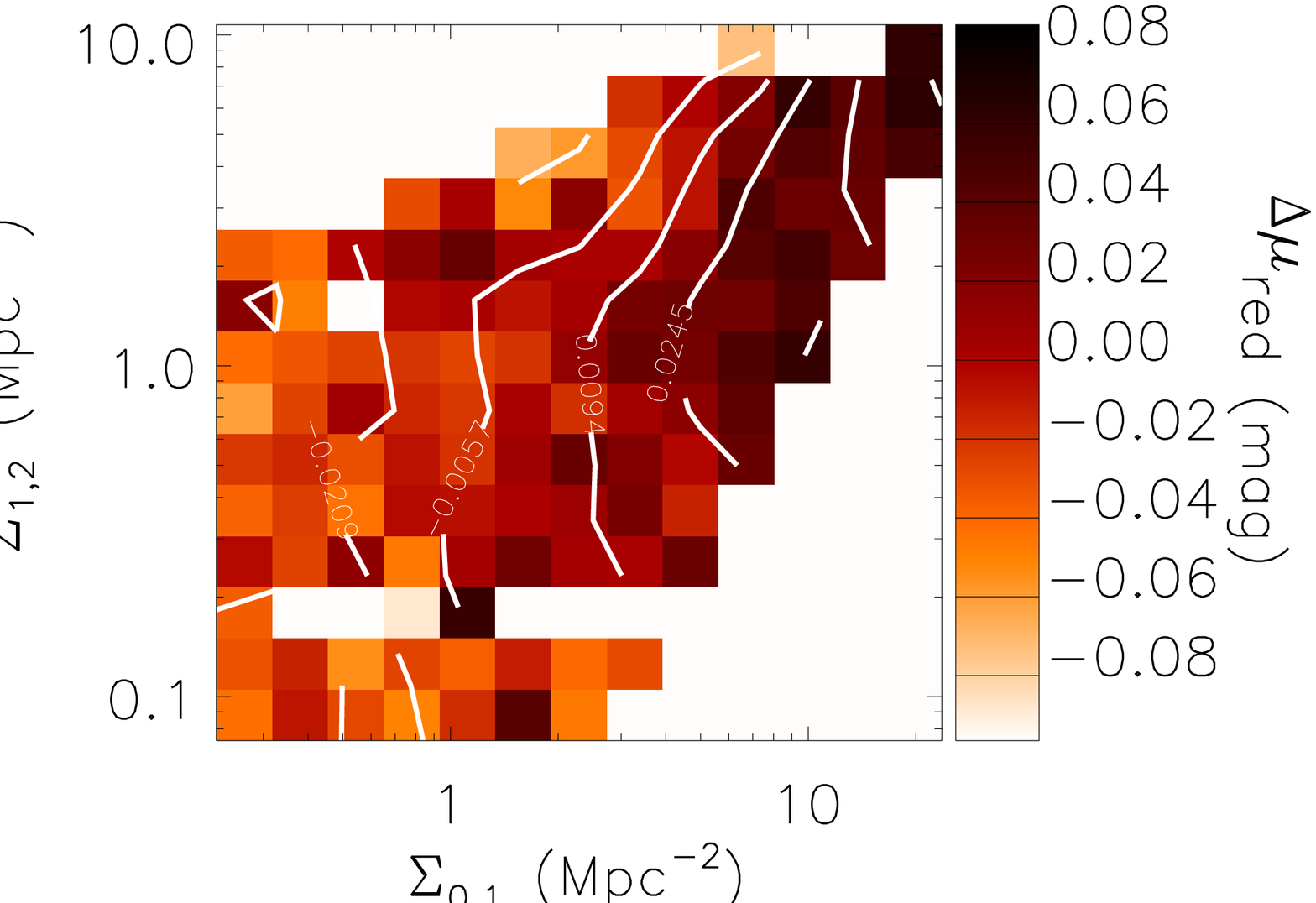}
                          \hspace{0.02\textwidth}
                          \includegraphics[width=0.3\textwidth]{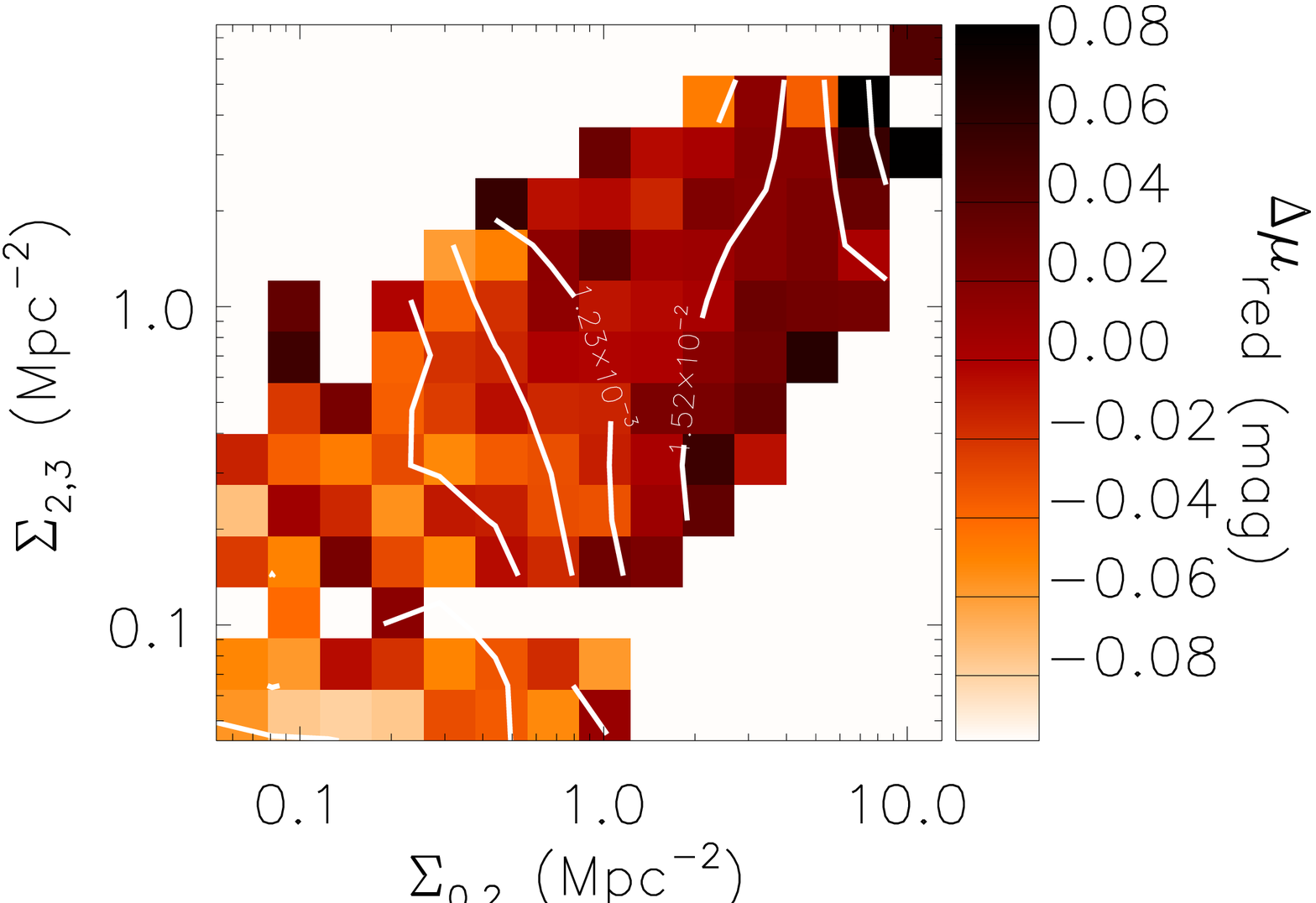}}
              \centerline{\includegraphics[width=0.3\textwidth]{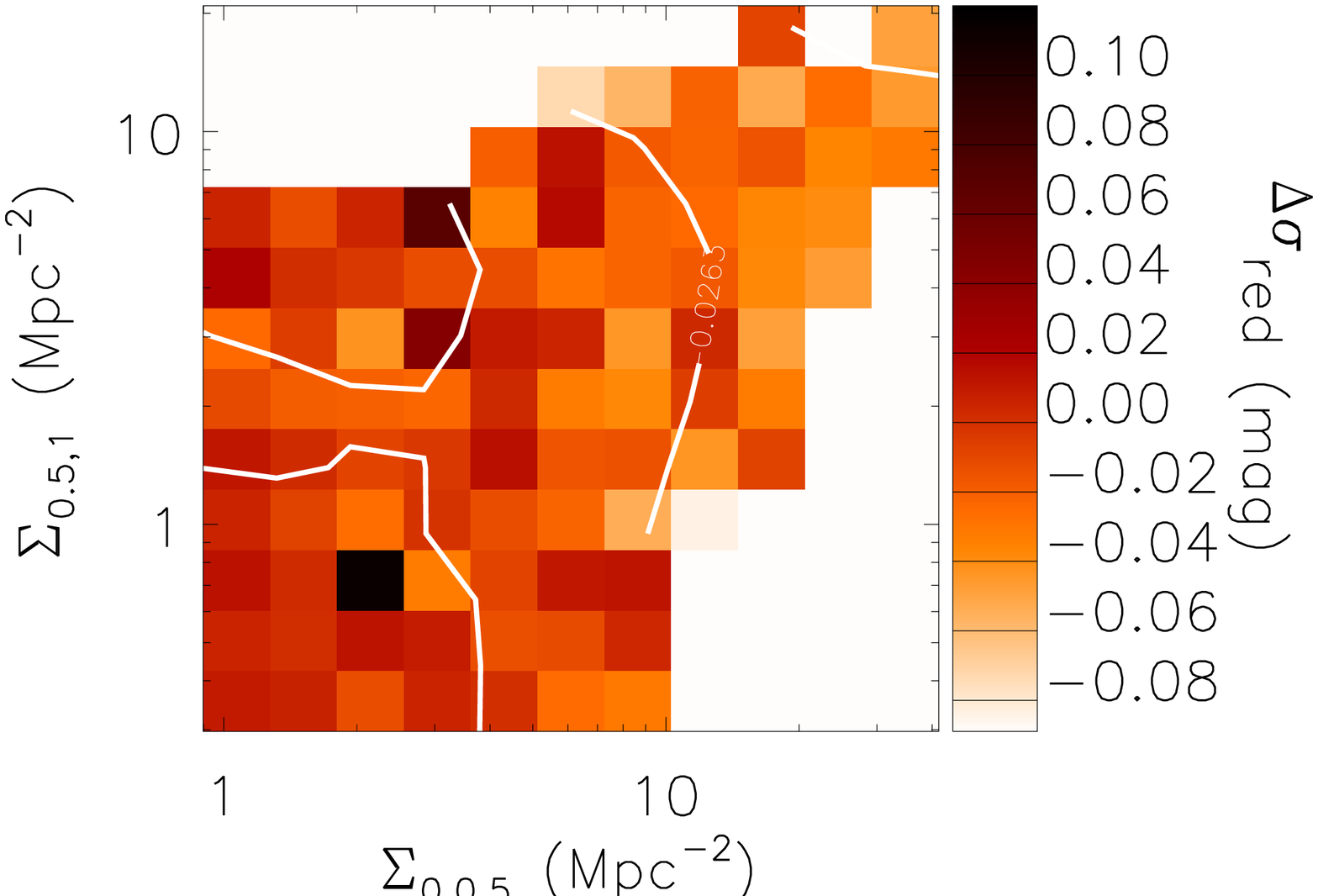}
                          \hspace{0.02\textwidth}
                          \includegraphics[width=0.3\textwidth]{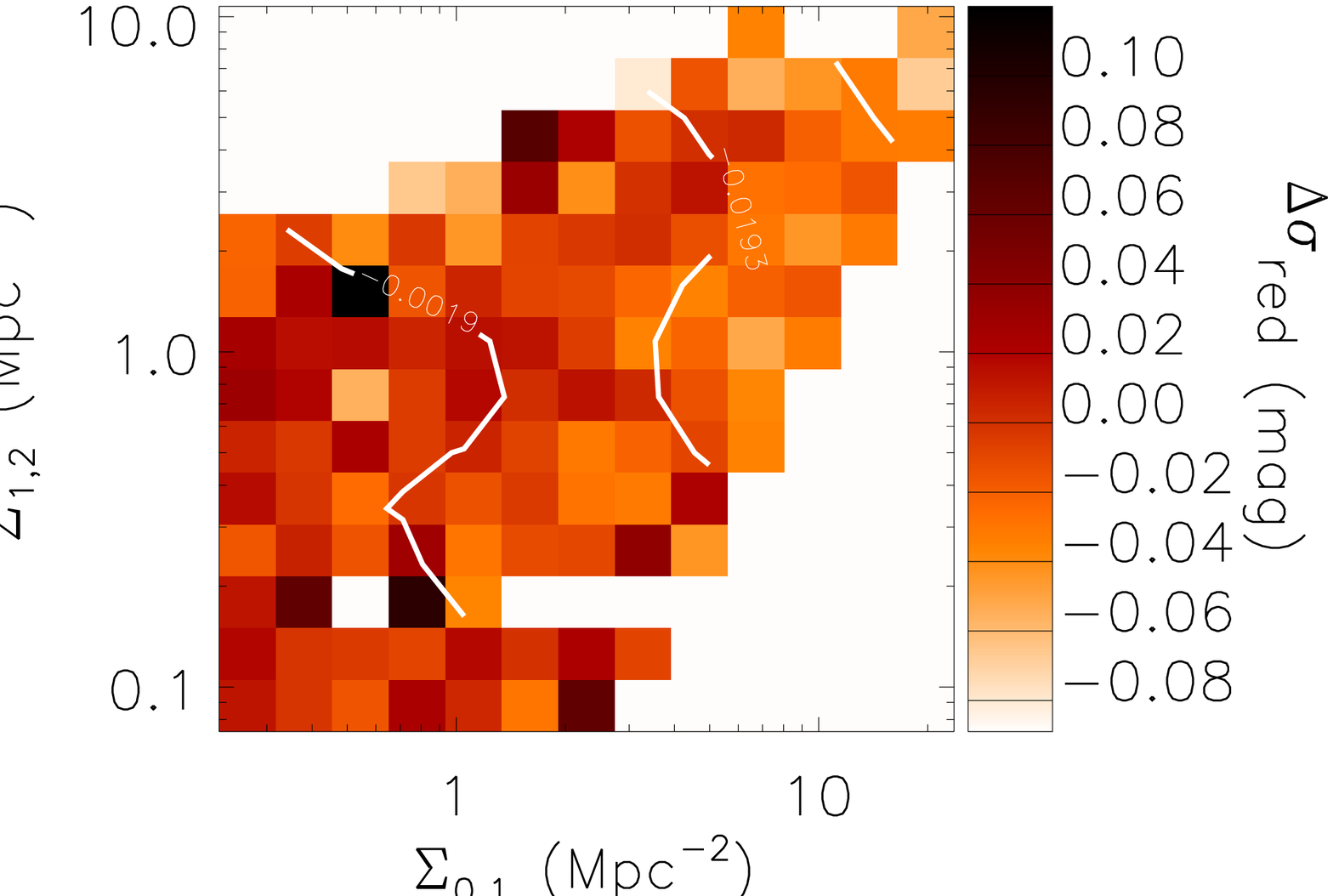}
                          \hspace{0.02\textwidth}
                          \includegraphics[width=0.3\textwidth]{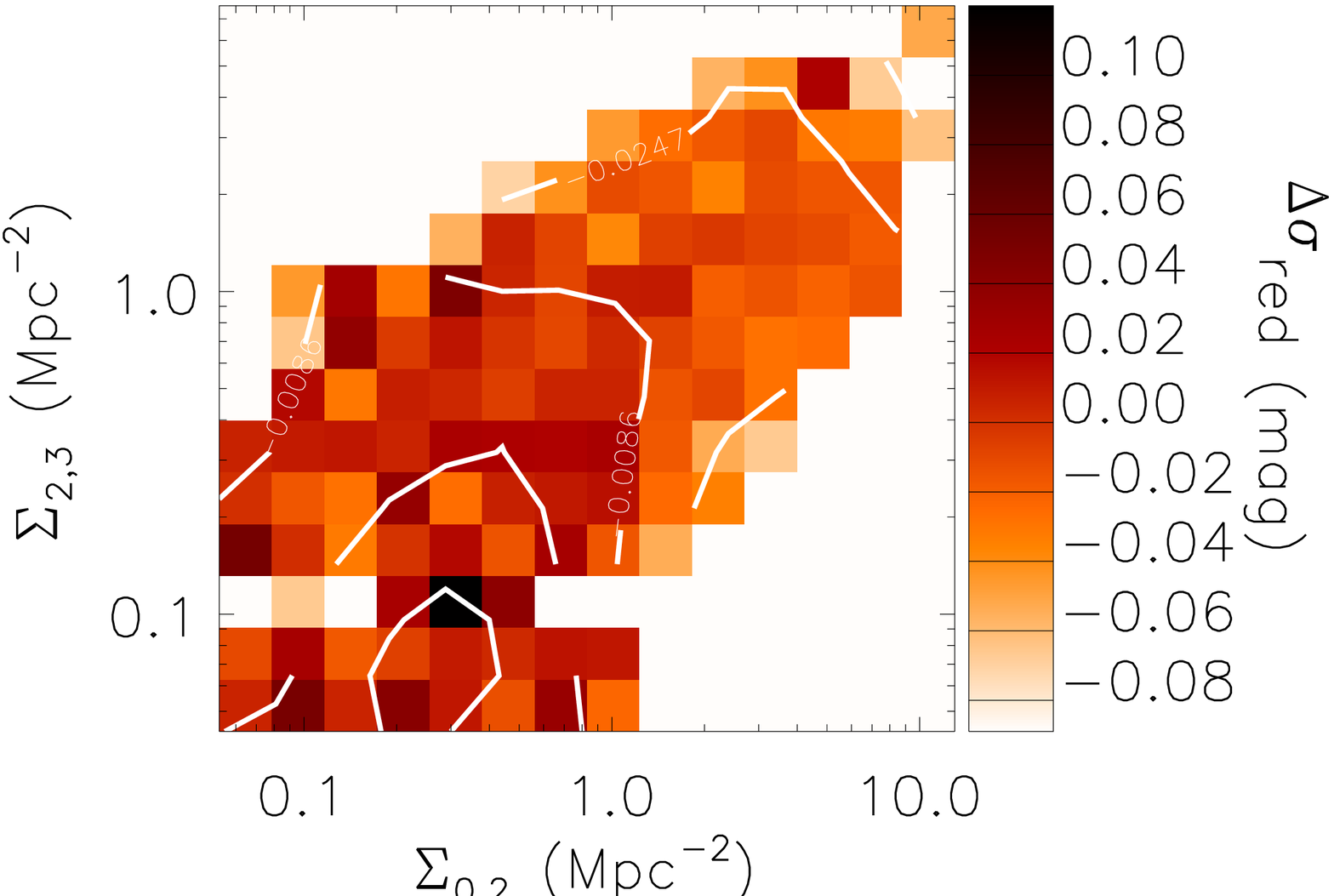}}
  \caption{The density dependence of the red peak parameters $\mur$ (above) and 
    $\sigmar$ (below), 
    for $\-21.5<\Mr<-20.0$ galaxies binned in two scales of density.
    {\it Left:} on scales $\dsca$ and $\dscc$; {\it Centre:} on scales $\dscb$ and $\dscd$; {\it Right:} on scales $\dscj$ and $\dsce$.
  }
  \label{figure:redpeakscale}
\end{figure*}

\begin{figure*}
              \centerline{\includegraphics[width=0.3\textwidth]{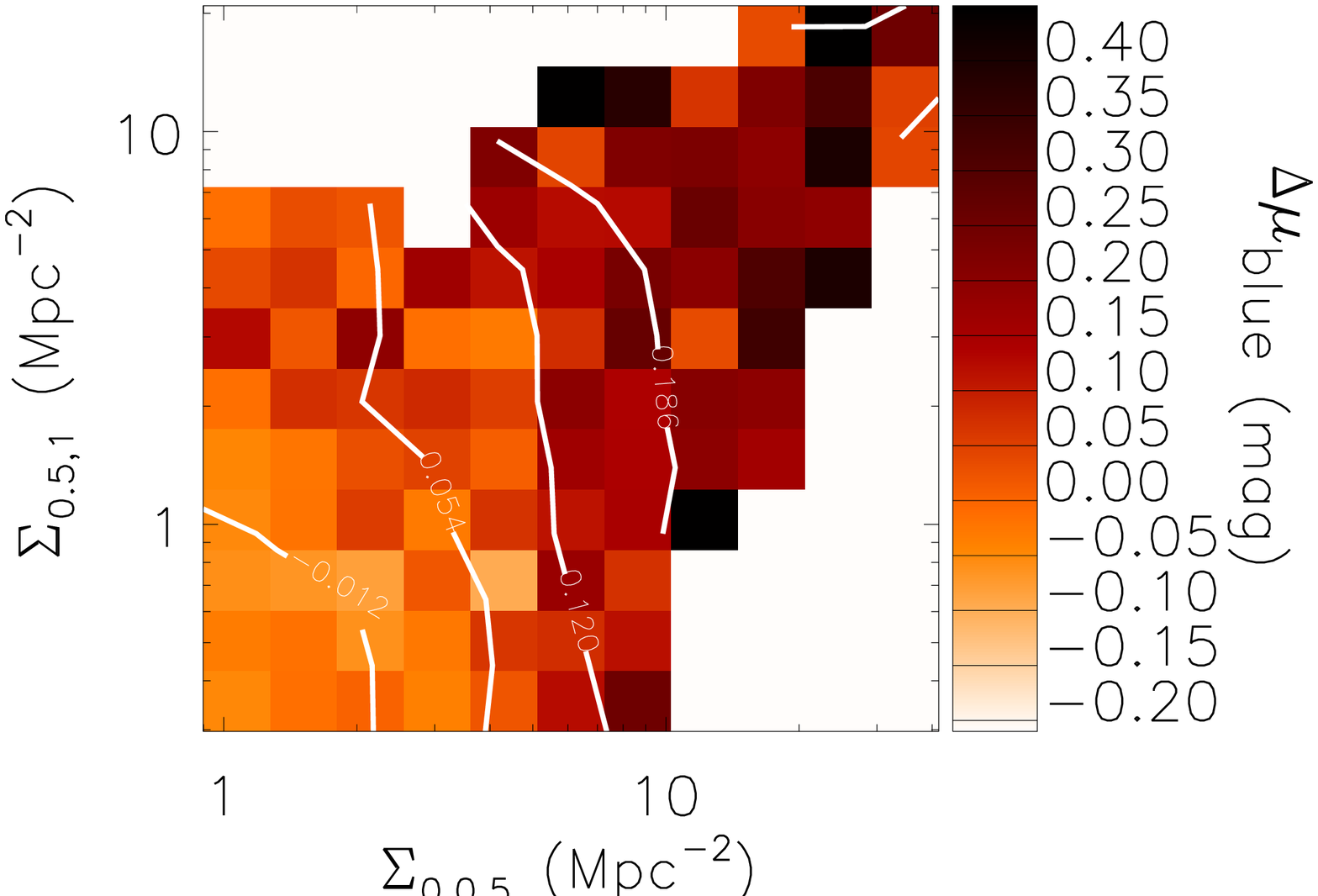}
                          \hspace{0.02\textwidth}
                          \includegraphics[width=0.3\textwidth]{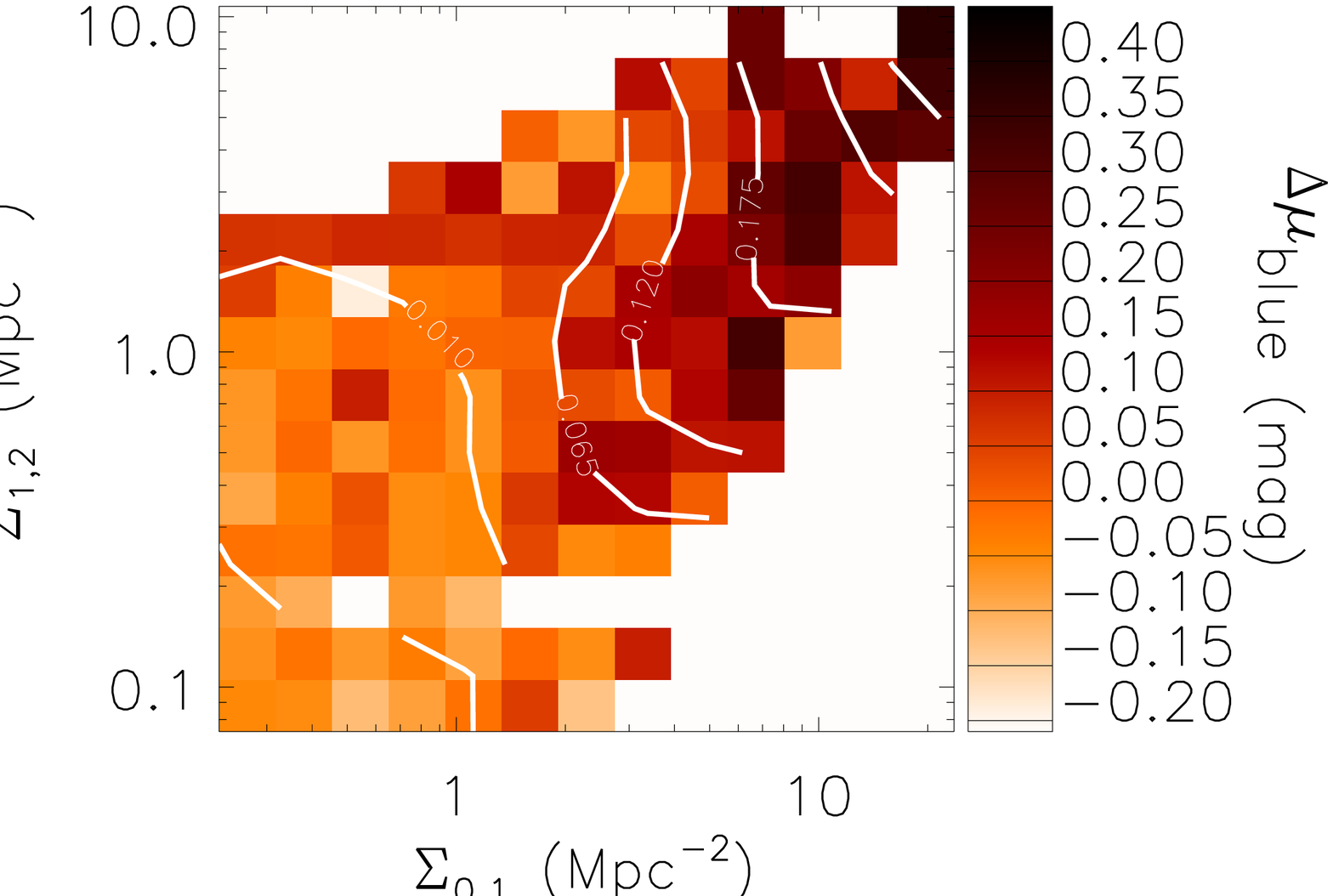}
                          \hspace{0.02\textwidth}
                          \includegraphics[width=0.3\textwidth]{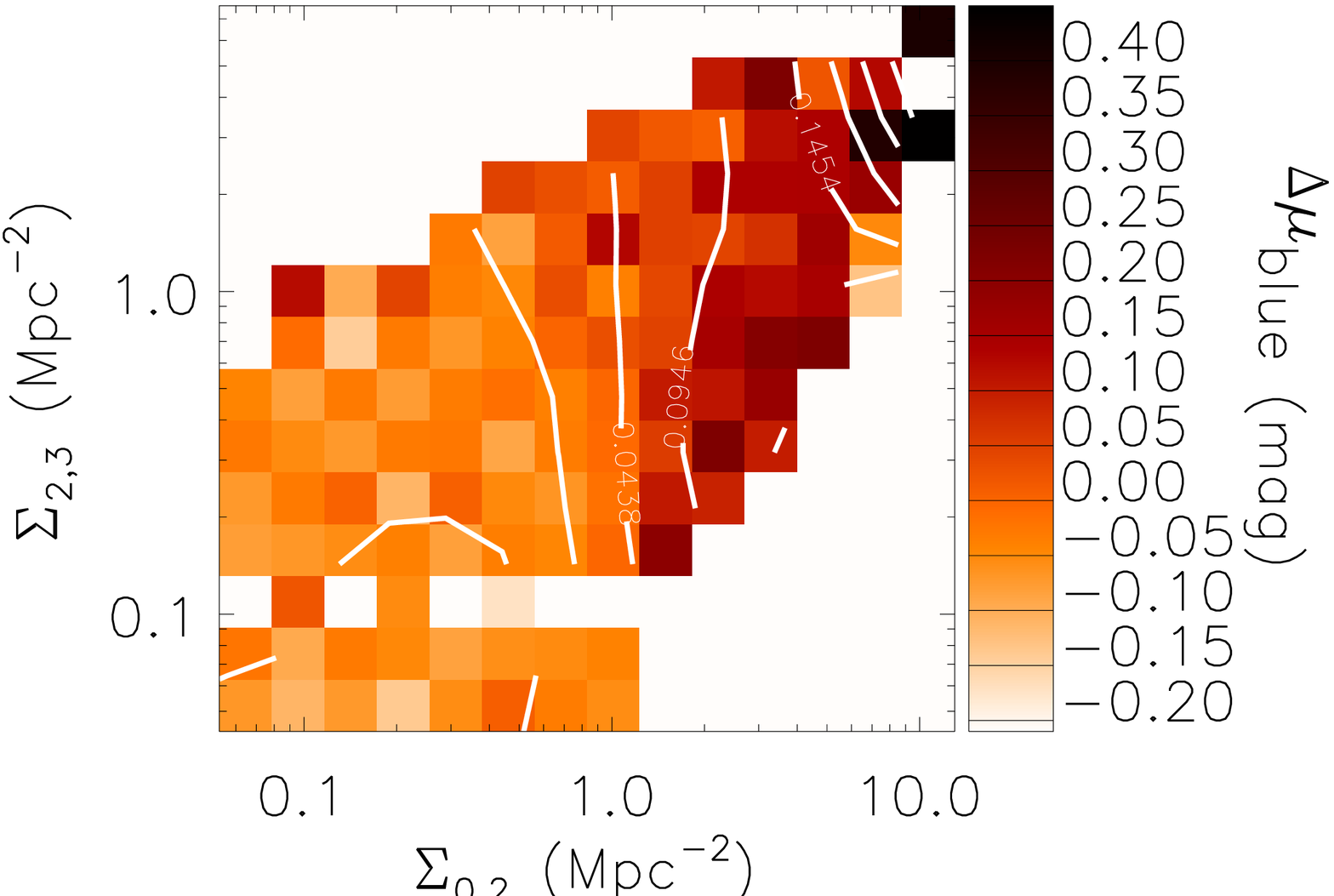}}
              \centerline{\includegraphics[width=0.3\textwidth]{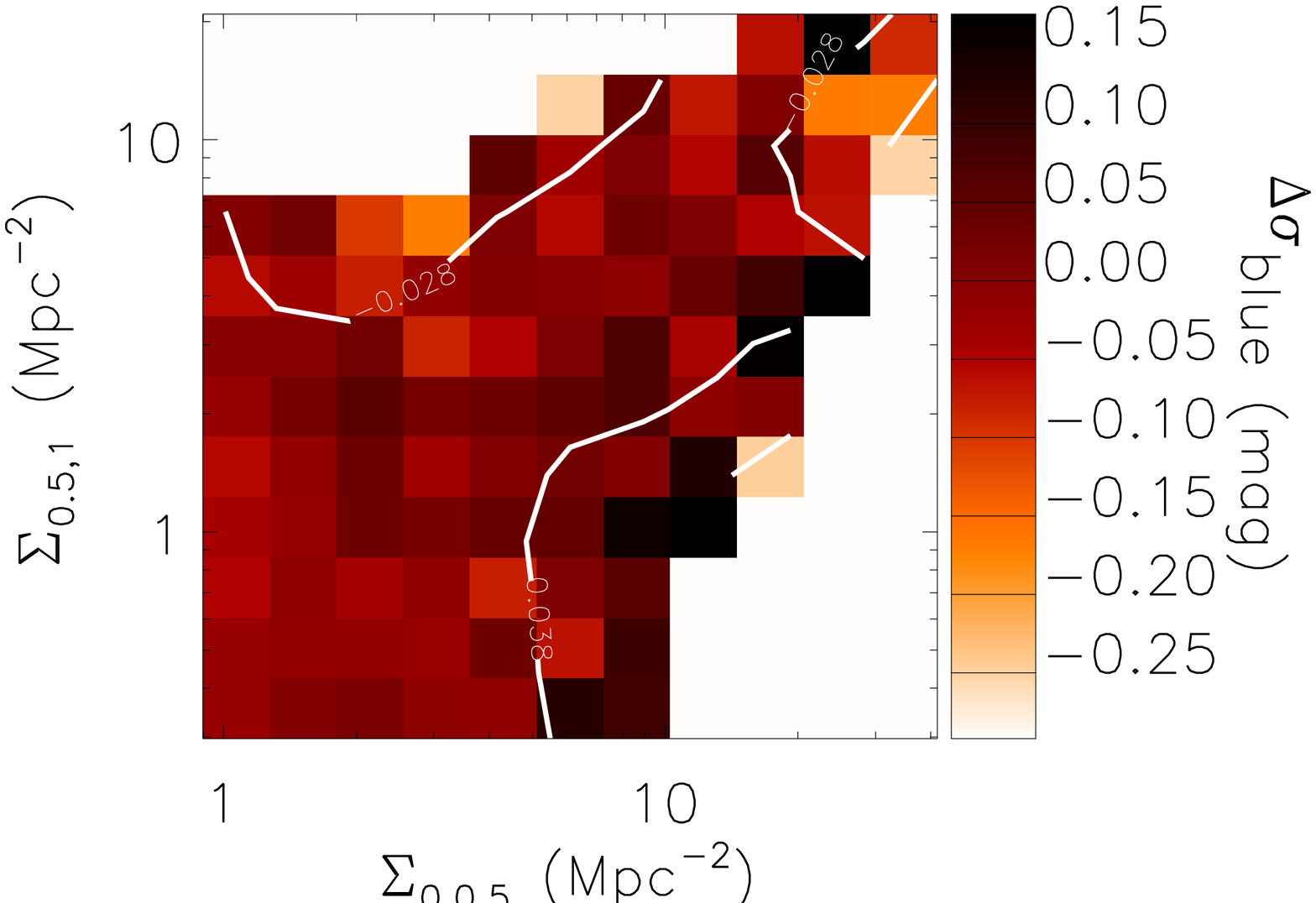}
                          \hspace{0.02\textwidth}
                          \includegraphics[width=0.3\textwidth]{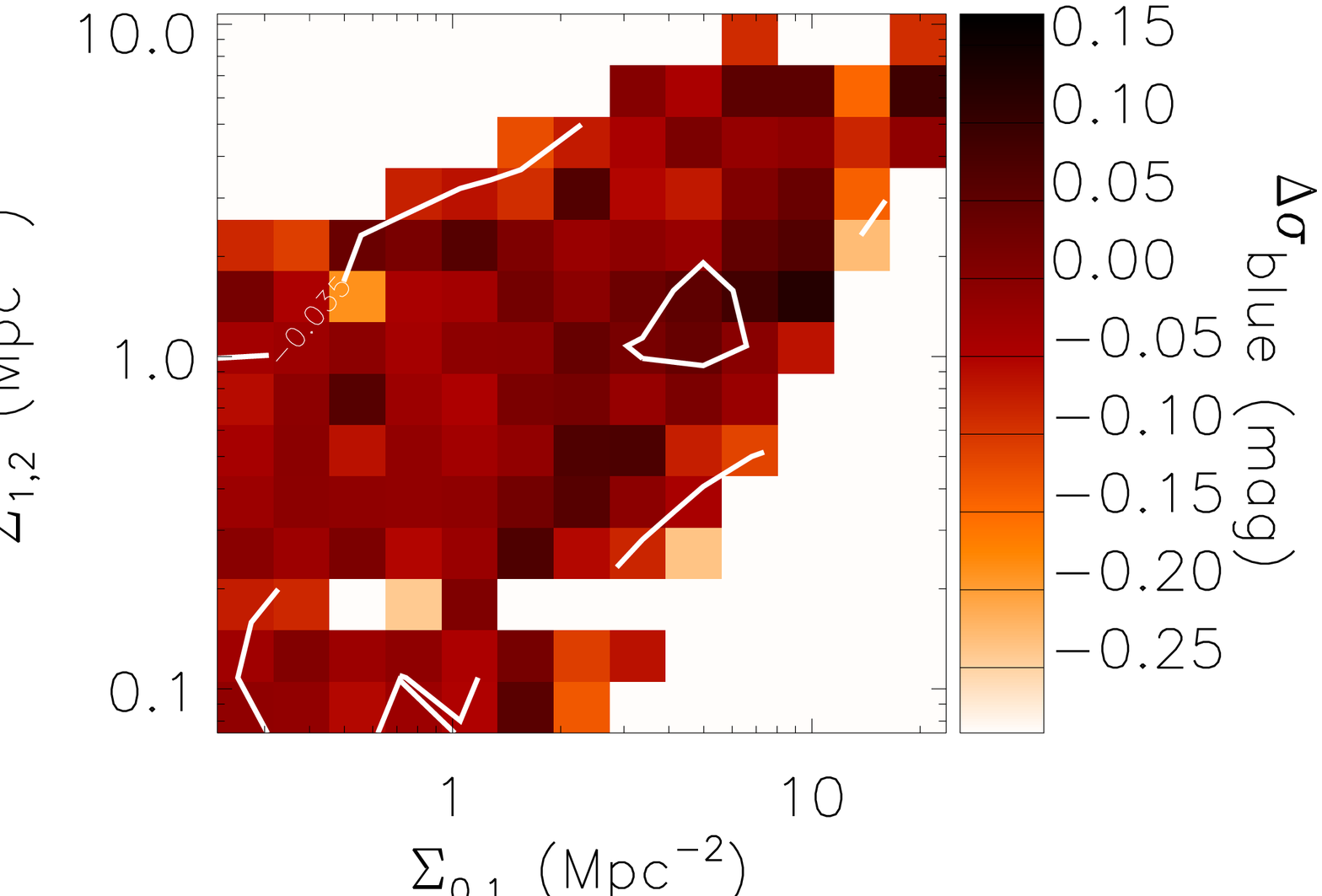}
                          \hspace{0.02\textwidth}
                          \includegraphics[width=0.3\textwidth]{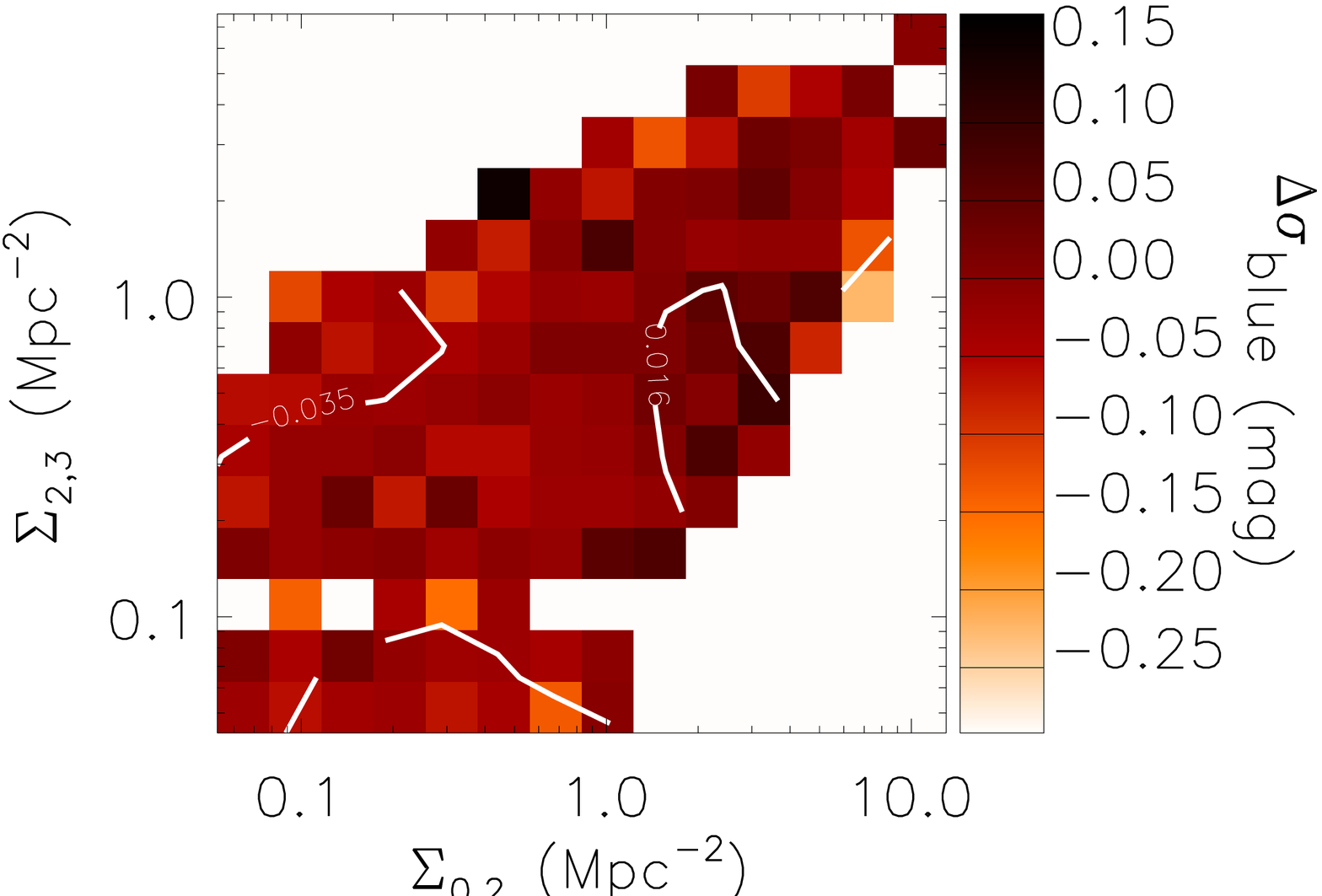}}
  \caption{The density dependence of the blue peak parameters $\mub$ (above) and 
    $\sigmab$ (below), 
    for $\-21.5<\Mr<-20.0$ galaxies binned in two scales of density.
    {\it Left:} on scales $\dsca$ and $\dscc$; {\it Centre:} on scales $\dscb$ and $\dscd$; {\it Right:} on scales $\dscj$ and $\dsce$.
  }
  \label{figure:bluepeakscale}
\end{figure*}

Figures~\ref{figure:fredscale},~\ref{figure:redpeakscale}
and~\ref{figure:bluepeakscale} examine how the relative fraction of
red galaxies ($\dfr$), the parameters describing the red peak
($\dmur$, $\dsigmar$) and the blue peak ($\dmub$, $\dsigmab$) depend
upon small and large scale density for our selected planes of scale. 
Colour-coded bins in this plane are totally independent of each other. 
Overplotted contours are computed for a smoothed map of the data,
using a gaussian smoothing kernel with $\sigma=1.5$pixels
\footnote{The smoothing kernel is computed within a $5\times5$pixel
  box, with pixels beyond the data range assigned the same value as
  the edge pixel.}.  Contours indicate the direction of change for
each parameter in density-density space, and thus the importance of
the two different scales.  Smoothing reduces small scale noise, but
note that local or abrupt features can be smoothed away.


We also quantify the significance of any dependence on large scale
density, at fixed small scale density. For each parameter and set of
scales, we compute the Spearman Rank Correlation Coefficient ($\rho$)
of the correlation with large scale density, and the probability, $\Pl$, 
of attaining the value of $\rho$ given a null hypothesis in which the
parameter depends only upon small scale density.  The method
used to compute $\Pl$ is described in Appendix~\ref{app:signiflarge}.
A value of $\Pl=0.5$ means no significance, while $\Pl=0$ is a high
significance anti-correlation and $\Pl=1$ is a high significance
positive correlation \footnote{$\Pl$ is computed with a numerical accuracy 
of 0.0003}. Subsections of the small scale density range are also
tested so that effects of large scale density can be detected which are 
local in density space.  The results are recorded in
Table~\ref{table:spearmantest}.


\begin{table*}
\begin{center}
\caption{For different choices of small and large scale density and a given range of small scale density 
(columns 1 and 2, units $\Mpc^{-2}$), we provide the Spearman Rank Correlation Coeffiecient $\rho$ which 
measures the correlation between each parameter and the large 
scale density (columns 3 to 7). $\Pl$, the probability that this coefficient is greater than would be expected 
in the absence of any intrinsic large scale correlation (i.e. any remaining correlation is purely a result of 
correlations between the parameter and small scale density, and between density on the two scales) 
is provided in columns 8 to 12. Computed using a monte-carlo technique, a value of $\Pl=1$ implies a highly 
significant positive correlation; $\Pl=0$ implies a highly significant negative correlation and $\Pl=0.5$ 
implies no correlation.}
\label{table:spearmantest}
\vspace{0.1cm}
\begin{tabular}{cccccccccccc}
\hline\hline
\noalign{\smallskip}
small scale & large scale & $\rho$($\mur$) & $\rho$($\mub$) & $\rho$($\sigmar$) & $\rho$($\sigmab$) & $\rho$($\fr$) & $\Pl$($\mur$) & $\Pl$($\mub$) & $\Pl$($\sigmar$) & $\Pl$($\sigmab$) & $\Pl$($\fr$)\\
\hline
$\dsca$ & $\dscc$ & 0.4444 & 0.5008 & -0.3799 & -0.2157 & 0.6308 & 0.8210 & 0.9960 & 0.2850 & 0.0800 & 1.0000\\
$0-2$ & & 0.4236 & 0.6096 & -0.3112 & -0.0276 & 0.4751 & 0.9933 & 1.0000 & 0.1387 & 0.7507 & 0.9770\\
$1-10$ & & 0.2527 & 0.4038 & -0.1888 & -0.1117 & 0.4645 & 0.9453 & 0.9990 & 0.3020 & 0.1937 & 1.0000\\
$2-10$ & & 0.1940 & 0.3806 & -0.1627 & -0.2279 & 0.4420 & 0.8073 & 0.9953 & 0.2860 & 0.0657 & 0.9947\\
$2-30$ & & 0.4083 & 0.4581 & -0.3538 & -0.3057 & 0.6232 & 0.5663 & 0.9750 & 0.2983 & 0.0353 & 0.9960\\
$4-30$ & & 0.3598 & 0.3767 & -0.3619 & -0.3675 & 0.5894 & 0.4470 & 0.9160 & 0.2193 & 0.0280 & 0.9773\\
\hline
$\dscb$ & $\dscd$ & 0.3166 & 0.5819 & -0.3642 & 0.0515 & 0.4190 & 0.1020 & 0.9910 & 0.4430 & 0.3257 & 0.5287\\
$0-2$ & & 0.1097 & 0.4768 & -0.1227 & 0.0008 & 0.1581 & 0.4170 & 0.9997 & 0.5893 & 0.1837 & 0.8313\\
$1-10$ & & 0.0381 & 0.4238 & -0.2040 & 0.0857 & 0.2067 & 0.0167 & 0.8117 & 0.5710 & 0.3850 & 0.2853\\
$2-10$ & & 0.0440 & 0.3318 & -0.2322 & 0.2158 & -0.0619 & 0.1313 & 0.7177 & 0.2760 & 0.7947 & 0.0273\\
$2-30$ & & 0.2206 & 0.3858 & -0.3392 & 0.0607 & 0.1786 & 0.2073 & 0.7137 & 0.2763 & 0.7840 & 0.0337\\
$4-30$ & & 0.1754 & 0.3537 & -0.2074 & 0.0456 & 0.0231 & 0.3583 & 0.7917 & 0.6233 & 0.7767 & 0.0337\\
\hline
$\dscj$ & $\dsce$ & 0.4951 & 0.5785 & -0.4314 & 0.1280 & 0.3713 & 0.8030 & 0.9487 & 0.0370 & 0.5063 & 0.0503\\
$0-2$ & & 0.3860 & 0.4945 & -0.3618 & 0.1099 & 0.1262 & 0.9280 & 0.9900 & 0.0170 & 0.4307 & 0.0937\\
$1-10$ & & 0.0990 & 0.2664 & -0.2232 & -0.2282 & 0.1356 & 0.0177 & 0.1920 & 0.7230 & 0.0727 & 0.0063\\
$2-10$ & & 1524 & 0.0527 & -0.0809 & -0.2761 & -0.1869 & 0.0100 & 0.1393 & 0.8033 & 0.1837 & 0.0000\\
\hline
\end{tabular}
\end{center}
\end{table*}

A common feature of all the panels in
  Fig. \ref{figure:fredscale},~\ref{figure:redpeakscale}
  and~\ref{figure:bluepeakscale} is that contours, where they exist, 
  are more closely aligned with the large scale
  axis than with the small scale one, indicating that the parameters
  react more strongly to smaller scales (even $<0.5\Mpc$) 
  than to the larger ones.

Nonetheless, there is a highly significant
positive correlation of $\dfr$ with $\dscc$, at fixed $\dsca$.  This
must reflect a transformation process for galaxies which relates to
the environment and is imprinted on scales beyond $0.5\Mpc$.  However
parameters defining the red peak ($\dmur$ and $\dsigmar$) show no
significant trend with $\dscc$ at fixed $\dsca$, other than $\dmur$ at
low density ($\dsca=0-2\Mpc^{-2}$).  A significant residual positive
correlation does exist for the blue peak position $\dmub$ with
$\dscc$, though this is weaker than the trend with $\dfr$.

On $>1.0\Mpc$ scales there is no significant correlation of $\dfr$
with $\dscd$ at fixed $\dscb$ other than a $\sim2\sigma$ level
anti-correlation at low density ($\dscb=2-10\Mpc^{-2}$).  A stronger
anti-correlation exists between the position of the red peak $\dmur$
and $\dscd$ at intermediate density ($\dscb=1-10\Mpc^{-2}$).  The blue
peak position $\dmub$ does correlate positively with $\dscd$, though
this is only significant at low density ($\dscb=0-2\Mpc^{-2}$).  The
lack of any positive correlation with densities above $1\Mpc$ scales,
at least for $\dfr$ and $\dmur$, and at higher densities typical of
groups and clusters, indicates that there is no longer any direct
impact of these environments on these galaxy properties.  This is
consistent with the results of \citet{Kauffmann04}, \citet{Blanton06}
and \citet{Blanton07}.  The anti-correlations are interesting however,
and deserve further attention.

At $>2\Mpc$ scales and at intermediate density
($\dscj=2-10\Mpc^{-2}$), the anti-correlation of $\dfr$ with $\dsce$
is even more significant.  For roughly the same range of density
($\dscj=1-10\Mpc^{-2}$) we also see a highly significant
anti-correlation of $\dmur$ with $\dsce$.  In other words, red
galaxies which have intermediate to large $0-2\Mpc$ scale densities
avoid regions overdense on $2-3\Mpc$ scales, and are otherwise bluer
on average. Blue galaxies, on the other hand, exhibit a positive
correlation of $\dmub$ with $\dsce$, but only at low density
($\dscj=0-2\Mpc^{-2}$), as seen on $1-2\Mpc$ scales.  The width of the
red peak $\dsigmar$ is also anti-correlated (in this case the same
direction as the trend with small scale density) with $\dsce$ at these
densities.


\subsubsection{Summary of observational results:}

To summarize the way in which the galaxy colour distribution traces
density on different scales:
\begin{itemize}
\item{The fraction of red galaxies $\fr$ increases dramatically with
    increasing local ($<0.5\Mpc$ scale) density. The properties of
    both peaks also change: the red peak becomes redder and narrower,
    and the blue peak becomes redder and broader.}
\item{The position and width of the red peak, $\mur$ and $\sigmar$
    show no residual dependence on $0.5-1.0\Mpc$ scales.}
\item{The fraction of red galaxies and the position of the blue peak,
    $\fr$ and $\mub$, both show residual positive correlation with
    density on $0.5-1.0\Mpc$ scales.  Other than some dependence of
    $\mub$ at very low densities, there are no significant positive
    correlations with densities computed on $>1.0\Mpc$ scales.}
\item{Nonetheless, at fixed smaller scales, there exist significant
    {\it anti-correlations} with large scale density ($\fr$ and $\mur$
    with $\dscd$, $\dsce$), typically at intermediate values of
    interior density.}
\item{Correlations with larger scale density exist at low values of
    small scale density ($\mur$ with $\dscc$ at low $\dsca$, $\sigmar$
    with $\dsce$ at low $\dscj$, $\mub$ with $\dscc$, $\dscd$, $\dsce$
    at low $\dsca$, $\dscb$, $\dscj$ respectively).  This suggests
    that relatively isolated galaxies react in some ways to their
    environments on large scales. However this result might be
    explained by the effects of correlated noise on density
    measurements.  We discuss this effect in
    Appendix~\ref{app:corrnoise}.}
\end{itemize}


 

\section{Multiscale density dependences in the halo model framework}\label{sec:interpretation}


\subsection{Dependence of halo properties on density}

To aid our interpretation of these results, we cross-correlate the
sample with the groups catalogue of \citet[][Y07]{Yang07} to examine
how halo mass and group-centric radius depend upon our density
measurements at different scales.  We emphasize that our approach has
been to compute an entirely model-independent, measurable quantity
(density) and that this exercise thus serves as an examination of the
halo model in an independent parameter space.

The Y07 catalogue is constructed from the SDSS Data Release 4 (DR4)
catalogue \citep[via the New York University Value Added Galaxy
Catalogue, NYU-VAGC,][]{Blanton05}, incomplete for our sample. A 
friends-of-friends
algorithm is applied to the survey for group detection, and the total
luminosity, $L_{19.5}$, of galaxies is computed for each group down 
to an $r$-band magnitude of $M_{r,0.1\mathrm{ev}}=-19.5$ (note that
in Y07 all luminosities are k- and evolution-corrected to $z=0.1$). A
halo mass is assigned to each group assuming a one to one relation in
rank order of halo mass with $L_{19.5}$. This is in turn used to
redefine group membership, which is iterated until a stable solution
is found. A halo is defined so long as a member galaxy with
$M_{r,0.1\mathrm{ev}}<-19.5$ makes it into the sample. At the upper
redshift boundary of our sample $z=0.08$, this limit ensures that the
group catalog is complete to $\log_{10}(\rm\,M_{\mathrm{halo}}/\Msol) = 11.6$ and
includes all galaxies with $\Mr < -20.434$ (mag as defined in 
Section~\ref{sec:sample}, corrected to $z=0$ rest-frame, 
no evolution correction).
This sample, after removal of galaxies due to
DR4 edge effects for which $\Mhalo$ could not be estimated, numbers
38733 galaxies.  Importantly, this catalogue provides halo masses for
individual galaxies, as well as groups with more than one known
member.

\begin{figure*}
     \centerline{\includegraphics[width=0.45\textwidth]{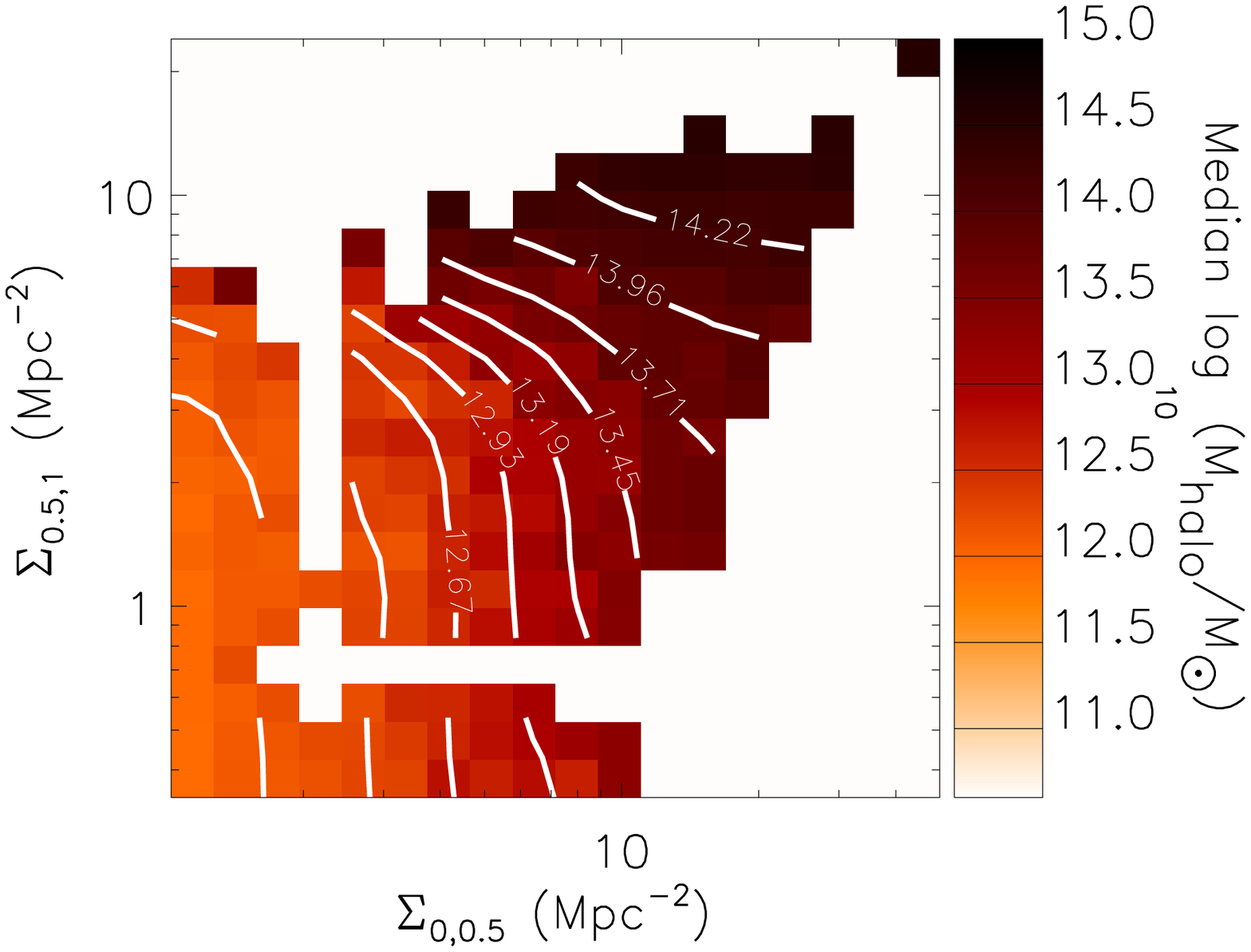}
                 \hspace{0.02\textwidth}
                 \includegraphics[width=0.45\textwidth]{ConstBinFit_doublegausspro_2D_fred_a_c.eps}}
     \centerline{\includegraphics[width=0.45\textwidth]{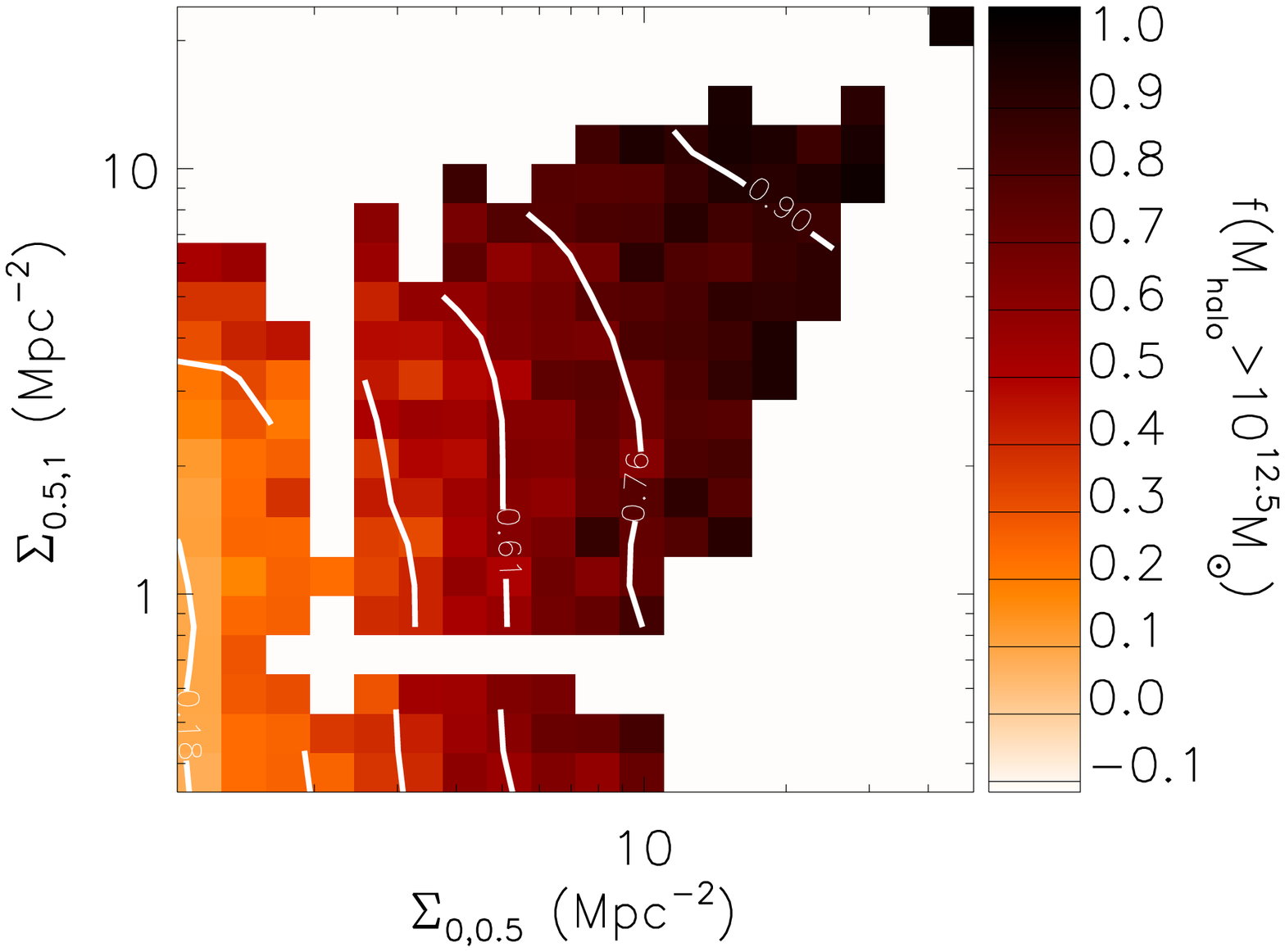}
                 \hspace{0.02\textwidth}
                 \includegraphics[width=0.45\textwidth]{ConstBinFit_doublegausspro_2D_mean_red_a_c.eps}}
     \centerline{\includegraphics[width=0.45\textwidth]{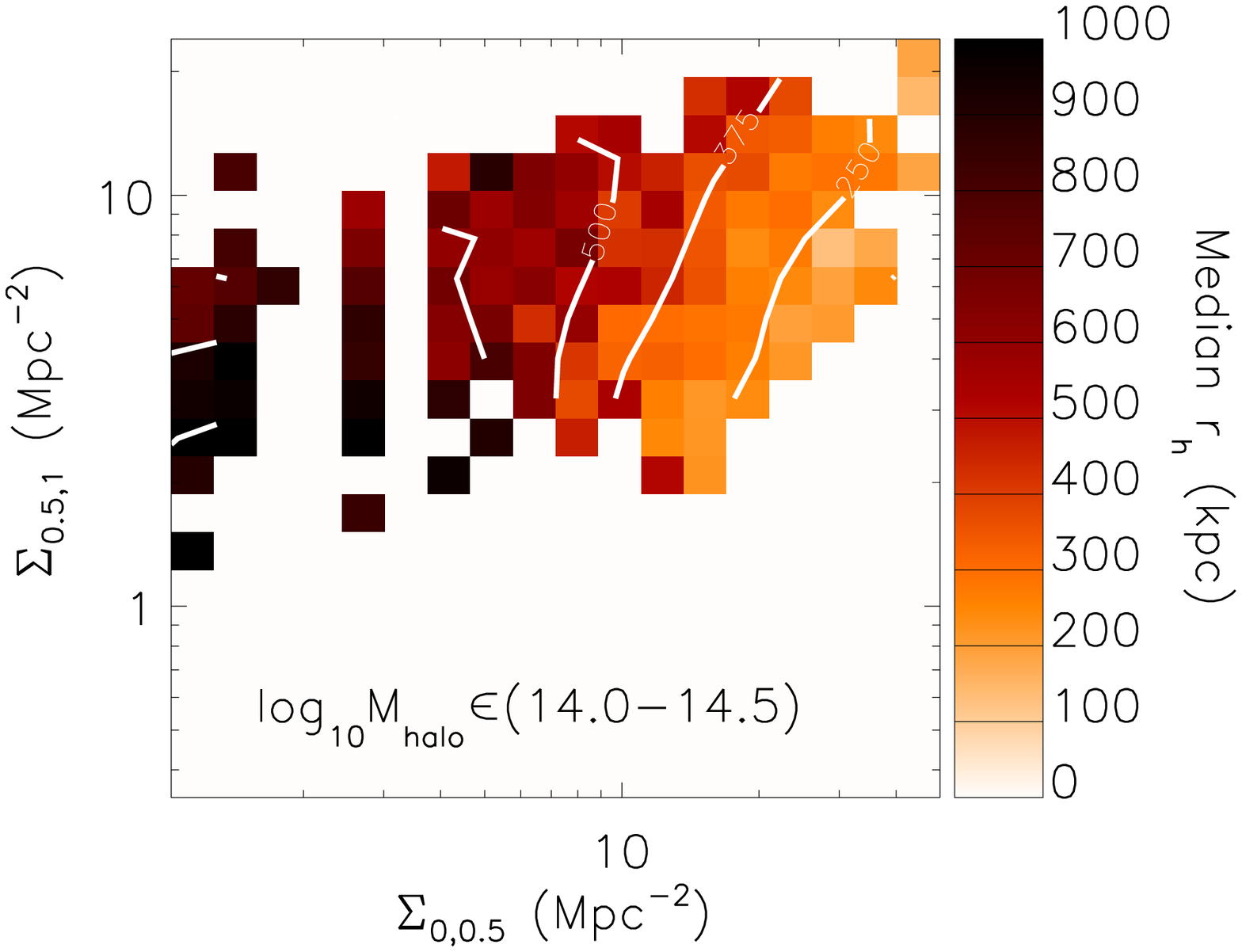}
                 \hspace{0.02\textwidth}
                 \includegraphics[width=0.45\textwidth]{ConstBinFit_doublegausspro_2D_mean_blue_a_c.eps}}
               \caption{A comparison between the dependence of halo
                 mass and halo-centric radius with density on scales
                 $\dsca$ and $\dscc$ (left) and the dependence of the
                 colour distribution parameters $\fr$, $\mub$ and
                 $\mur$ on $\dsca$ and $\dscc$ (right). The top,
                 centre and lower panels on the left illustrate the
                 behaviour of the median halo mass, the fraction of
                 galaxies inside haloes $>10^{12.5}\Msol$, and the
                 median halo-centric radius in haloes of
                 $10^{14.0-14.5}\Msol$, taken from the Y07 catalogue.}
  \label{figure:group051Mpc}
\end{figure*}
\begin{figure*}
     \centerline{\includegraphics[width=0.45\textwidth]{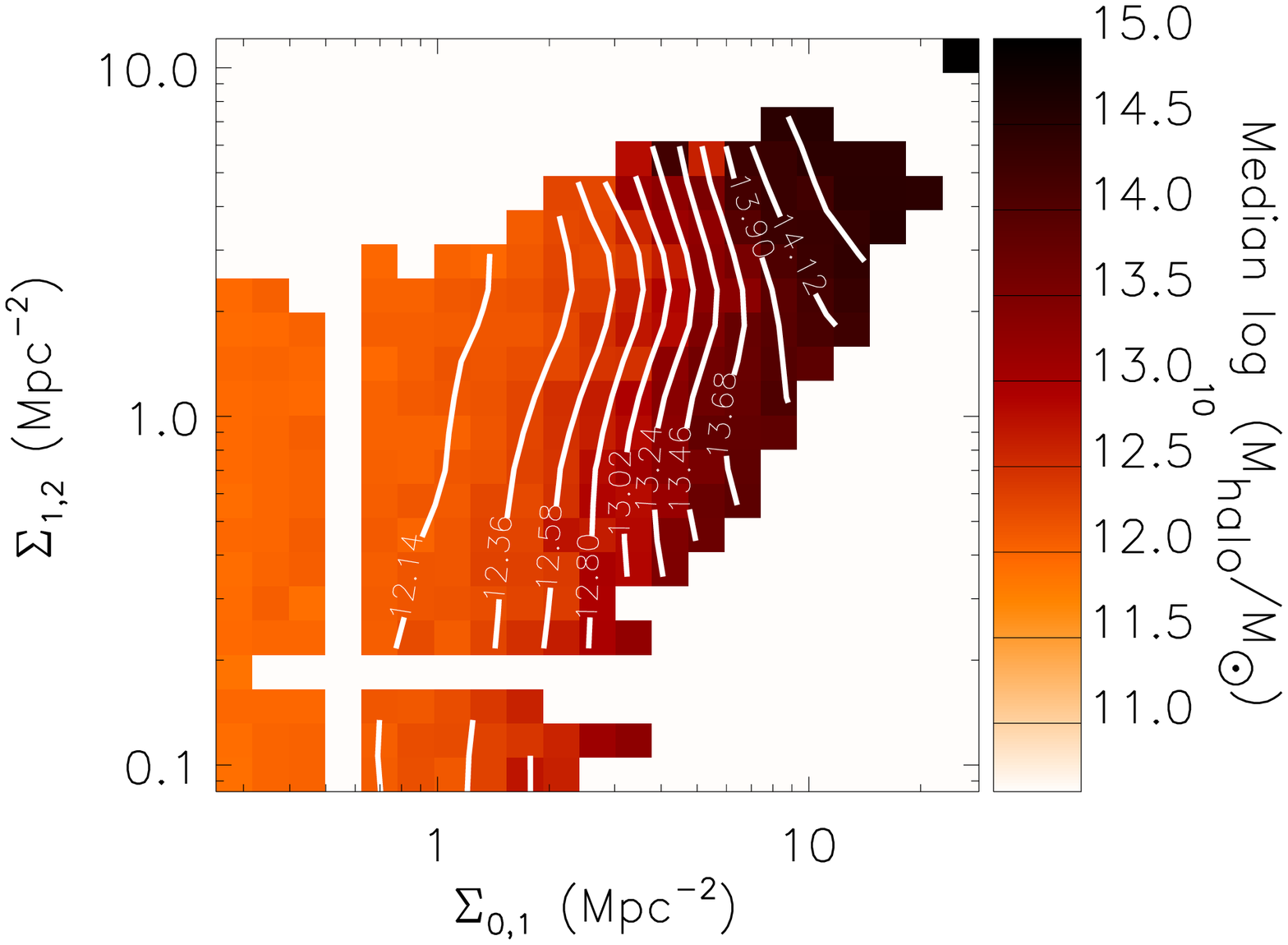}
                 \hspace{0.02\textwidth}
                 \includegraphics[width=0.45\textwidth]{ConstBinFit_doublegausspro_2D_fred_b_d.eps}}
     \centerline{\includegraphics[width=0.45\textwidth]{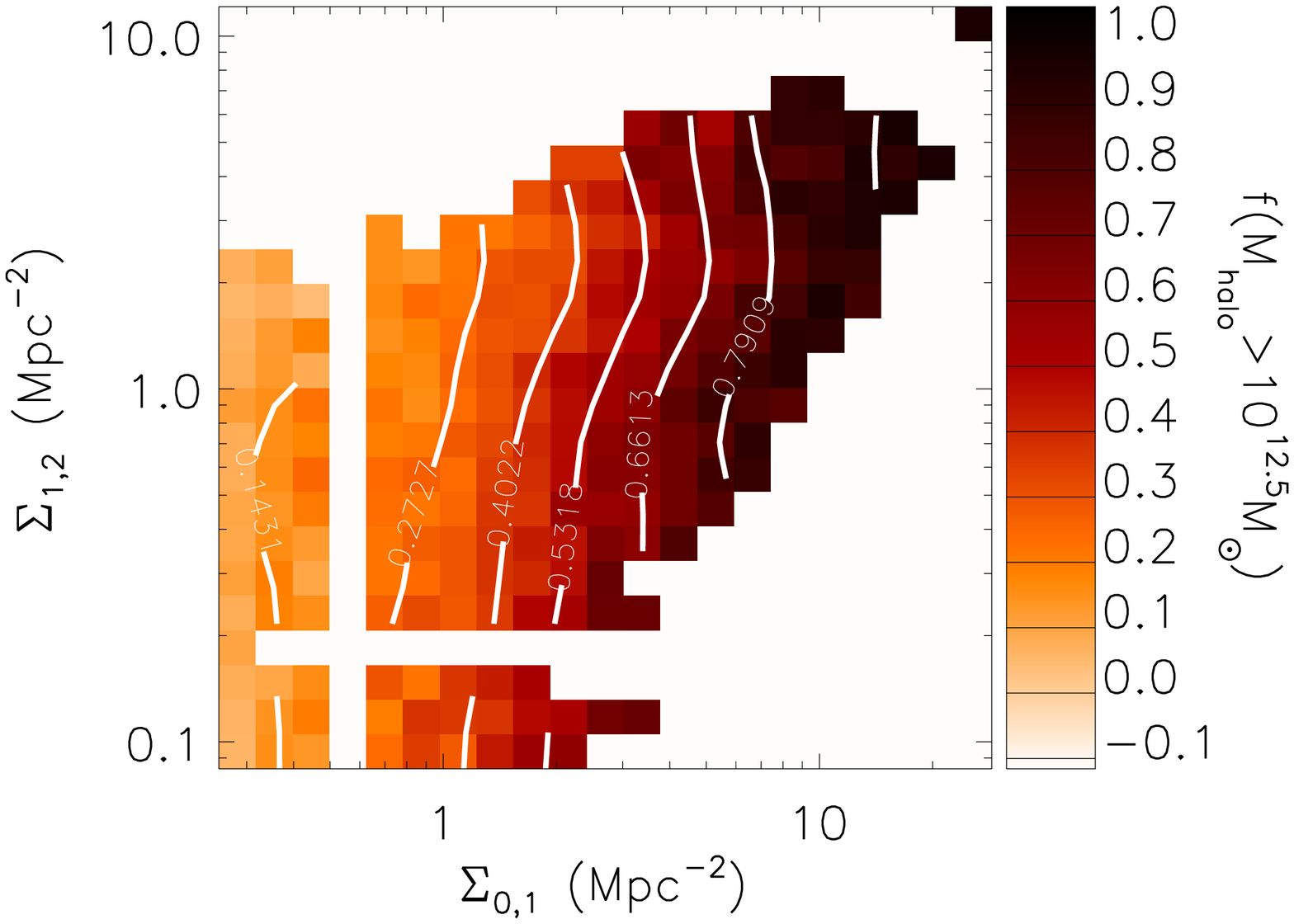}
                 \hspace{0.02\textwidth}
                 \includegraphics[width=0.45\textwidth]{ConstBinFit_doublegausspro_2D_mean_red_b_d.eps}}
     \centerline{\includegraphics[width=0.45\textwidth]{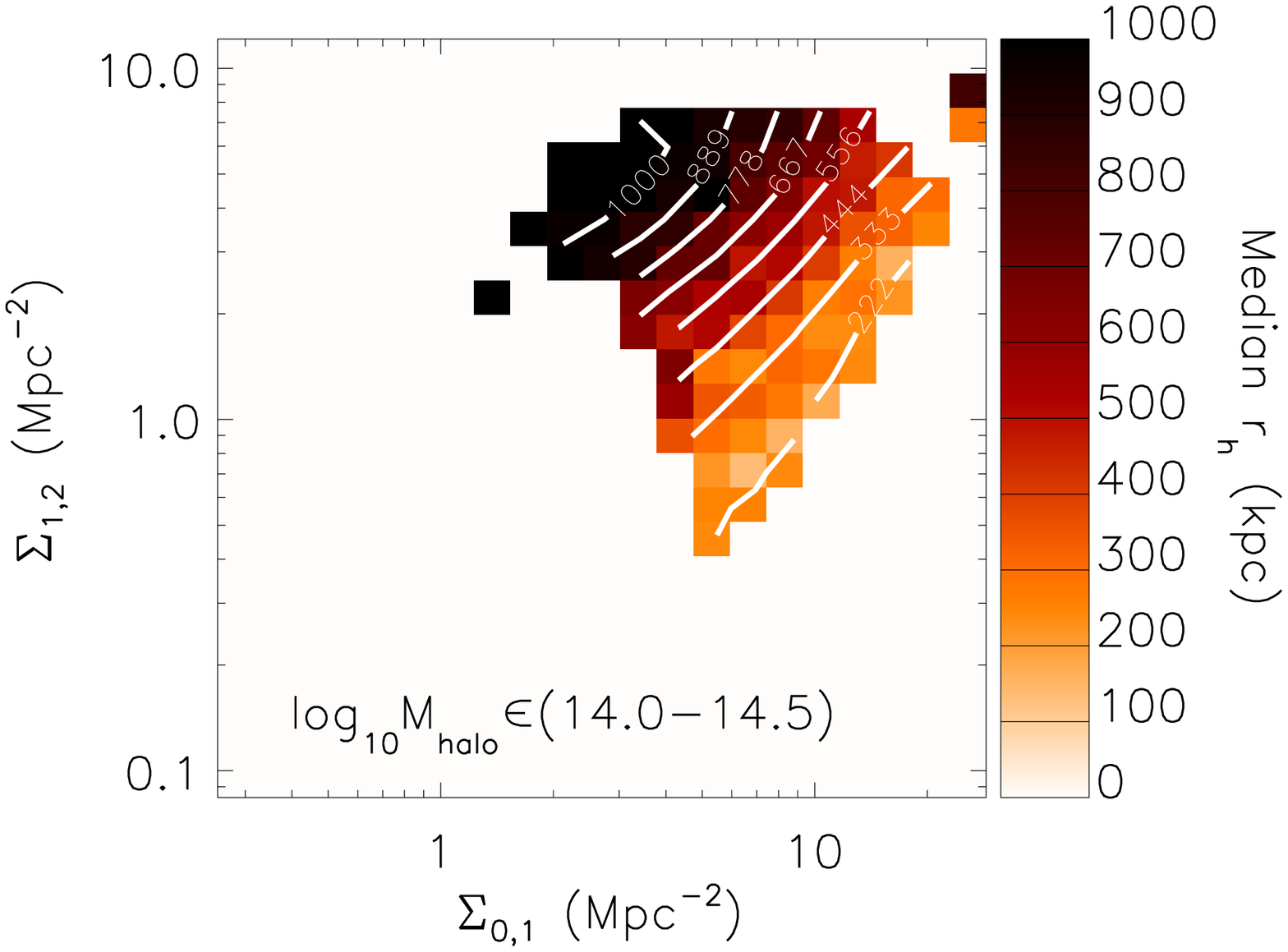}
                 \hspace{0.02\textwidth}
                 \includegraphics[width=0.45\textwidth]{ConstBinFit_doublegausspro_2D_mean_blue_b_d.eps}}
   \caption{As Figure~\ref{figure:group051Mpc}, but on density scales $\dscb$ and $\dscd$.}
   \label{figure:group12Mpc}
 \end{figure*}
\begin{figure*}
     \centerline{\includegraphics[width=0.45\textwidth]{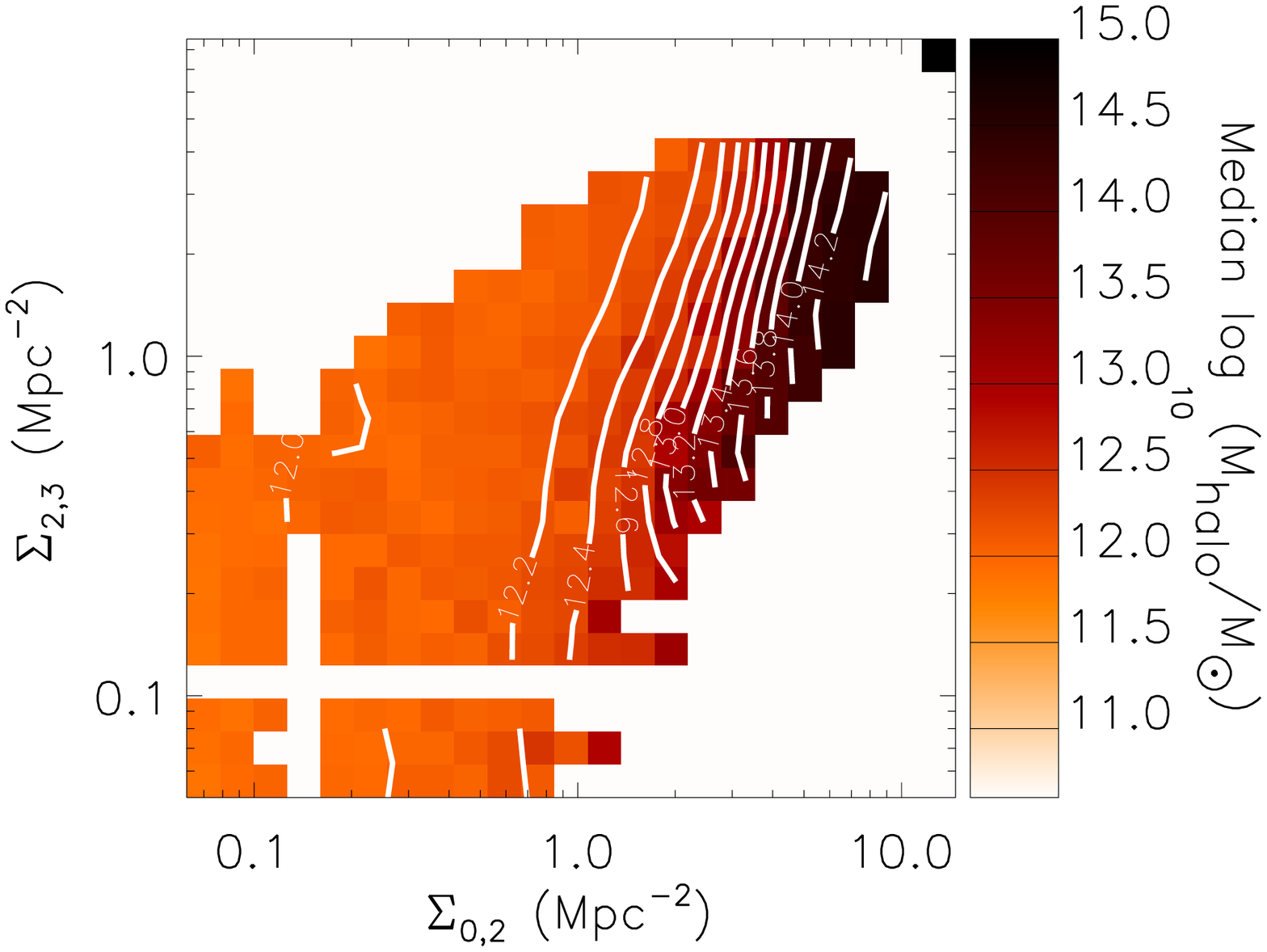}
                 \hspace{0.02\textwidth}
                 \includegraphics[width=0.45\textwidth]{ConstBinFit_doublegausspro_2D_fred_j_e.eps}}
     \centerline{\includegraphics[width=0.45\textwidth]{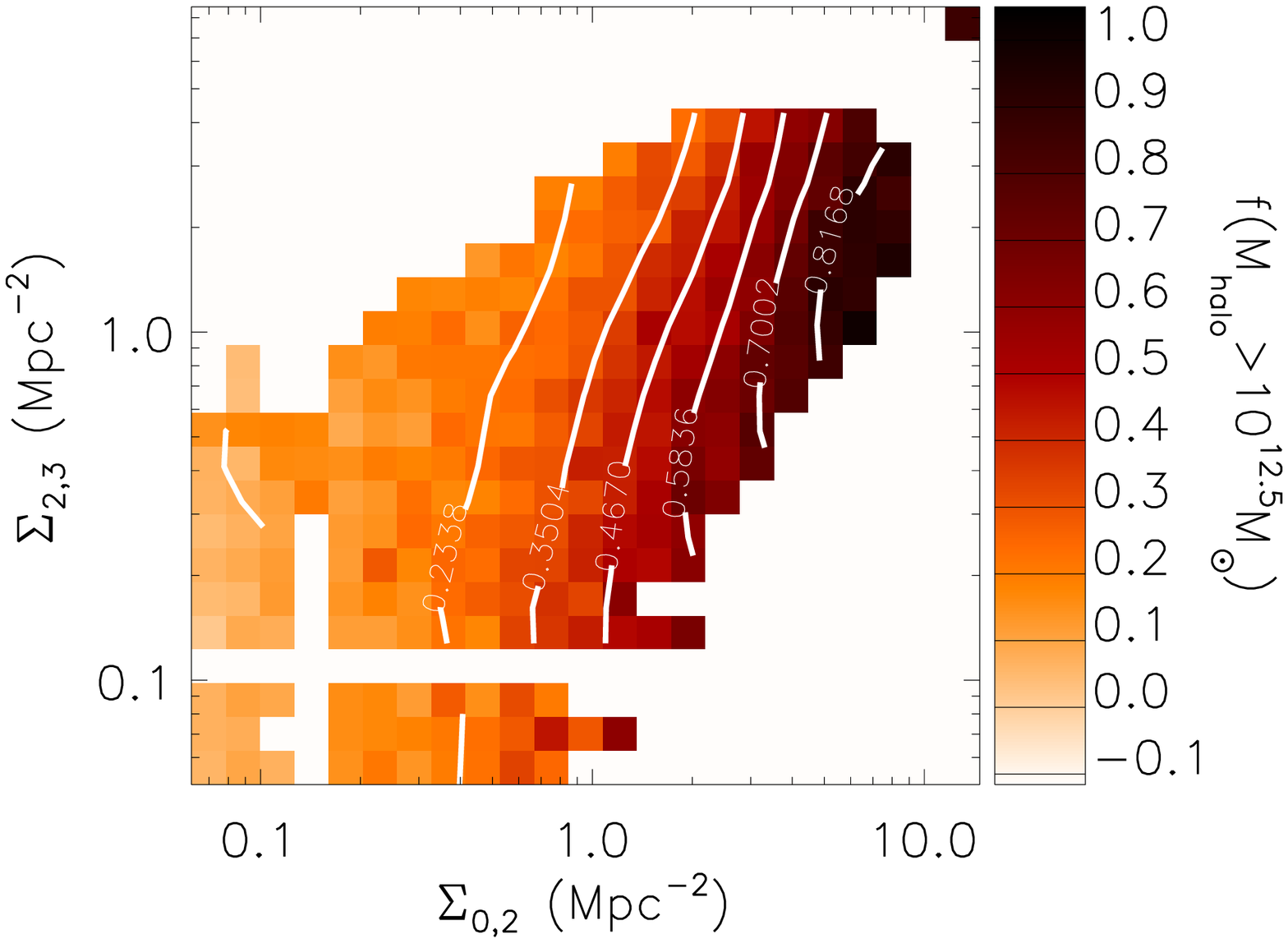}
                 \hspace{0.02\textwidth}
                 \includegraphics[width=0.45\textwidth]{ConstBinFit_doublegausspro_2D_mean_red_j_e.eps}}
     \centerline{\includegraphics[width=0.45\textwidth]{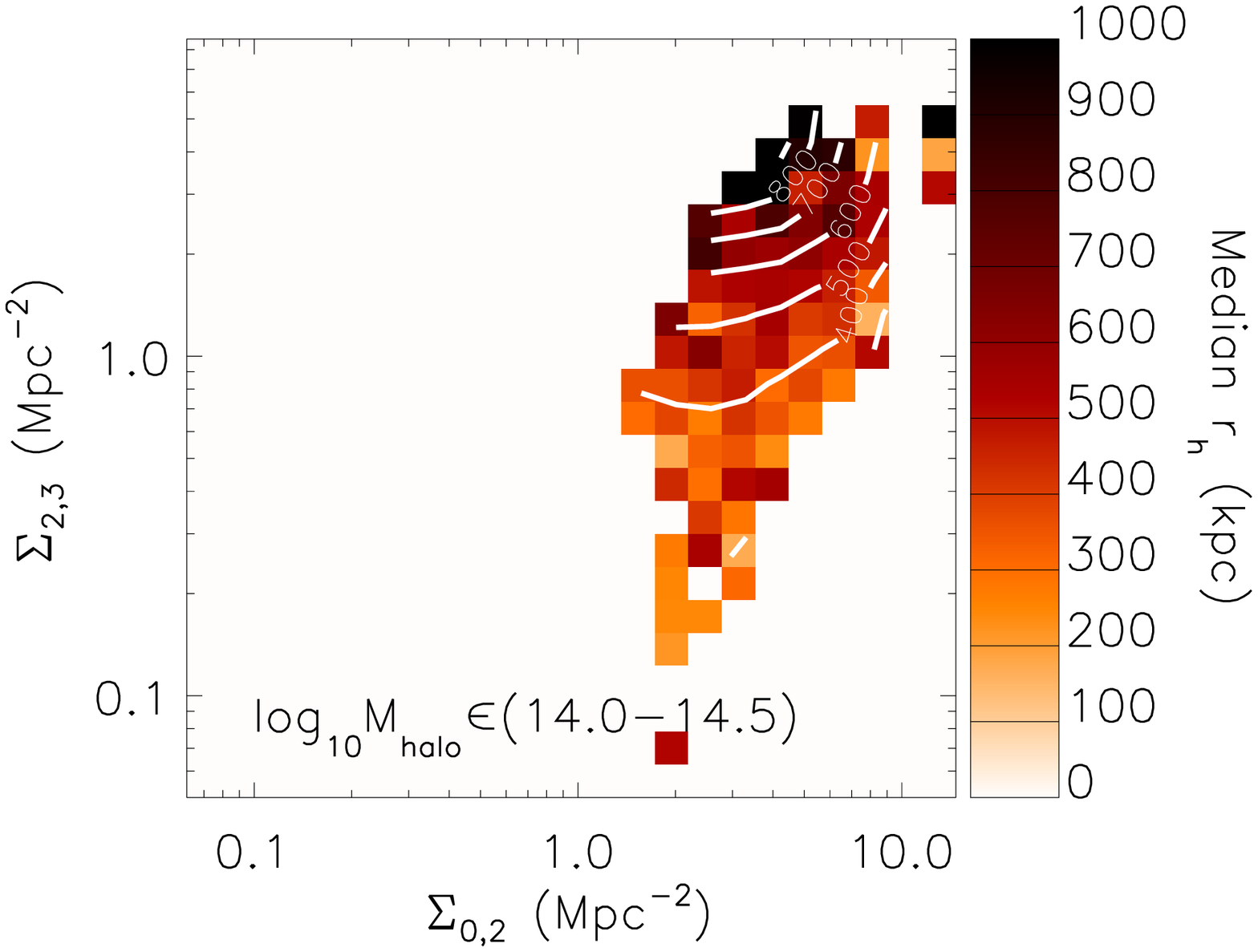}
                 \hspace{0.02\textwidth}
                 \includegraphics[width=0.45\textwidth]{ConstBinFit_doublegausspro_2D_mean_blue_j_e.eps}}
   \caption{As Figure~\ref{figure:group051Mpc}, but on density scales $\dscj$ and $\dsce$.}
   \label{figure:group23Mpc}
 \end{figure*}

Figures~\ref{figure:group051Mpc},~\ref{figure:group12Mpc}
and~\ref{figure:group23Mpc} illustrate how the halo masses $\Mhalo$
and halo-centric radii $\rh$ correlate with density computed in our
three choices of small versus large scale density plane ($\rdiv =
0.5$,$1$,$2\Mpc$).  The left panels illustrate (from top to bottom)
the dependence on density of the median halo mass, the fraction of
galaxies in haloes of $>10^{12.5}\Msol$ (both restricted to grid
pixels with $\geq 25$ galaxies), and the median halo-centric radius
for galaxies in massive $10^{14-14.5}\Msol$ haloes (for $\geq 5$
galaxies).  Each panel is also overplotted with contours, computed for
a grid smoothed in exactly the same way as those in
Figures~\ref{figure:fredscale} to~\ref{figure:bluepeakscale}.

For direct visual comparison, the right hand panels reproduce the
equivalent dependence of $\dfr$, $\dmur$ and $\dmub$ on density, as
already seen in Figures~\ref{figure:fredscale}
to~\ref{figure:bluepeakscale}.


The median halo mass and the fraction of galaxies in $>10^{12.5}\Msol$
haloes both depend mostly on the smallest scale densities. In more
massive haloes the halo diameter exceeds $\rdiv = 0.5\Mpc$ ($2r_{200}
\sim 0.4\Mpc$ for $10^{12.5}\Msol$ haloes) and thus $\dscc$ becomes
more important, particularly at higher density where more massive
haloes dominate.  On $\dscb$ versus $\dscd$ scales ($\rdiv=1\Mpc$),
the contours are closer to vertical, and on $\dscj$ versus $\dsce$
scales ($\rdiv=2\Mpc$) both parameters anti-correlate with $\dsce$ at
fixed $\dscj$.  This results mainly from the way halo centric radius
$\rh$ tracks density on different scales.  For $10^{14-14.5}\Msol$
haloes the contours tracking the change in $\rh$ with density are
almost vertical with $\rdiv=0.5\Mpc$, but moving to larger scales
become diagonal for $\rdiv=1\Mpc$ (larger median radius at smaller
$\dscb$ but larger $\dscd$), and almost horizontal with $\rdiv=2\Mpc$
(radius positively correlates with $\dsce$).  The reason for this is
fairly obvious: a galaxy at the centre of its halo will have high
densities within annuli up to the size of the halo
($r_{200}\sim1.1$,$2.5\Mpc$ for $\Mhalo=10^{14}$,$10^{14.5}\Msol$) and
low densities on larger scales.  A galaxy in the outskirts of this
halo will experience the opposite, since the dense halo core is
located within the larger scale annulus.  Moving to smaller halo
masses than $10^{14-14.5}\Msol$, this pattern simply shifts to smaller
scales.

At the point of accretion, a galaxy moves from the infalling halo (low
$\Mhalo$), to the more massive halo (high $\Mhalo$ and high $\rh$),
but the change in density will not be instantaneous.  Therefore at
high density on large scales there are not only more galaxies with
large $\rh$ within the massive halo, but also more galaxies assigned
instead to an infalling halo of lower mass.  This can account for the
anti-correlations of both median halo mass and the fraction of
galaxies in $>10^{12.5}\Msol$ haloes with large scale density.

This effect is further amplified by the depth of the cylinder,
required to account for redshift-space distortions (the ``finger of
God effect'').  By integrating to $\pm1000\kms$ along the line of
sight, the contribution of galaxies from outside a massive halo is
increased, including galaxies from the fore- and background.


\subsection{Comparison with the dependence of galaxy properties on density}

Now we make a qualitative comparison of the density dependence of
galaxy properties with the density dependence of the assumed embedding
halo to examine whether and how the halo model can explain the way the
colour distribution of galaxies traces density on different scales.

Both $\Mhalo$ and $\rh$ trace mostly $<0.5\Mpc$ scales in the $\dsca$
versus $\dscc$ plane.  Therefore any such dependence of galaxy
properties can be explained by invoking either trends with halo mass
or halo-centric radius.  On larger scales (larger $\rdiv$) this is no
longer the case.  The primary dependence on small scales (contours
close to vertical) of $\dfr$, $\dmur$ and $\dmub$ more closely
resemble trends with halo mass than with radius.

Indeed $\dfr$ trends in particular are very well matched qualitatively
to the trends seen for the fraction of galaxies in $>10^{12.5}\Msol$
haloes, including the roughly equal spacing of contours.  Trends on
all scales are very comparable, but the anti-correlation with $\dsce$
at fixed $\dscj$ is particularly striking.  The anti-correlation for
haloes traces the accretion of galaxies onto the halo in the infall
regions, and the reproduction of this trend for the fraction of red
galaxies is intriguing.  It provides strong circumstantial evidence
that galaxies remain blue until they have experienced some truncation
mechanism which is intimitely linked to the accretion time onto a
massive halo.  This is fully consistent with a simple model in which
galaxies become red at some interval after their incorporation into a
halo of mass $\gtrsim10^{12-13}\Msol$, which can also match total red
fractions as a function of halo mass and redshift \citep{McGee09}.
However the qualitative dependence on density is not strongly
sensitive to the choice of limiting halo mass: similar looking trends
are produced with limits in the range $\sim10^{12.5-13.5}\Msol$, with
contours becoming slightly more horizontal with larger limiting mass.

$\mur$ does not trace density on $>0.5\Mpc$ scales in the same way as
the median halo mass or the fraction of galaxies in haloes above a
given threshold.  Instead the positive correlations with density are
all on local $<0.5\Mpc$ scales, and $\mur$ anti-correlates with
density on $1.0-2.0\Mpc$ scales.  The lack of positive density
dependence scales greater than $0.5\Mpc$ can be related to subhalo or
radial effects, in that the colour may be imprinted at a time when the
galaxy's $<0.5\Mpc$ scale subhalo was not part of the current
$>0.5\Mpc$ scale main halo.  Alternatively, if the colour of red
galaxies is influenced by some combination of halo mass and
halo-centric radius such that in massive haloes galaxies with larger
$\rh$ are likely to be bluer than those in the centre, then it is also
possible to qualitatively reproduce the observed trends.

The colour of a red (assuming passive) galaxy can be driven by either
age (luminosity weighted time since stars were formed), or metallicity
of the stars.  If age is the driving factor then the subhalo
hypothesis is hard to reconcile with observations: i.e. If galaxies
all became red within such small subhaloes then we cannot explain the
trend of $\fr$ with $>0.5\Mpc$ scale density.  Halo-centric radius, on
the other hand, should correlate with the time since a galaxy was
accreted onto the halo, and provides a possible solution.  If
metallicity is the driving factor, the galaxy's environment at the
time it was forming stars should be more important: i.e. the subhalo.
Thus we suggest that where $\mur$ is concerned, the relative
importance of halo-centric radius versus local density is equivalent
to the relative importance of age versus metallicity.  The
anti-correlation of $\mur$ with density on $1-2\Mpc$ scales suggests
that the average colour of red galaxies is bluer outside more massive
haloes ($\gtrsim2\times10^{13}\Msol$).  This is consistent with both
hypotheses (red galaxies belonging to smaller infalling haloes are
younger and metal poorer).

We note that these trends describe only the influence
of environment on red galaxies in the luminosity range
$-21.5\leq\Mr\leq-20.0$, and that the way in which both age
and metallicity depend upon their halo mass seems to be a
function of stellar mass \citep{Pasquali10}. Indeed, the local
density of red sequence galaxies correlates with both age
and metallicity, even at fixed luminosity and colour 
\citep{Cooper10}, whilst there is no clear correlation with larger
scale (6Mpc) density \citep[][who also show thec contribution of 
rejuvenated galaxies decreases with density]{Thomas10}. 
Thus this complex picture has to be taken with the caveat
that not all red galaxies are passive, and not all passive red
galaxies have comparible star formation histories.

The blue peak colour $\mub$ depends on $0-1\Mpc$ scale densities in
the same way as $\fr$, and therefore a pure halo mass dependence can
be similarly invoked.  $\mub$ probably traces the level of recent star
formation in a typical star forming galaxy, and so this means that
such galaxies experience partial truncation of their fuel supplies
inside massive haloes, consistent with observations of truncated gas
disks or anemic spirals in the Virgo cluster \citep[e.g.][and
references
therein]{Giovanelli85,Koopmann04,Gavazzi06,Chung09}.
It is currently unclear whether this is necessarily a step on the path
to total truncation and transition to a red colour.

$\mub$ exhibits residual large scale dependence at low values of small
scale density, no matter which radius $\rdiv$ is chosen to separate
small from large scales. This dependence cannot be pure halo mass
dependence, because the median halo mass has such residual dependence
on large scale only if $\rdiv = 0.5\Mpc$.  This might merely be a
result of correlated noise in the density plane (see
Appendex~\ref{app:corrnoise}), but there is no reason this should
apply to certain parameters and not others.  Otherwise the suggestion
is that relatively isolated blue galaxies are redder when they live in
relatively overdense larger scale environments.  On such scales
(including $>2.0\Mpc$) this might relate to an earlier formation time
of a galaxy living in a large scale overdensity coupled with a
declining star formation history, or to a more efficient cooling of
hot gas onto galaxies in more isolated environments.

 We conclude that the halo model can provide an excellent qualitative
 description of the dependence of the parameters governing the galaxy
 colour distribution, in particular $\fr$, $\mur$ and $\mub$, on
 simple annular density measurements on a range of scales from
 $0-3\Mpc$.  The data is consistent with a simple picture in which the
 density dependence of the fraction of red galaxies $\fr$ and the peak
 colour of blue galaxies $\mub$ is governed mainly by the fraction of
 galaxies assigned to haloes above some threshold halo mass
 ($10^{12.5}\Msol$ provides a reasonable but not unique description in
 this case).  This suggests that the gas supply regulating star
 formation can be partially or completely suppressed as a function of
 halo mass.  The halo model is particularly good at explaining the
 anti-correlation of $\fr$ with large scale density.  This can be
 explained if galaxies not yet or recently incorporated into a
 relatively massive halo have not experienced the blue to red
 transition.  The lack of positive correlations with density beyond
 $0.5\Mpc$ suggests that the red peak colour $\mur$ traces only local
 density within subhaloes (explicable if $\mur$ traces metallicity) or
 with additional dependence on halo-centric radius (since if
 suppression time is linked to the time a galaxy was accreted onto a
 halo, and this is correlated with radius, then a red galaxy will be
 older and thus also redder towards the halo centres).

\section{Conclusions}\label{sec:conclusions}

We have developed a new multiscale method to characterize the
dependence of galaxy properties on environment.  This provides a
model-independent, rich parameter space in which to evaluate galaxy
properties.

In this paper we have examined the dependence of the $u-r$ colour
distribution for galaxies in the luminosity range
$-21.5\leq\Mr\leq-20$ on multiscale density.  This is parameterized
using the double gaussian model, with parameters $\fr$ (the fraction
of red galaxies), $\mur$ and $\sigmar$ (the position and width of the
red peak) and $\mub$ and $\sigmab$ (the position and width of the blue
peak).

We confirm and extend known trends with small scale ($\lesssim 0.5 - 1
\Mpc$) density:
\begin{itemize}
\item{Galaxies in denser environments are more likely to be red
    ($\fr$), implying an almost complete truncation of star
    formation.}
\item{If they are still blue (star forming), then they are also likely
    to be relatively redder in denser environments ($\mub$) although
    with greater scatter ($\sigmab$), implying partial but varied
    truncation of gas supplies, possibly relating to truncated gas
    disks such as those observed in the Virgo cluster.}
\item{Red Galaxies are redder ($\mur$) and with less scatter
    ($\sigmar$) in denser environments.  This implies that they either
    form earlier and/or are more metal-rich, and that this formation
    path is also less varied than galaxies in less dense environments.
    This depends upon the environment of a galaxy during its epoch of
    star formation.}
\end{itemize}

On larger scales:
\begin{itemize}
\item{No parameter correlates positively and significantly with
    density on scales $>1\Mpc$, excluding trends seen for relatively
    isolated galaxies where correlated noise in the measurement of
    density might be important.  This confirms the results of
    \citet{Kauffmann04}, \citet{Blanton06} and \citet{Blanton07}.}
\item{Although all parameters correlate most strongly with density on
    $<0.5\Mpc$ scales, a residual positive correlation of $\fr$ and
    $\mub$ with $0.5-1\Mpc$ scale density at fixed smaller scale
    density implies that total or partial truncation of star formation
    can relate to a galaxy's environment on these scales. A $1\Mpc$
    diameter halo corresponds to a mass of $\sim2\times10^{13}\Msol$
    and a typical crossing time on the order of $4\Gyr$, and one or
    both of these parameters must relate to a truncation mechanism
    which is active on these scales.}
\item{On $>2\Mpc$ scales $\fr$ {\it anti-correlates} with density at
    fixed smaller scale density, and $\mur$ anti-correlates with
    density on $>1\Mpc$ scales.  In other words galaxies are more
    likely to have ongoing star formation, or if not then to be
    relatively blue, when the large scale density is relatively high
    for the density on smaller scales (they live at projected
    distances $\sim1-3\Mpc$ from a local overdensity).}
\end{itemize}

We interpret these trends qualitatively using the halo model.  To make
this comparison we utilize the halo catalogue of Y07 which is
constructed using a friends of friends group finder, and under the
assumptions that galaxies live in haloes, and that the mass of these
haloes has a one to one relationship with the total galaxy light.  The
properties of embedding haloes has been cross-correlated with our
galaxy catalogue to examine how halo properties trace multiscale
density.  By comparing this dependence to that of galaxy colours, we
find:

\begin{itemize}
\item{Qualitatively, the multiscale dependence of $\fr$ and $\mub$ is
    remarkably comparable to that seen in the fraction of galaxies
    living in haloes above a given threshold mass in the range
    $\sim10^{12.5-13.5}\Msol$.  This is consistent with a scenario in
    which galaxies can experience truncation of their star formation
    at some time after they are accreted onto a halo.}
\item{The halo model offers a simple and yet profound explanation for
    anti-correlations with larger (particularly $>2\Mpc$) scale
    density.  For galaxies at large radial distances from the centre
    of massive haloes the halo core is included in measurements of
    density on large scales.  Thus this region of parameter space
    traces the accretion region of haloes (effectively marking the
    crossover from the one to the two halo term in the correlation
    function).  Anti-correlations on these scales therefore trace the
    accretion region of haloes.  That many such galaxies are still
    blue implies that the influence of environment is felt at or after
    accretion onto a massive halo.}
\item{$\mur$ correlates positively with density only on $<0.5\Mpc$
    scales, implying that the red colour is set by physics within
    subhaloes of this size. In this case we suggest metallicity
    effects must drive these trends, since we know that age must
    correlate with larger scale densities: i.e. galaxies become red
    due to the influence of $0.5-1\Mpc$ scales ($\fr$ correlates
    positively with $\dscc$). However we cannot rule out age effects
    if the $0.5-1\Mpc$ scale dependence of $\mur$ is negated due to a
    combination of halo mass and radial trends.}
\end{itemize}

Whilst we have restricted ourselves in this paper to qualitative
comparisons within the context of the halo model, we advocate the use
of multiscale density to quantitatively examine physically motivated
models (from simple halo occupation models to semi-analytically
populated haloes).  This provides a uniquely rich and observationally
motivated parameter space to examine the dependence of galaxy
properties on environment.


\section*{Acknowledgments}


Funding for the SDSS has been provided by the Alfred P. Sloan
Foundation, the Participating Institutions, the National Science
Foundation, the U.S. Department of Energy, the National Aeronautics
and Space Administration, the Japanese Monbukagakusho, the Max Planck
Society, and the Higher Education Funding Council for England. The
SDSS Web Site is http://www.sdss.org/.  The SDSS is managed by the
Astrophysical Research Consortium for the Participating
Institutions. The Participating Institutions are the American Museum
of Natural History, Astrophysical Institute Potsdam, University of
Basel, University of Cambridge, Case Western Reserve University,
University of Chicago, Drexel University, Fermilab, the Institute for
Advanced Study, the Japan Participation Group, Johns Hopkins
University, the Joint Institute for Nuclear Astrophysics, the Kavli
Institute for Particle Astrophysics and Cosmology, the Korean
Scientist Group, the Chinese Academy of Sciences (LAMOST), Los Alamos
National Laboratory, the Max-Planck-Institute for Astronomy (MPIA),
the Max-Planck-Institute for Astrophysics (MPA), New Mexico State
University, Ohio State University, University of Pittsburgh,
University of Portsmouth, Princeton University, the United States
Naval Observatory, and the University of Washington.
This research has made use of software provided by the US National Virtual Observatory, which is sponsored by the
National Science Foundation.
We thank the anonymous referee whose excellent report helped us in particular to improve the clarity of this paper.  
We also thank Stefanie Phleps for the useful discussion and to everyone who has commented on this work, with relevance not only to this, but also to future papers.


\appendix

\section{Poisson sampled neighbours in a correlated density field}\label{app:corrnoise}

The number of neighbours brighter than a given luminosity limit, and within $1000~\kms$ 
of a given galaxy, is clearly not going to provide the perfect measure of its environment, 
either in the sense of ``that which best correlates with the galaxy's properties" 
or in the sense of ``that which traces the underlying dark matter density field". 
In most senses this is unimportant: it merely provides a measurable tracer of environment at the position of 
each galaxy and is the best that we can do empirically, without assuming a halo model. 

\citet{Blanton06} has suggested that a Poisson-sampled galaxy field can lead to correlated 
noise in the small versus large scale density plane and therein to false correlations with large scale density. 
Effectively, sampling of rare, quantized tracers such as bright galaxies 
leads to large errors sampled from a Poisson distribution, 
particularly on the smallest scales and the lowest densities where these errors are largest. 
Assuming some perfect, underlying measure of environment exists, 
measurements with large errors might have residuals which 
correlates with the large scale density measurement, which has a higher signal to noise. 
In reality this is highly complex, relating to our observing galaxies at random positions along their orbits; 
projection effects; the correlated measurements of density for neighbouring galaxies 
and the less than $100\%$ deterministic nature of galaxy formation which allows a galaxy to enter our 
luminosity selected sample.
Modelling all these effects and their influence on the error vector in our density planes 
is well beyond the scope of this paper. 

Nonetheless, we do not believe that correlated noise significantly influences the departure from vertical 
contours, since we do have some plots with contours which appear almost vertical at all densities, as confirmed using the statistical test outlined in Appendix~\ref{app:signiflarge}. 
If correlated noise has any influence, it is most likely at small density, where $N/\sqrt(N)$ is smaller, 
and thus we shall treat any possible influence of large scales at small density with caution.

\section{Significance of parameter dependence on Large Scales}\label{app:signiflarge}

To test the significance of residual large scale dependencies, 
we setup a null hypothesis: that only smaller scale density matters. 
Under this assumption, the predicted influence of large scale density can be tested using 
bootstrap catalogues, resampled from the real one as follows: 
The galaxies in each bin of luminosity are ranked in order of their smaller scale density. 
For every galaxy in the sample, one of the twenty nearest in rank is selected to replace it, 
regardless of its large scale density. 
This effectively shuffles the large scale information whilst 
preserving the small scales (imposing vertical contours in the small versus large scale density plane).
The usual fitting analysis is applied to the resulting 
bootstrap catalogue, and the 
residual parameter set ($\dfr$, $\dmur$, $\dsigmar$, $\dmub$, $\dsigmab$) is computed at each point 
in the 2D density grid. 
Three thousand such catalogues are created, resulting in three thousand such parameter sets for comparison with the true parameter set. 


The Spearman Rank Correlation Coefficient, $\rho$ \citep{SpearmanTest} tests the strength of correlation 
between the rank of one variable and another, and therefore does not assume a linear correlation between the  
parameters themselves. 
Computing this coefficient for the correlation between each derived parameter and the large scale 
density for all bootstrap catalogues defines the expected 
distribution of $\rho$ in the case that the null hypothesis were true (i.e. no large scale dependence). 
Therefore the significance that additional dependence on large scales exists 
(on top of that induced by correlations with small scale density, averaged over the whole 2D density grid) 
can be quantified as the frequency of coefficients determined from the set of bootstrap 
catalogues which are more extreme than that computed for the real data. 
We define $\Pl$ as the fraction of bootstrap samples for which the real data 
coefficient is greater than the bootstrap one. 
A value of $\Pl$ close to one infers additional positive correlation on large scales, 
whilst values close to zero infer additional negative correlation. 

\end{document}